\documentclass{IEEEtran}

\usepackage[printonlyused, nolist]{acronym}
\usepackage{subfigure}
\usepackage{balance}
\usepackage{amsmath, bm} % packages for maths and bold symbols
\usepackage{cite}        % better citations
\usepackage{todonotes}
\usepackage{enumerate}
\usepackage{epstopdf}

\usepackage{booktabs}

\usepackage[crop=pdfcrop]{pstool}

\graphicspath{{figures/}}

\newcommand{\fs}{\footnotesize}   % font size 8 in SVG images

\pdfoutput=1

\title{Towards 1\,Gbps/UE in Cellular Systems: \\ Understanding Ultra-Dense Small Cell Deployments}
\author{
\IEEEauthorblockN{David L\'opez-P\'erez$^1$, Ming Ding$^2$, Holger Claussen$^1$, and Amir H. Jafari$^1$$^3$}\\
\IEEEauthorblockA{\textit{ $^1$ Bell Laboratories, Alcatel-Lucent, Republic of Ireland}\\
\textit{$^2$ National ICT Australia (NICTA), Australia}\\
\textit{$^3$ University of Sheffield, United Kingdom}}
}

\begin{document}
\maketitle
\begin{abstract}
% The future
Today's heterogeneous networks comprised of mostly macrocells and indoor small cells
will not be able to meet the upcoming traffic demands.
Indeed, it is forecasted that at least a 100$\times$ network capacity increase will be required to meet the traffic demands in 2020.
% three paradigms
As a result,
vendors and operators are now looking at using every tool at hand to improve network capacity.
In this epic campaign, three paradigms are noteworthy,
i.e., network densification, the use of higher frequency bands and spectral efficiency enhancement techniques.
This paper aims at bringing further common understanding
and analysing the potential gains and limitations of these three paradigms,
together with the impact of
idle mode capabilities at the small cells as well as the user equipment density and distribution in outdoor scenarios.
% focus on one of the three paradigms, i.e., network densification
Special attention is paid to network densification and its implications when transitioning to ultra-dense small cell deployments.
% let's revisit the following conclusions later.
Simulation results show that network densification with an average inter site distance of 35\,m can increase the cell-edge UE throughput by up to 48$\times$,
while the use of the 10\,GHz band with a 500\,MHz bandwidth can increase the network capacity up to 5$\times$.
The use of beamforming with up to 4 antennas per small cell base station lacks behind with cell-edge throughput gains of up to 1.49$\times$.
% energy efficiency
Our study also shows how network densifications reduces multi-user diversity,
and thus proportional fair alike schedulers start losing their advantages with respect to round robin ones.
The energy efficiency of these ultra-dense small cell deployments is also analysed,
indicating the need for energy harvesting approaches to make these deployments energy-efficient.
% top-ten challenges
Finally, the top ten challenges to be addressed to bring ultra-dense small cell deployments to reality are also discussed.

\end{abstract}
\section{Introduction}
\label{introduction}

\emph{``As old as I have become,
many developments my eyes have seen.
When I was a young man,
changes used to take place from time to time,
every now and then.
Nowadays, they occur every so often ...
life itself seems to change every day."}
-- a 97-years-old man commented,
making reference to the increasing pace in the development of communication systems and applications through the last century.
% [Ming]: It's poetic. I like it!
% [David]: Thanks :)
% [Holger]: Do you have the reference? - Answered: David's grandfather

These comments may seem exaggerated at first sight,
but they may not be so,
in light of the advancements seen in the telecommunication industry
since Graham Bell carried out the first successful bi-directional telephone transmission in 1876~\cite{GrahamBell}.
In this period of time,
society has witnessed long distance communications being freed from wires
and being operated through air at the speed of light;
wireless and mobile communications made available to over 6 billion users worldwide~\cite{MobiForge2014};
new types of communications and social interactions are emerging through the Internet and social networking~\cite{WWW}~\cite{SocialNetworks2013};
and many other breakthroughs that have certainly changed our every day lives.

These developments,
although of great importance to the 97-years-old man,
will probably appear as small steps towards a new era to future generations.
This new era of communications,
still in its first infancy,
will continue to change the world in unpredictable and fascinating manners.

Even though there is uncertainty on how such future advancements will look like,
it is expected that they follow the same trends as previous communication systems and technology breakthroughs,
and they require more and more capacity, bits per second (bps),
as time goes by.
Voice services~\cite{GSM} were the killer applications at the beginning of this century,
demanding tens of kbps per \ac{UE},
while high quality video streaming~\cite{Wu01streamingvideo} is the most popular one today,
needing tens of Mbps per \ac{UE}~\cite{1497859}.
Future services such as augmented reality, 3D visualisation and online gaming may use multiple displays requiring hundreds of Mbps each,
resulting in total sum of up to 1\,Gbps per \ac{UE},
and who knows what else tomorrow will bring?

\bigskip

In view of such significant future traffic demands,
the mobile industry has set its targets high,
and has decided to improve the capacity of today's networks
by a factor of 100$\times$ or more over the next 20 years
-- 1000$\times$ the most ambitious~\cite{qualcom2012}.

%Figure of the triangle
\begin{figure}[t]
  \centering
  \def\svgwidth{8.1cm}
  
\immediate\write18{inkscape -z -D --file=figures/triangle.svg
--export-pdf=figures/triangle.pdf --export-latex}
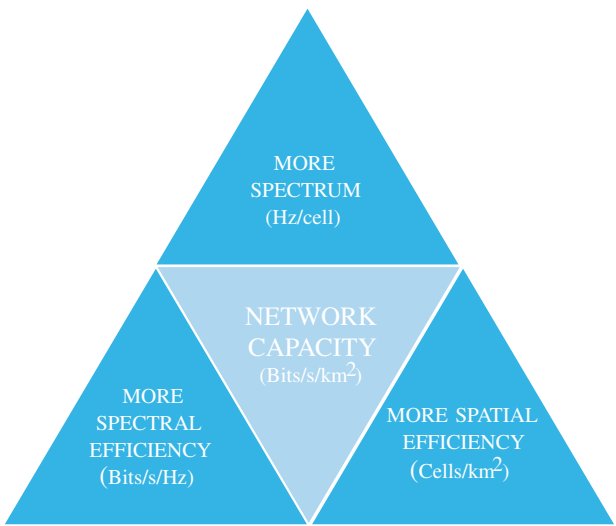

  \caption{Existing paradigms to improve network capacity: more spectrum, more spectral efficiency and more spatial efficiency.}
  \label{fig:the_triangle}
  \vspace{-0.5cm}
\end{figure}
% [Ming]: In Fig.1, how about changing "WIRELESS CAPACITY" to "NETWORK CAPACITY" and changing the units of the three factors to bps/Hz, Hz/cell and cells/km^2 so that the product of the three factors becomes xxx\,bps/km^2, which might make our description easier to understand.
% [David]: Done.
% [Ming]: Wonderful! It would be nice if the "2" in "km2" could be shown as a superscript.
% [David]: Done.
% [Holger]: change Sec -> s ?
% [David]: Done.

In order to achieve this goal,
vendors and operators are currently looking at using every tool they have at hand,
where the existing tools can be classified within the following three paradigms as illustrated in Fig.~\ref{fig:the_triangle}:
\begin{itemize}
\item
Enhance spatial reuse
through network densification,
i.e., \acp{HetNet} and small cells~\cite{HetNetbook}~\cite{Claussen2008}~\cite{Chandrasekhar2008a}~\cite{Andrews:12a}.
\item
Use of larger bandwidths,
exploiting higher carrier frequencies,
both in licensed and unlicensed spectrum~\cite{TR36808}~\cite{TR36889}~\cite{mmWavebook}.
% [Ming]: ~\cite{LTE-U}: ETSI MCC, "Draft Report of 3GPP workshop on LTE in unlicensed spectrum," 3GPP workshop on LTE in unlicensed spectrum, Sophia Antipolis, France, June, 2014.
%[David]: Citations added.
% [Ming]: Emphasise unlicensed carrier frequencies for the sake of the following discussion.
%[David]: Done.
\item
Enhance spectral efficiency
through multi-antenna transmissions~\cite{MIMObook}, cooperative communications~\cite{CooperativeCommunicationsbook}, dynamic TDD techniques~\cite{6353682}~\cite{DavidLopez2014homoDynamicTdd}~\cite{DavidLopez2014hetnetDynamicTdd}, etc.
%i.e., beamforming and spatial multiplexing.
\end{itemize}
% [Ming]: Add more techniques of enhancing spectral efficiency.
% [David]: Done.

% [Ming]: Regarding references [16], [17], [18], [52], [93], please correct my name from D. Ming to M. Ding, thanks!
% [David]: Done

This toolbox provides vendors and operators with choice and flexibility to improve network capacity,
but leads to new enigmas on how each operator should make the investment to meet the capacity demands.
Different operators with different financial means, customer market segmentation, existing network assets and technical expertise may require different solutions.
Network densification complicates network deployment as well as backhauling and mobility management,
while higher carrier frequencies suffer from larger path losses,
and usually require more expensive equipment.
Most spectral efficiency enhancement technologies depend on a tight synchronisation as well as relatively complex signal processing capabilities,
and may be compromised due to inaccuracies in \ac{CSI}.
\ac{CAPEX} and \ac{OPEX} are also major concerns since the price per bit should be kept at minimum.
To make things more complex,
network deployments strategies should consider that these three paradigms have their own fundamental limitations and thus cannot be infinitely exploited.
Finding the right portfolio of tools to meet the key requirements is vital for both vendors and operators.

In this light, this paper aims at bringing further common understanding and
analysing the potential gains and limitations of network densification,
in combination with the use of higher frequency bands and spectral efficiency enhancement techniques.
The impact of idle mode capabilities at the small cells as well as \ac{UE} density and distribution are also studied
since they are foreseen to have a large impact~\cite{IdleMode2010}.
The objective of this paper is to shed new light on the Pareto set of network configurations in terms of
small cell density, frequency band of operation and number of antennas per small cell \ac{BS} that can meet future traffic demands,
achieving an average throughput exceeding 1\,Gbps per \ac{UE}.

The rest of the paper is organised as follows:
Section~\ref{sec:hetnet} addresses the important roles of small cells,
and depicts our view on future small cell deployments~\acp{HetNet}.
Section~\ref{sec:todaysSmallCells} presents the state of the art of current small cell technologies together with its drawbacks,
and motivates the need for uncoordinated dense small cell deployments.
Section~\ref{sec:systemModel} presents the system model to be used in this paper to analyse ultra-dense small cell deployments.
Sections~\ref{sec:densification},~\ref{Sec:higherFrequecnyBands} and~\ref{Sec:MIMO} elaborate on the three paradigms to enhance network performance,
i.e.,  network densification together with the usage of higher carrier frequencies and multi-antenna transmissions, respectively.
Section~\ref{Sec:scheduling} investigates the impact of network densification in small cell \ac{BS} schedulers.
Section~\ref{Sec:costAndEnergyEfficiency} studies the impact of network densification on the energy efficiency.
Section~\ref{sec:Differences} highlights  the main differences between regular HetNets and  ultra-dense HetNets,
while Section~\ref{sec:Challenges} summarises the challenges in ultra-dense small cell deployments.
Finally, Section~\ref{sec:conclusion} draws the conclusions.
% [David]: Finish later
% [Ming]: I have provided a draft. Please check the above paragraph.
% [David]: Thanks!

\section{Small Cells in \acp{HetNet}}
\label{sec:hetnet}

Given the different approaches to enhance network capacity,
it may be worth understanding how network capacity has been improved in the past and which have been the lessons learnt
to make sure the best choices are taken.
To this end, Prof. Webb analysed the different methods used to enhance network capacity from 1950 to 2000~\cite{Webb2007}.
According to his study, the wireless capacity has increased around a 1 million fold in 50 years.
The breaking down of these gains is as follows:
15$\times$ improvement was achieved from a wider spectrum,
5$\times$ improvement from better \ac{MAC} and modulation schemes,
5$\times$ improvement by designing better coding techniques,
and an astounding  2700$\times$ gain through network densification and reduced cell sizes.
According to this data,
it seems obvious that if we are looking for a 1000$\times$ improvement in network performance,
network densification through ultra-dense small cell deployments is the most appealing approach,
and today's networks have already started going down this path.

In order to meet the exponentially increasing traffic demands~\cite{Cisco_TrafficUpdate_2015},
mobile operators are already evolving their networks from the traditional macrocellular-only networks to \acp{HetNet}~\cite{5876496, Lopez_perez2011HetNet},
in which small cells reuse the spectrum locally and provide most of the capacity
while macrocells provide a blanket coverage for mobile \acp{UE}.
Currently, small cells are deployed in large numbers.
Indeed, according to recent surveys, in 2012,
the number of small cell \ac{BS} was already larger than that of macrocell \acp{BS}~\cite{SmallCellForum2012}.
These small cell deployments are mainly in the form of home small cells,
known as femtocells~\cite{Chandrasekhar2008a, Claussen:08a, Lopez-Perez2009a},
but many operators have also already started to deploy outdoor small cell solutions to complement their macrocellular coverage~\cite{6166483}.

\begin{figure*}[t!]
  \centering
	\includegraphics[width=3.5in]{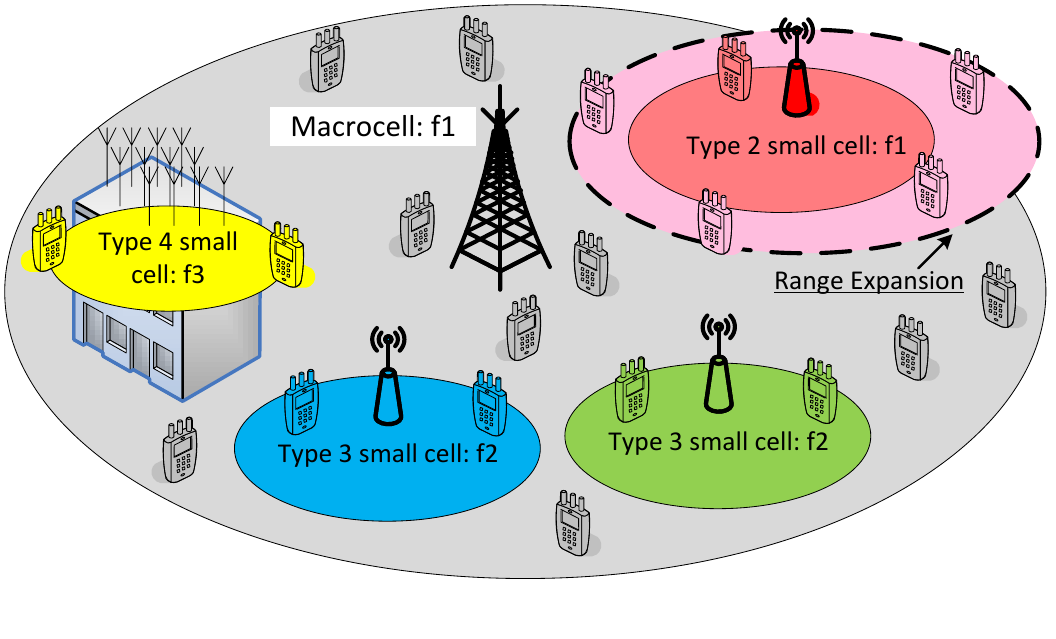}
  	\caption{Cell tier types.}
  	\label{fig:cell_tier_types}
\end{figure*}

\begin{table*}
\begin{centering}
{\footnotesize \caption{\label{tab:cell_tier_types} Cell tier types and their characteristics.}}
\vspace{-0.1cm}
\par\end{centering}{\footnotesize \par}
\centering{}{\footnotesize }%
\scalebox{0.93}{
\begin{tabular}{|l|l|l|l|l|}
\hline
Cell Type No. & Cell Type & Spectrum & Relationship with the Macrocell Tier & Typical Use Case\tabularnewline
\hline
\hline
1 & low-frequency macrocell tier & around 1$\sim$2GHz, licensed & - & umbrella coverage\tabularnewline
\hline
2 & low-frequency small cell tier & around 1$\sim$2GHz, licensed & co-channel deployment, CRE \& ABS & capacity enhancement in hotspots\tabularnewline
\hline
3 & mid-frequency small cell tier & around 5GHz, unlicensed & non-co-channel deployment, dual connectivity & high traffic offloading\tabularnewline
\hline
4 & high-frequency small cell tier & $>$10GHz, unlicensed & non-co-channel deployment, dual connectivity & very high traffic offloading\tabularnewline
\hline
\end{tabular}}
\vspace{-0.1cm}
\end{table*}

Most of the existing small cell deployments,
particularly femtocells,
are configured to transmit on a dedicated carrier different from that of the macrocells~\cite{Claussen:08a}.
While this avoids inter-tier interference,
it also limits the available radio spectrum that each cell can access,
and is less efficient than co-channel deployments,
in which small cells and macrocells share the same frequency bands~\cite{Hobby:09a}.
However, while co-channel operation provides better frequency utilisation,
the additional inter-tier interference can result in coverage and handover issues for mobile \acp{UE}~\cite{Lopez-Perez2010a}~\cite{6384454}.
This interference issue is particularly severe in femtocell deployments with \ac{CSG} access,
in which \acp{UE} cannot connect to the strongest cell;
the later thus  becoming a strong interferer~\cite{Lopez-Perez2010a}.

In order to take advantage of the benefits of both orthogonal and co-channel network deployments
-- interference mitigation and spectrum reuse, respectively --
and further enhance the capacity of future networks,
we anticipate that future networks will be comprised of different small cell tiers with different types of small cell \acp{BS}.
These different types of small cell \acp{BS} will be targeted at different types of environments and traffic.
Fig.~\ref{fig:cell_tier_types} and Table~\ref{tab:cell_tier_types} summarises our classification of future network tiers.
Note, that to make the framework more complete, we treat the macrocell tier as a special case of the small cell tiers  in~Table~\ref{tab:cell_tier_types}.

%[David]: To be filled by Ming with his 4 types of small cell BS. One table would be good.
%The low-frequency macrocell tier (for umbrella coverage )
%The low-frequency small cell tier, co-channel deployment with the macrocell tier (for capacity enhancement in hotspots)
%The mid-frequency small cell tier, non-co-channel deployment with the macrocell tier (for high traffics in specific locations)
%The high-frequency small cell tier, non-co-channel deployment with the macrocell tier (for very high traffics and specific \acp{UE})
%[Ming]: I add a table in the following. Please check it.
%[David]: As in your slides, it would be nice to add a bit more information and insert some explanations into the text.
%[Ming]: Done. Please check the added text.
%[David]: Wonderful!

\emph{Cell Type~1} is essentially the conventional macrocell tier that provides an umbrella coverage for the network.
This topic will not be covered in this paper since it is out of our scope.
However, it is important to mention that a noteworthy enhancement for Cell Type~1 that is currently being investigated is the \ac{3D} \ac{MU} \ac{MIMO} transmission,
which is expected to enhance the indoor penetration and spectral efficiency of macrocells, especially for the scenario with high-rise buildings~\cite{6525612}~\cite{6966194}.
%[David]: Commenting on remote radio heads and V-RAN?
%[Ming]: I commented a little bit on the high rise scenario.
%[David]: Thanks

\emph{Cell Type~2} features today's state of the art co-channel deployments of small cells with the macrocell tier.
Cell Type~2 is conceived as an add-on to Cell Type~1 for capacity enhancement through cell splitting gains in hotspots.
\ac{LTE} Release 10 features such as \ac{CRS} and \ac{eICIC} are critical for this small cell type
to provide efficient macrocell off-loading and cope with the inter-tier interference issue~\cite{Lopez_perez2011HetNet}.
Besides, advanced receivers with interference cancellation capabilities are an enhancement for small cell \acp{UE} to remove the residual interference from CRS~\cite{6692396}. Moreover, multi-cell cooperation among nodes with different power levels has been proved to be beneficial in recent works~\cite{6692101}~\cite{6965874}.
The eICIC and another important characteristics such as \ac{CRE} and \ac{ABS} of Cell Type~2 will be introduced in the following section,
together with its drawbacks and need for new cell types.
% Which characteristics
%[David]: This opens up the opportunity to review CRE and ABS and motivates the needs for Cell Type~3
%[Ming]: Agreed.

\emph{Cell Type~3} features dense orthogonal deployments of small cells with the macrocell tier,
which are  envisaged to be the workhorse for network capacity boosting through extensive spatial reuse in the near future.
Due to its small size,
Cell Type~3 is not appropriate to support mobile \acp{UE} and is targeted at static \acp{UE},
which represent a vast majority of the \ac{UE} population
with more than 80\,\% of today's data traffic carried indoors~\cite{Sorrells2014}.
In order to ensure a smooth inter-working between Cell Type~3 and Cell Type 1,
\ac{DC} is a promising technology currently being investigated in the LTE framework~\cite{LTE-R12, 6477646, 6515050, 6825019},
where a given \ac{UE} may use radio resources provided by at least two different network points (Master and Secondary \acp{BS}) connected with non-ideal backhaul.
The typical usage of DC is the splitting of traffic flows~\cite{RP132069}.
In more detail, it is beneficial to let macrocells provide the voice service for a UE and outsource the UE's data service to small cells. In addition, DC also improves the robustness of the mobility management since UEs are now connected to two cell tiers~\cite{RP132069}.
%[Holger]: add Anna's/your paper as reference?
%[David]: added
Note that \ac{DC} is not suitable for Cell Type~2 because its compatibility with \ac{eICIC} is challenging.
%[David]: Ming to comment on this
%[Ming]: Done. Please check the explanation below. 
%[David]: Fantastic.
In essence, eICIC tries to make macrocells invisible to small cell UEs in certain subframes using the ABS mechanism to eliminate the inter-tier interference. 
On the other hand, DC tries to maintain both connections, one to the macrocells and one to the small cells. 
Hence, eICIC is mainly used in the co-channel deployment where the macrocell to small cell interference is a major issue, 
while DC is mainly suitable for the orthogonal deployment where the traffic flow splitting or the mobility management is a major issue.
Other new emerging technologies in Cell Type~3 include dynamic small cell idle modes~\cite{IdleMode2010}~\cite{5978418}~\cite{Razavi:12a}, dynamic \ac{TDD} transmission~\cite{6353682}~\cite{DavidLopez2014homoDynamicTdd}~\cite{DavidLopez2014hetnetDynamicTdd} (which have attracted a lot of momentum in both the academia and the industry in the last years), high-order MIMO techniques~\cite{6692392}~\cite{6965928}, etc.
%[David]: Ming to provide reference for high-order MIMO techniques
%[Ming]: Added.
%[David]: Thanks.

\emph{Cell Type~4}, finally, will push the technology frontier even further by using a very wide spectrum in high-frequency bands,
e.g., mmWave~\cite{6894456}~\cite{6515173}~\cite{6824746}, %~\cite{MicJournal},
and exploiting massive MIMO techniques to achieve large beam forming and spatial multiplexing~\cite{6375940}~\cite{6736761}, etc.

%%%FIGURE Offloading
\begin{figure}[t]
    \centering
    \def\svgwidth{3.5in}
\immediate\write18{inkscape -z -D --file=figures/offloading.svg
--export-pdf=figures/offloading.pdf --export-latex}
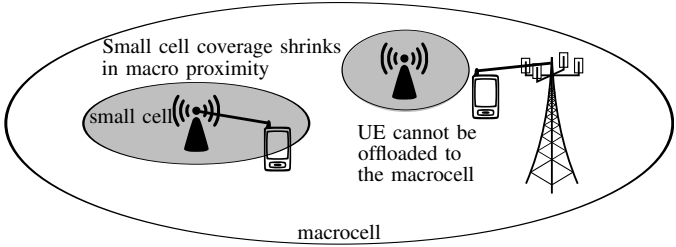

    \caption{Small cell coverage shrinks in the proximity of a macrocell BS leading to poor offloading.}
    \label{fig::offloading}
\end{figure}

%[Holger]: there is a typo in the figure "iin"
%[David]: fixed 

%%%FIGURE UL interference
\begin{figure}
    \centering
    \def\svgwidth{3.5in}
\immediate\write18{inkscape -z -D --file=figures/ulInterference.svg
--export-pdf=figures/ulInterference.pdf --export-latex}
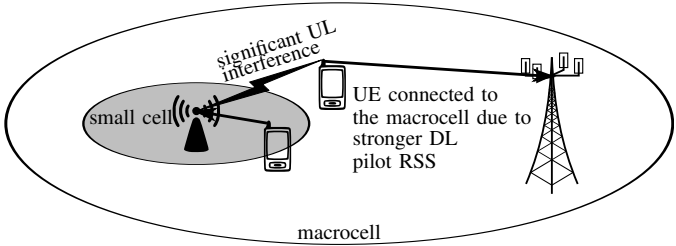

    \caption{Macrocell UE jamming the UL of a nearby small cell.}
    \label{fig::ulInterf}
\end{figure}

\section{Why Are Today's Small Cells Not Practical to Meet Future Capacity Demands?}
\label{sec:todaysSmallCells}

In contrast to \ac{CSG},
open access helps to minimise inter-tier interference since \ac{UE} are always allowed to connect to the strongest cell,
thus avoiding the \ac{CSG} interference issue~\cite{Lopez-Perez2010a}. %access control mechanisms
However, in a co-channel deployment of small cells with the macrocell tier,
being attached to the cell that provides the strongest pilot \ac{RSS} may not be the best strategy
since \acp{UE} will tend to connect to macrocells rather than to small cells,
even if they are at a shortest path loss distance.
This is due to the large difference in transmission power between both types of \acp{BS}.
As shown in Fig.~\ref{fig::offloading},
the closer the small cell \ac{BS} is to the macrocell \ac{BS},
the smaller the small cell coverage is
due to macrocell \ac{BS} power dominance.
This results in a poor macrocell off-load~\cite{5876496}~\cite{Lopez_perez2011HetNet}. %surveys on HetNet issues and features
Moreover, due to this server selection procedure based on pilot \ac{RSS},
\acp{UE} connected to macrocells will also severely interfere with all small cells located in their vicinity in the UL.
This is shown in Fig.~\ref{fig::ulInterf}.
Note that due to the lower path loss,
if a macrocell \ac{UE} would connect to the small cell with the smallest path loss,
this \ac{UE} would transmit with a much lower \ac{UL} power.
This would allow load balancing as well as \ac{UL} interference mitigation,
thus improving network performance.

\begin{figure}[t]
    \centering
    \def\svgwidth{3.5in}
\immediate\write18{inkscape -z -D --file=figures/rangeExpansion.svg
--export-pdf=figures/rangeExpansion.pdf --export-latex}
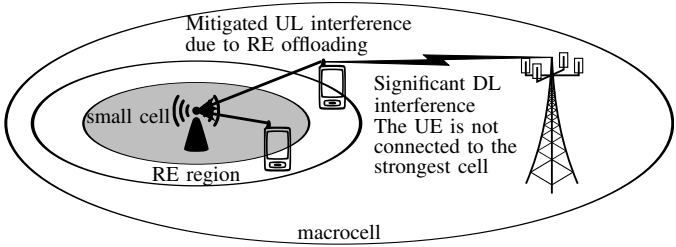

    \caption{Range expansion mitigates UL interference and facilitates offloading at the expense of increasing DL interference.}
    \label{fig::rangeExpansion}
\end{figure}

In order to address these problems arising from the significant power difference between co-channel \acp{BS} in \acp{HetNet},
new cell selection methods that allow \ac{UE} association with cells that do not necessarily provide the strongest pilot \ac{RSS} are necessary.
In this regard,
\ac{CRE} has been proposed in the \ac{3GPP} for increasing the \ac{DL} coverage footprint of small cells
by adding a positive cell individual offset to the pilot \ac{RSS} of the small cells during the serving cell selection procedure~\cite{R1-100701_Qualcomm}~\cite{6166483}. % range expansion
\ac{CRE} mitigates UL interference and facilitates offloading.
With a larger \ac{REB},
more \acp{UE} are offloaded to the small cells,
at the cost of increased co-channel \ac{DL} interference for the range-expanded \acp{UE} (see Fig.~\ref{fig::rangeExpansion}).
Without \ac{ICIC},
range-expanded \acp{UE} are not connected anymore to the strongest cell,
\ac{CRE} has been shown to degrade the throughput of the overall network but improve the sum capacity of the macrocell \acp{UE} due to offloading.
In~\cite{Andrews_Globecom2011_RE}, %drawback of range expansion
closed form analytical expressions of outage probability with \ac{CRE} in \acp{HetNet} corroborate that
\ac{CRE} without \ac{ICIC} degrades the outage probability of the overall network.

In order to address this issue,
the use of \ac{eICIC} schemes has been proposed to guarantee the proper operation of \ac{CRE}~\cite{5876496}~\cite{Lopez_perez2011HetNet}. %surveys on HetNet issues and features
In such schemes,
special attention is given to the mitigation of inter-cell interference in the control channels transmitted in the \ac{DL}.
\acp{UE} may declare radio link failure under severe interference,
and experience service outage due to the unreliable control channels.

\begin{figure}[t]
	\begin{center}
	\def\svgwidth{3.5in}
\immediate\write18{inkscape -z -D --file=figures/pico_abs_concept.svg
--export-pdf=figures/pico_abs_concept.pdf --export-latex}
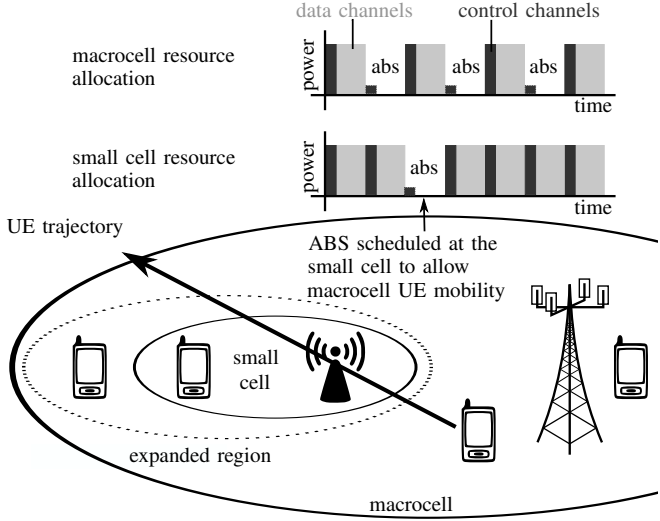

	\caption{ABS concept. ABS scheduled at the macro BS to helps pico range-expanded UEs. ABS scheduled at small cells to help mobile macro UEs.}
	\label{fig:abs_concept}
	\end{center}
\end{figure}

Among the proposed \ac{eICIC} schemes,
time-domain eICIC methods have received a lot of attention,
particularly \acp{ABS}~\cite{TS36133}.
In an \ac{ABS}, no control or data signals but only reference signals are transmitted,
thus significantly mitigating interference since reference signals only occupy a very limited portion of the whole subframe.
As shown in Fig.~\ref{fig:abs_concept},
\acp{ABS} can be used to mitigate interference problems in open access small cells that implement \ac{CRE}.
A macrocell can schedule \acp{ABS}
while small cells can schedule its range-expanded small cell \acp{UE} within the subframes that are overlapping with the macrocell \acp{ABS}.
\acp{ABS} can also be used to mitigate interference problems in \ac{CSG} small cells.
\ac{CSG} small cells can schedule \acp{ABS}
while macrocells can schedule their victim macrocell \acp{UE} located nearby a \ac{CSG} small cell within the subframes that are overlapping with the small cell \acp{ABS}.
Moreover, \acp{ABS} can be scheduled at the small cell \ac{BS} to allow fast moving macrocell \acp{UE} to move through it.
In order to minimise the capacity loss at the \acp{BS} scheduling \acp{ABS},
the agreement in~\cite{R2-111701} generalises \acp{ABS} from completely blank data and control symbols
to the transmission of such symbols at a reduced-power level,
implying that the aggressing \ac{BS} may also be able to  transmit information.
In~\cite{IGuvenc2011}, the network performance improvement due to \ac{CRE} and \acp{ABS} was analysed showing the merits of \acp{ABS}.
Moreover, analytic expressions for average capacity and 5th percentile throughput where derived in ~\cite{Merwaday2014}, while considering \acp{ABS} and power reduced subframes
as a function of \ac{BS} densities, transmit powers and interference coordination parameters in a two-tier HetNet scenario. The results confirm the benefits of power reduced subframes over \acp{ABS}.

%%%FIGURE Range expansion

%[Holger]: place font labels differently to not cover the macrocell in the figure - hard to read
%[David]: fixed

%%%FIGURE

In order to enable efficient interference mitigation,
however,
\ac{REB} and \ac{ABS} schemes (reduced power subframes also included in this category) should be able to dynamically adapt to the number of small cells at various geographical locations and different traffic conditions.
The larger the \ac{REB} for a given small cell \ac{BS},
the more expanded-region \acp{PUE} connect to it,
and thus a larger macrocell  \ac{ABS} duty cycle (ratio of ABS to non-ABSs) is needed to provide a desired quality of service to these expanded-region \acp{PUE}.
In contrast, the larger the \ac{ABS} duty cycle,
the lower the macrocell performance may be due to subframe blanking.
A dynamic adaptation of \ac{ABS} patterns can be realised through X2 backhaul inter-BS coordination,
where neighbouring macrocell \acp{BS} agree on a given \ac{ABS} pattern, 
and then  each macrocell \ac{BS} informs its overlaid small cells \acp{BS} of which subframes it will use for scheduling \acp{MUE} and which ones will be blanked for interference mitigation.
%[Amir]:Can you rewrite the above sentence please.
%[David]: Done
In~\cite{Deb2013}, a mathematical framework to efficiently compute \ac{UE} association and corresponding \acp{REB} together with \ac{ABS} duty cycles is proposed.
The solution is provably within a constant factor of the optimal solution,
scales linearly with the number of cells
and is also amenable to distributed optimisation.
However, it requires extensive input information from each \ac{UE} with respect to its best macrocell and small cell \acp{BS},
which may be difficult to obtain since a \ac{UE} cannot be simultaneously connected to multiple cells.

In addition, for an efficient interference mitigation,
inter-macrocell \ac{BS} coordination should be considered~\cite{6692277}.
If neighbouring macrocells do not coordinate their \ac{ABS} patterns and they are not fully time aligned,
this will cause additional interference fluctuations in the network,
resulting in less efficient link adaptation and radio-aware packet scheduling.

From the previous discussion,
it can be derived that
a joint optimisation of \acp{REB}, \ac{ABS} patterns, power reduction factors in reduced power subframes, scheduling thresholds and frequent inter-\ac{BS} coordination may be required to achieve a good performance in a co-channel deployment of small cells with the macrocell tier.
The complexity of these optimisation procedures will be aggravated with the number of cells,
and thus we anticipate that  Cell Type~2 is not suitable for ultra-small cell deployments,
where network planning should be completely avoided.
This suggests that co-channel deployments of small cells with the macrocell tier should be aimed at hotspot locations with a reasonable low number of small cells (less than 10 per macrocell sector),
while ultra-dense deployments should use dedicated carriers to circumvent interference problems and the planning stage.
This gives rise to Cell Type~3, the focus of this paper.

\section{System Model}
\label{sec:systemModel}

Due to the insufficient capacity provided by Cell Type~1 and~2 to meet the forecasted mobile traffic demands,
in this paper, our focus is on Cell Type~3,
i.e., non-co-channel mid-frequency small cell deployments with the macrocell tier.
This cell type has the potential to significantly enhance network performance through a high network densification and the usage of relatively high frequency bands,
while avoiding interference and coordination issues with the macrocell tier,
designed to support fast moving \acp{UE}.
The relatively small size of antennas at mid-frequency bands around 10\,GHz is also an appealing feature
in order to exploit multi-antenna transmission technique gains.
%[Ming]: The following paragraphs set the scene for our paper. I agree with the current approach of only considering key practical factors and adopting ideal assumptions for other factors. Please note that my following suggestions are based on this basic principle.
%[David]:Agreed.

In the following, our system model to analyse ultra-dense small cell networks is introduced.

Consider a dense network of small cells.
In this system model,
a 500-by-500\,m scenario is used,
and small cells are placed outdoors in a uniform hexagonal grid with different \acp{ISD} of 200, 150, 100, 75, 50, 35, 20, 10 or 5\,m,
which result in 29, 52, 116, 206, 462, 943, 2887, 11548 or 46189 small cell \acp{BS} per square km deployed in the scenario, respectively.

Two scenarios for \ac{UE} distribution are considered,
with three different \ac{UE} densities of 600, 300 or 100 active \acp{UE} per square km:
% [Ming]: It would be great if we indicate how many UEs and how many cells are deployed in the considered scenario. I suppose the UE numbers should be 150, 75, and 25. And I guess the cell numbers are roughly 18, 36, ..., 2500, 10000?
%[David]: Done.
\begin{itemize}
\item
Uniform: \acp{UE} are uniformly distributed within the scenario.
\item
Non-uniform: Half of the \acp{UE} are uniformly distributed within the scenario,
while the other half are uniformly distributed within circular hot spots of 40\,m radius with 20 \acp{UE} each.
Hot spots are uniformly distributed, and the minimum distance between two hotspot centres is 40\,m.
\end{itemize}
Note that 300 active \acp{UE} per square km is the density usually considered in dense urban scenarios, such as Manhattan~\cite{819497}.

In terms of frequency bands,
four carrier frequencies are considered, i.e., 2.0, 3.5, 5.0 and 10\,GHz,
where the available bandwidth is 5\,\% of the carrier frequency,
i.e., 100, 175, 250 and 500\,MHz, respectively.
% [Ming]: I suppose a more realistic percentage might be 10% considering that at most eight 20MHz carriers can be supported for each LTE-A UE and LTE-U is now coveting the 25 Wi-Fi channels (20MHz each) in 5GHz. But I'm OK with the assumption of 5% because this value is not a big deal.
% [David]: In today's LTE-A, at 2GHz, we have up to 5x20MHz=100MHz, which is 5% of the carrier frequency. Moreover, I think it will be hard that an operator benefits from the 25 WiFi channels in the 5GHz band for itself, since other operators and devices will operate in the ISM band. Let's keep 5%.
% [Ming]: Agreed. Just a further note. The bandwidth from the UE's perspective is at most 100MHz in LTE. Having said that, the bandwidth from the operator's perspective could be much larger than that, depending on the investment on spectrums.
% [David]: Agreed. Since we were targeting the study of 1 UE per cell, then I thought that 100MHz should be bandwidth at 2GHz.
%[Ming]: Agreed.

In terms of antennas implementation and operation,
each small cell \ac{BS} has 1, 2, or 4 antennas deployed in a horizontal array (see Fig.~\ref{fig:dipole}),
while the \ac{UE} has only 1 antenna.
% [Ming]: I guess 8 Tx antennas and non-codebook based precoding might be the future, considering the 8-layer DMRS-based transmission in LTE-A and the 8-layer MIMO solely based on the explicit CSI feedback in 802.11ac. I'm fine with the current assumptions by the way. We can enhance the simulations by considering ideal MRT later.
% [David]: We had this discussion and we are not clear whether 8 antennas will be used in the near future at small cells. It seems ok for macrocells today but not for small cells according to our RF guys.
%[Ming]: Agreed.
% [David]: With 4 antennas, there are 16 possible antennas patterns. We select the antenna pattern that maximises the UE received signal strength. This is some sort of ideal MRT, isn't it? I am not sure here.
%[Ming]: I suppose the codebook based MRT is a quantized beamformer from the ideal MRT. In other words, a quantized MRT beamformer is selected from a codebook while an ideal MRT is calculated from ideal CSI. This is my understanding and we can continue this discussion.
%[David]: This is my understanding too. I have updated the text.
%[Ming]: Good! Agreed!
The standardised \ac{LTE} code book beamforming is adopted,
targeted in this case at maximising the received signal strength of the intended \ac{UE}
(quantised \ac{MRT} beamforming with no inter-\ac{BS} coordination required).
It is important to note that the power per antenna remains constant 
and that  beamforming is only applied to the data channels and not to the control channels,
which define the small cell coverage,
% [Ming]: The assumption of single-antenna UEs should be mentioned here. Besides, have we considered MU-MIMO in the simulations?
% [David]: Done. Spatial multiplexing considering SU-MIMO as well as MU-MIMO is an extension of the work that I want to do with Amir and I would be glad to carefully discuss this with you because there are some interesting issues here. Indeed, I want to analyse the issue that you mention below: smaller cell sizes results in higher LOS and then spatial multiplexing suffers. There is a discussion later.
% [Ming]: Agreed. We don't need to touch this issue in the paper.

The transmit power of each active \ac{BS} is configured such that it provides a \ac{SNR} of 9, 12 or 15\,dB at the targeted coverage range,
which is $\frac{\sqrt{3}}{2}$ of the \ac{ISD}~\cite{Claussen:08a}.
%[Ming]: Nice touch on the DL power setting!!! Just FYI, in LTE and Wi-Fi, the parameter is set to a much larger value due to implementation considerations, such as poor receiver structures and/or decoding algorithms.
%[David]: We had a discussion on this and since we are not simulating indoor UEs, we decided to use the selected values. Let's keep them.
%[Ming]: Agreed.

\acp{UE} are not deployed within a 0.5\,m range of any \ac{BS},
and all \acp{UE} are served by the \ac{BS} from which they can receive the strongest received pilot signal strength,
provided that the pilot \ac{SINR} is larger than -6.5\,dB.

A \ac{BS} with no associated \ac{UE} is switched off.
This is the idle mode capability presented in~\cite{Ashraf:10a},
and adopted in this paper.

With regard to scheduling,
we envision that because of the use of mid- to high-frequency bands and due to the importance of the \ac{LOS} component in small cells,
the time spread of the \ac{CIR} will be very small,
typically in the order of several $\mu$s,
and hence the impact of multi-path fading will become less significant in the future.
Therefore, multi-path fading is not considered in our analysis,
and a round robin scheduler designed for simple implementation is adopted,
This argument is supported by the analysis in Section~\ref{Sec:scheduling},
where the impact of network densification in small cell \ac{BS} schedulers is investigated~\cite{2015Jafari}.
%[Ming]: Should we mention here that multi-path fading is not considered in this paper? Let me provide a justification for this simplification. "Consider the mid/high-frequency carriers and the importance of LoS connections in the future, the time spread of the \ac{CIR} will be very small, typically in the order of several us, and hence the impact of multi-path fading will become less significant in the future." Please note that in some cases multi-path fading is in fact beneficial to the system and thus the large CDD technology (that deliberately creates multi-path fading) has been adopted in both Wi-Fi and LTE. But I suppose we don't need to mention the CDD thing in this paper.
%[David]: Done. Agreed.

For details on path loss, antenna gain, shadow fading, \ac{SINR} computation and capacity mapping,
please refer to Appendix~\ref{sec:appendix}.
The modelling of multi-path fast fading for the analysis in Section~\ref{Sec:scheduling} is presented in Appendix~\ref{sec:appendix2}.
% [Ming]: Please go to Appendix A to see my further suggestions and comments on the system model.

For the sake of clarity,
it is also important to mention that in the legend of the figures of this paper,
\emph{i} indicates the \ac{ISD} of the scenario in meter,
\emph{d} indicates the density of \acp{UE} per square km,
\emph{ud} indicates the UE distribution ($ud$ = 0 uniform; $ud$ = 1 non-uniform)
\emph{s} indicates whether the idle mode capability of the small cell \ac{BS} is activated or deactivated,
\emph{sm} indicates the idle mode index of the idle mode used in the energy efficiency analysis,
\emph{f} indicates the carrier frequency in GHz
\emph{a} indicates the number of antennas, and
\emph{t} indicates the \ac{SNR} target of the small cell \ac{BS} at its cell-edge,
which is located at $\frac{\sqrt{3}}{2}$ of the \ac{ISD}.
%[Holger] u - uniform/non-uniform user distribution missing - conflict with user index u in Eq. 1
%r_cellWithAtLeastaUE_ueDistribution-0-freq-2-smallcellOnOff-1-smallCellAnts-1_Axis_logx
%[David]: changed to $ud$

\begin{figure*}[t!]
  \centering
	\subfigure[Average number of active \acp{BS} in the network per square km. ]{\label{fig:activeCells}
	\includegraphics[width=3.5in]{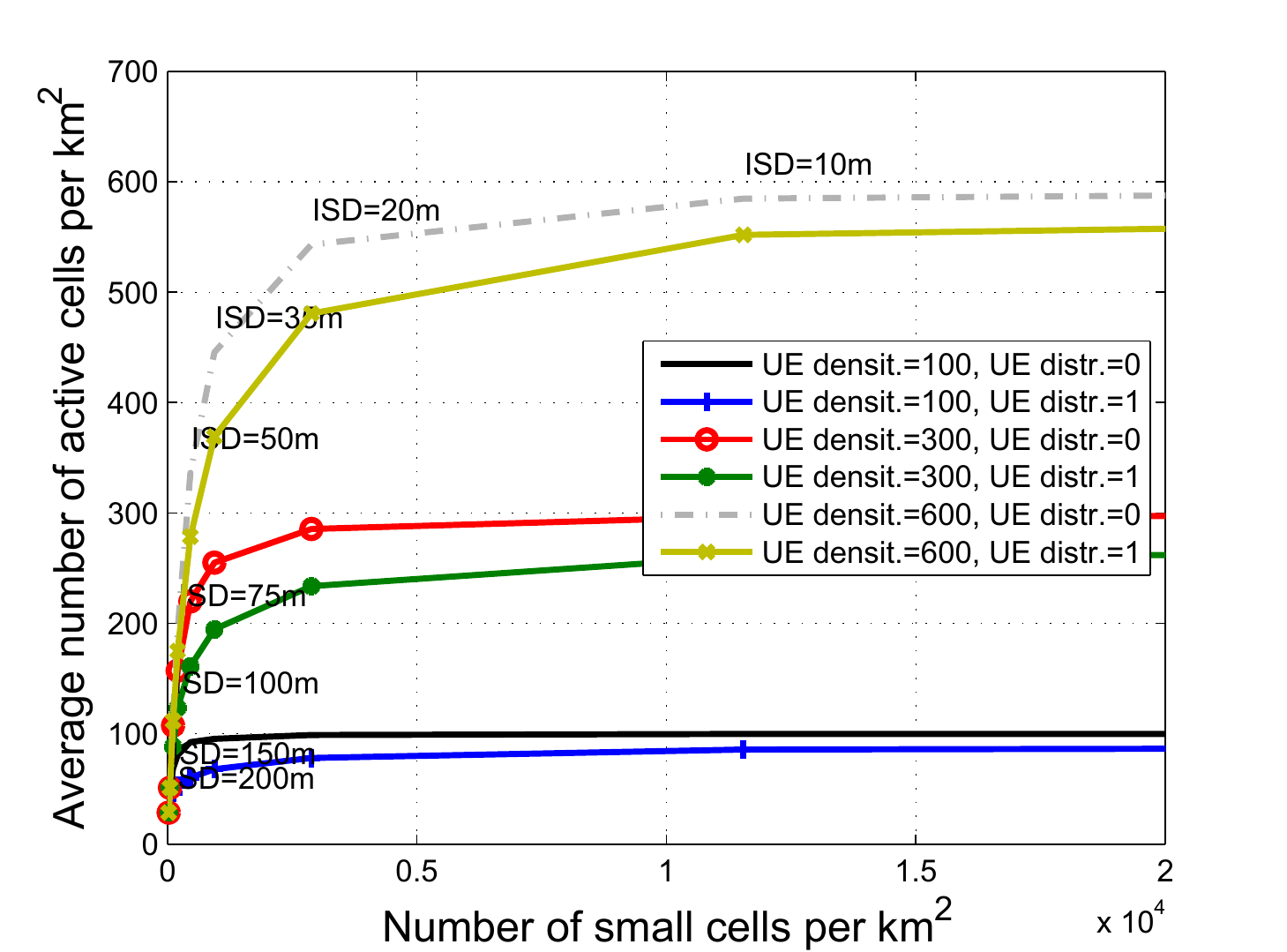}}
	\subfigure[Average number of active \acp{UE} per active \ac{BS}.]{\label{fig:uesPerCell}
	\includegraphics[width=3.5in]{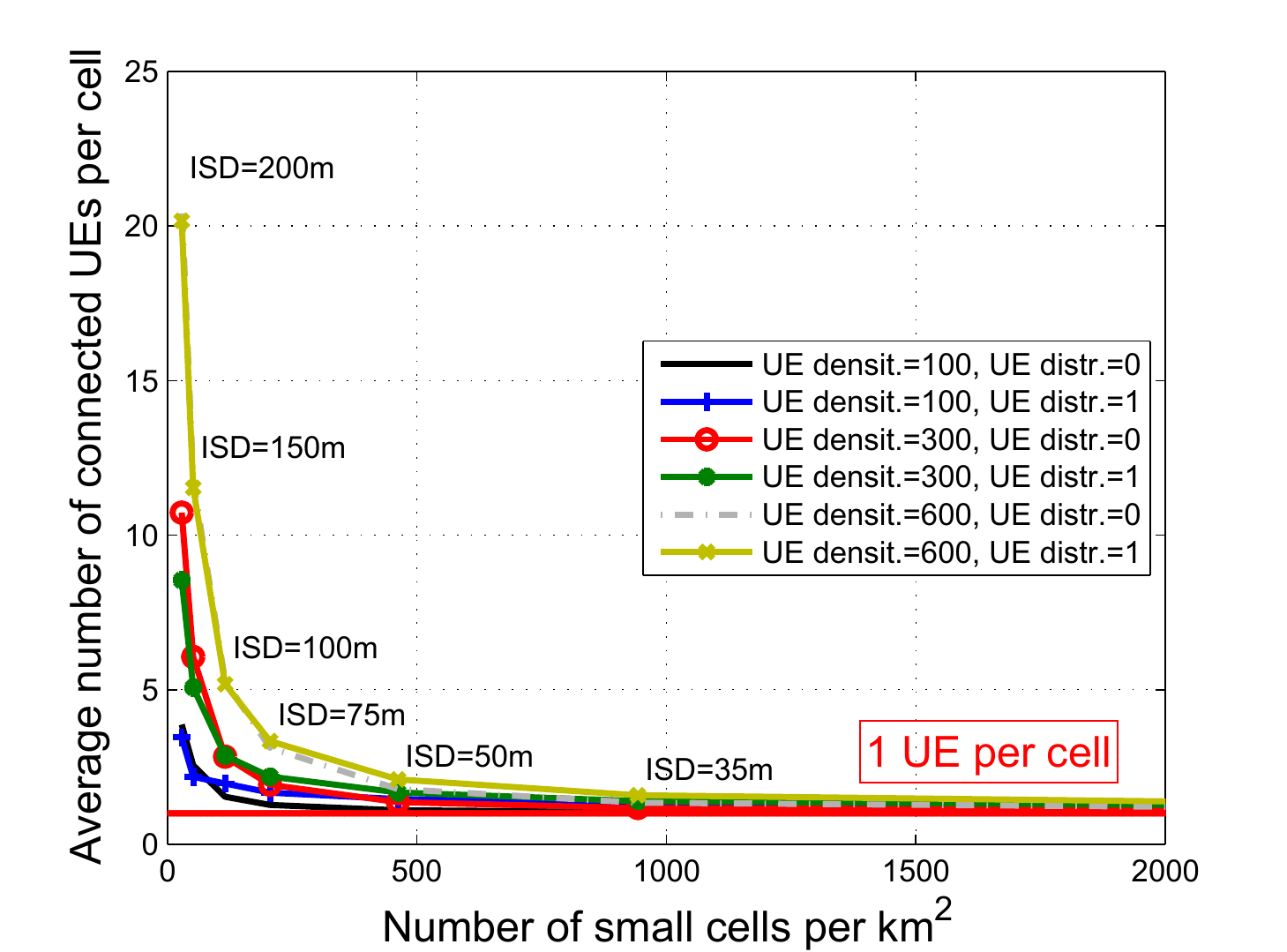}}
  \caption{Average number of active \acp{BS} in the network per square km and average number of active \acp{UE} per active \ac{BS}.
  The \ac{UE} densities are 600, 300 and 100 active \acp{UE} per km$^2$.
  The UE distributions are uniform and non-uniform within the scenario.
  The rest of the parameters are $s=1$, $f=2$\,GHz, $a=1$ and $t=$12\,dB.}
  %In the non-uniform distribution,
  %half of the \acp{UE} are uniformly distributed in the scenario,
  %while the other half are non-uniformly distributed within circular hot spots of 40\,m radius with 20 \acp{UE} each.}
  \label{fig:activeCellsAndUes}
\end{figure*}
%[Ming]: I recommend showing the x-axis in log-scale in this figure and the following figures so that the figures would be easier to read. Also the tendency of the curves (weaker than log, log, loglog) would be easier to identify.
%[David]:We had a discussion on this and Holger prefer the linear scale ... :)
%[Ming]: I see, you've considered this before :) Can you send me some of the .fig files? I'd like to tweak the axis and see the results.
%[David]: Done.
%[Ming]: I've received the .fig files. Two example figures with log(x) axis, respectively adapted from Fig. 2a and Fig. 7a, have been provided in this email. Please discuss with Holger and let me know whether you find them more reader-friendly. This is just a suggestion and feel free to keep the figures as they are right now.

%[Ming]: We need to explain the legend somewhere in the paper.
%[David]: Done.

%[Ming]: I believe that the curves in these two figures can be approximately calculated in closed-form expressions :) After I join NICTA, I will spend several days to derive the formula and compare my results with the valuable numerical results shown here.
%[David]: Cool. I will be happy to help.
%[Ming]: Fantastic!

%[Ming]: New comment. The label of the y-axis should be "Average ..."
%[David]: Done

\section{Network Densification}
\label{sec:densification}

Network densification has the potential to linearly increase the capacity of the network with the number of deployed cells
through spatial spectrum reuse,
and is considered to be the key enabler to provide most of the capacity gains in future networks.

In order to better understand the implications of network densification on network capacity,
let us define network capacity based on the framework developed by Claude Shannon~\cite{Shanon1948} as
\begin{equation}
C {\rm [bps]} = \sum_m^M \sum_u^{U_m} B_{m,u} {\rm [Hz]}  \log_2 (1 + \gamma_{m,u} {\rm [\cdot]}),
\label{eq:Shannon}
\end{equation}
%[Ming]: U => U_m ?
%[David]: Done
where $\{1, \dots, m, \dots, M\}$ is the set of \acp{BS} deployed in the network,
$\{1, \dots, u, \dots, U_m\}$ is the set of \acp{UE} connected to \ac{BS} $m$,
$B {\rm [Hz]}$ is the total available bandwidth,
and $B_{m,u} {\rm [Hz]}$ and $\gamma_{m,u} {\rm [\cdot]}$ are the bandwidth granted to and the \ac{SINR} experienced by
\ac{UE} $u$ when connected to \ac{BS} $m$.
This model assumes Gaussian interference.
%[Ming]: I guess \gamma should denote SINR, not SNR, right?
%[David]: Shannon model is for SNR. The model can be defined with SINR only if the interference is Gaussian. For consistency, I use SINR and indicate that the model assume a Gaussian interference.
%[Ming]: Agreed.

At the network level,
network densification increases the number of geographically separated \acp{BS} $M$
that can simultaneously reuse the available bandwidth $B$,
thus improving spatial reuse and linearly increasing network capacity with $M$.

At the cell level,
a consequence of network densification is cell size reduction,
which directly translates into a lower number of \acp{UE} $U_m$ connected per \ac{BS} $m$ and thus a larger bandwidth $B_{m,u}$ available per \ac{UE}.
In this way, network capacity linearly increases with the number of offloaded \acp{UE}.

Moreover, at the cell level too,
the average distance between a \ac{UE} and its serving \ac{BS} reduces,
while the distance to its interfering \acp{BS} does not necessarily reduce at the same pace assuming idle mode capabilities.
This leads to an increased UE signal quality $\gamma_{m,u}$,
and thus the network capacity logarithmically increases with the $\gamma_{m,u}$.
%[Ming]: I'm not sure that $\gamma_{m,u}$ increases, if $\gamma_{m,u}$ means SINR. Suppose that two adjacent UE1 and UE2 are originally served by BS1 and their SINRs are both approximately 10\,dB. The sum capacity is about 2*B/2*log_2(1+10)=3.46B. With network densification, let us assume that UE2 is now associated with a new BS2. The bandwidth of the two UEs will be doubled, but their SINRs will reduce to nearly 0dB when BS1 and BS2 transmit simultaneously. Thus, the sum capacity becomes 2*B*log_2(1+1)=2B. So I think it might not be accurate to state that "network densification increases $M$ and improves both $B_{m,u}$ and $\gamma_{m,u}$".
%[David]: You brought an interesting case here. However, we should consider that network densification with optimal idle mode capability tends to bring the UE closer to the serving cell BS and further from the interfering BS. As a result the SINR increases. This was my thinking that may not apply in all cases. However, this phenomena is backed up by the simulation results which show that even after reaching 1 user per cell the network throughput continues increasing because the UE is closer to the serving cell BS and further from the interfering BSs.
%[Ming]: Agreed. We can further look into this issue later.
%[Holger]: I have rewritten the paragraph above to make this clear.
%[David]: Thanks! Much better now

As can be derived from the above discussion,
network densification increases $M$ and in turn improves both $B_{m,u}$ and $\gamma_{m,u}$,
resulting in an increase of the capacity
-- see (\ref{eq:Shannon}).
However, the impact of \ac{UE} density and distribution should not be forgotten.
If network densification is taken to an extreme,
and the number of deployed \acp{BS} is larger than the number of existing active \acp{UE},
this ultra-dense small cell deployment may reach a fundamental limit
in which the number of active \acp{UE} per cell is equal or lower than one, i.e., $U_m\leq1$.
At this point,
the bandwidth $B_{m,u}$ available per \ac{UE} cannot be further increased through cell splitting,
and thus network densification can only enhance network capacity  at a lower pace in a logarithmic manner
by bringing the network closer to the \ac{UE} and improving the \ac{UE} signal quality $\gamma_{m,u}$,
which may not be as cost-effective.
As a result,
\emph{one \ac{UE} per cell} may be the operational sweet spot from a densification view point,
In the following, we discuss the \emph{one \ac{UE} per cell} concept.

\subsection{Idle mode capability and the 1 UE per cell concept}

One important advantage of having a surplus of cells in the network is that a large number of them could be switched off if there is no active \ac{UE} within their coverage areas,
which reduces interference to neighbouring \acp{UE} as well as energy consumption.

% Reduced number of active UEs per active BS
Provided that a surplus of cells exists and as a result of an optimal idle mode capability~\cite{Ashraf:10a},
the network would have the key ability of adapting the distribution of active \acp{BS} to the distribution of active \acp{UE},
and thus the number of active cells, transmit power of the network, and interference conditions
would strongly depend on the \ac{UE} density and distribution.
%[Challenge]: Optimal idle mode capability

%[Ming]: I suppose we adopt a simple UE association/cell on-off scheme based on the best RSRP criterion. Should we say a few words about our assumption here?
%[David]: Done. I have added a comment in the system model section. I agree that more sophisticated models that achieve a better energy efficiency can be implemented. However, they require network coordination, or at least inter-BS coordination.
%[Ming]: Agreed.

%Results related to reduced number of active UEs per active BS
According to our system model,
Fig.~\ref{fig:activeCellsAndUes} shows
the average number of active \acp{BS} in the network per square km and the average number of active \acp{UE} per active \ac{BS},
with the aforementioned idle mode capability (a \ac{BS} with no associated \ac{UE} is switched off).
Different \ac{UE} densities and distributions are considered.
%[Holger]:  Could we explain here what we mean with optimal idle mode capability?
%[David]: I have rewritten this sentence, since the word optimal was abused and provided a more accurate definition.
From Fig.~\ref{fig:activeCells}, the following conclusions can be extracted.
The lower the \ac{UE} density,
the lower the average number of active \acp{BS} in the scenario and the lower the average number of active \acp{UE} per active \ac{BS}.
With regard to \ac{UE} distribution,
a uniform distribution requires more active \acp{BS} than a non-uniform distribution to provide full coverage to a given \acp{UE} density.
This is because \acp{UE} are more widely spread in the former.
In contrast,
due to the more active \acp{BS},
the uniform distribution results in a lower number of active \acp{UE} per active \ac{BS} for a given \acp{UE} density,
which tends to provide a better \ac{UE} performance at the expense of an increased number of deployed active cells,
and thus cost.
This indicates the importance of understanding the \ac{UE} density and distribution in specific scenarios to realise efficient deployments.
%[Challenge]: Understanding the \ac{UE} density and distribution

\begin{figure*}[t]
  \centering
	\subfigure[Transmit power per active \ac{BS}.]{\label{fig:txPowerCells}
	\includegraphics[width=3.5in]{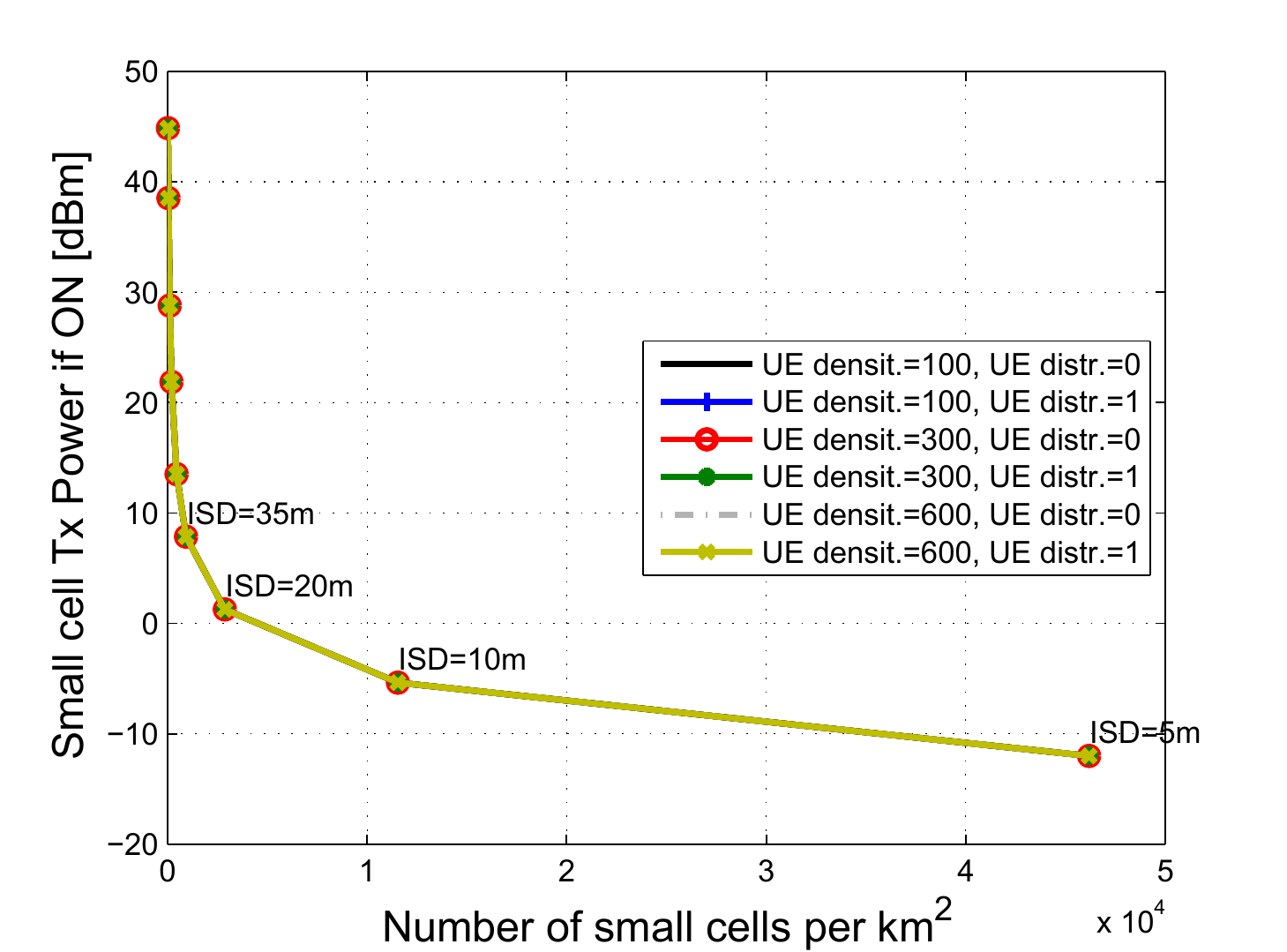}}
	\subfigure[Transmit power of the network per km$^2$.]{\label{fig:txPowerNetwork}
	\includegraphics[width=3.5in]{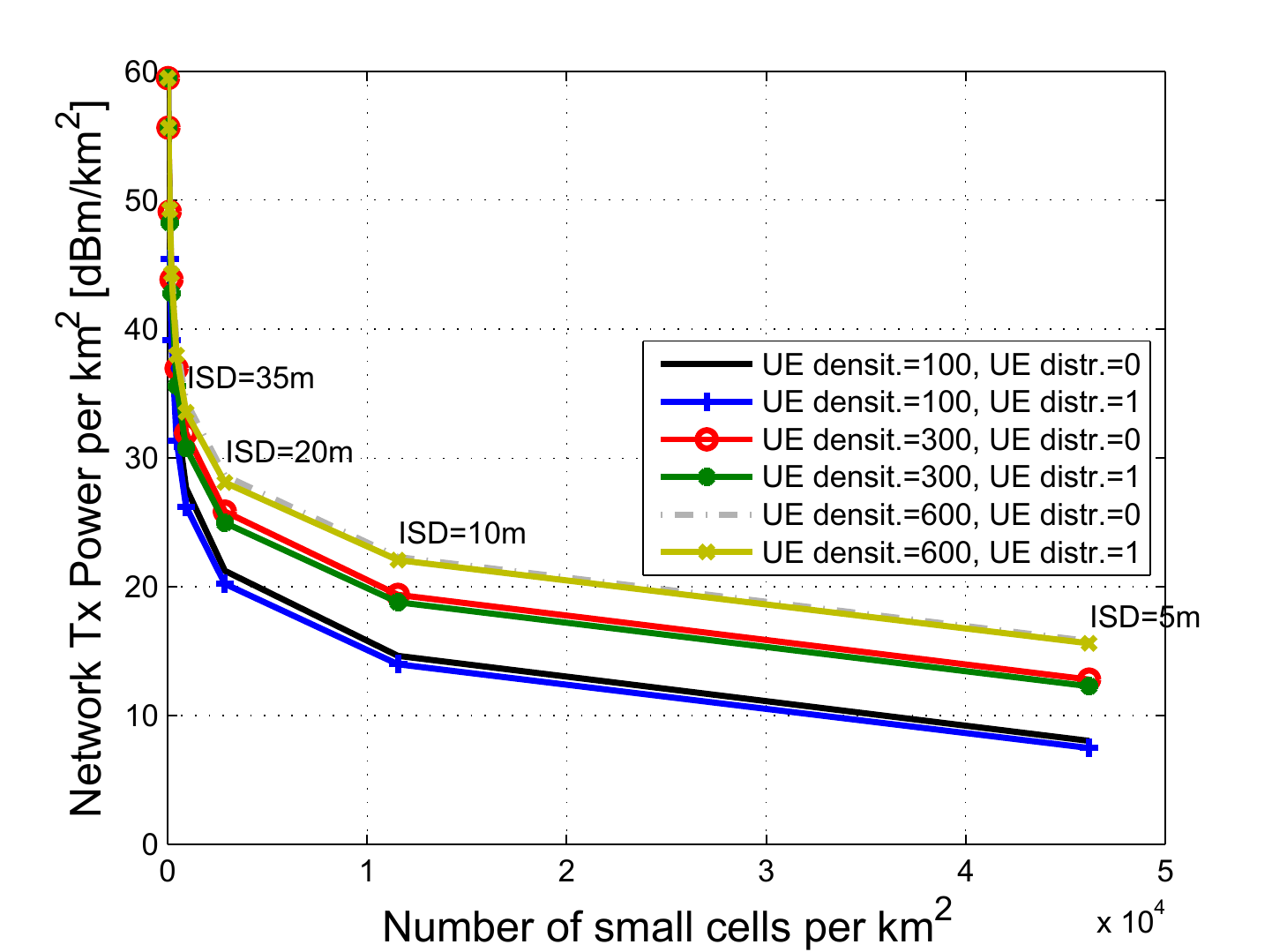}}
  \caption{Transmit power per active \ac{BS} and transmit power of the network per km$^2$.
  The \ac{UE} densities are 600, 300 and 100 active \acp{UE} per km$^2$.
  The UE distributions are uniform and non-uniform within the scenario.
  The rest of the parameters are $s=1$, $f=2$\,GHz, $a=1$ and $t=$12\,dB.
  It is important to note that in Fig.~\ref{fig:txPowerCells} there are 6 overlapping curves. This is because the power used by the cell if it is activated does not depend on the UE density. }
  %as described in Fig.~\ref{fig:activeCellsAndUes}.}
  \label{fig:txPower}
\end{figure*}

An important result that can be extracted from Fig.~\ref{fig:uesPerCell} is that
an \acp{ISD} of 35\,m can already achieve an average of 1.1 active \acp{UE} per active \ac{BS} or smaller,
approaching the fundamental limit of spatial reuse.
Reaching such limit has implications for the network and \ac{UE} performance,
as it has been qualitatively explained before,
and as it will be quantitatively shown in Section~\ref{Sec:higherFrequecnyBands}.
Moreover, it also has implications for the cost-effectiveness of the deployment,
since densifying further than the 1 \ac{UE} per cell sweet spot requires an exponential increase in investment to achieve a diminishing logarithmic capacity gain through signal quality enhancement.
In other words, when the network is dense enough,
a large number of cells have to be added to the network to enhance the \ac{UE} throughput in a noticeable manner,
and this may not be desired since the operator may have to pay exponentially more money to carry on with the deployment.
%as it will be analysed in Section~\ref{Sec:costAndEnergyEfficiency}.
%It is worth indicating that
%although this ultra-dense small cell deployments with \acp{ISD} equal or smaller than 35\,m are technically and financially challenging today,
%they may be feasible in dense urban Manhattan like scenarios in the years to come,
%where wired backhaul is expected to be widely available.
%[Challenge]: Cost-effective deployment

\begin{figure*}[t]
  \centering
	\subfigure[100 ISD \& idle mode capability deactivated.]{\label{fig:snirPlot100mOFF}
	\includegraphics[width=3.5in]{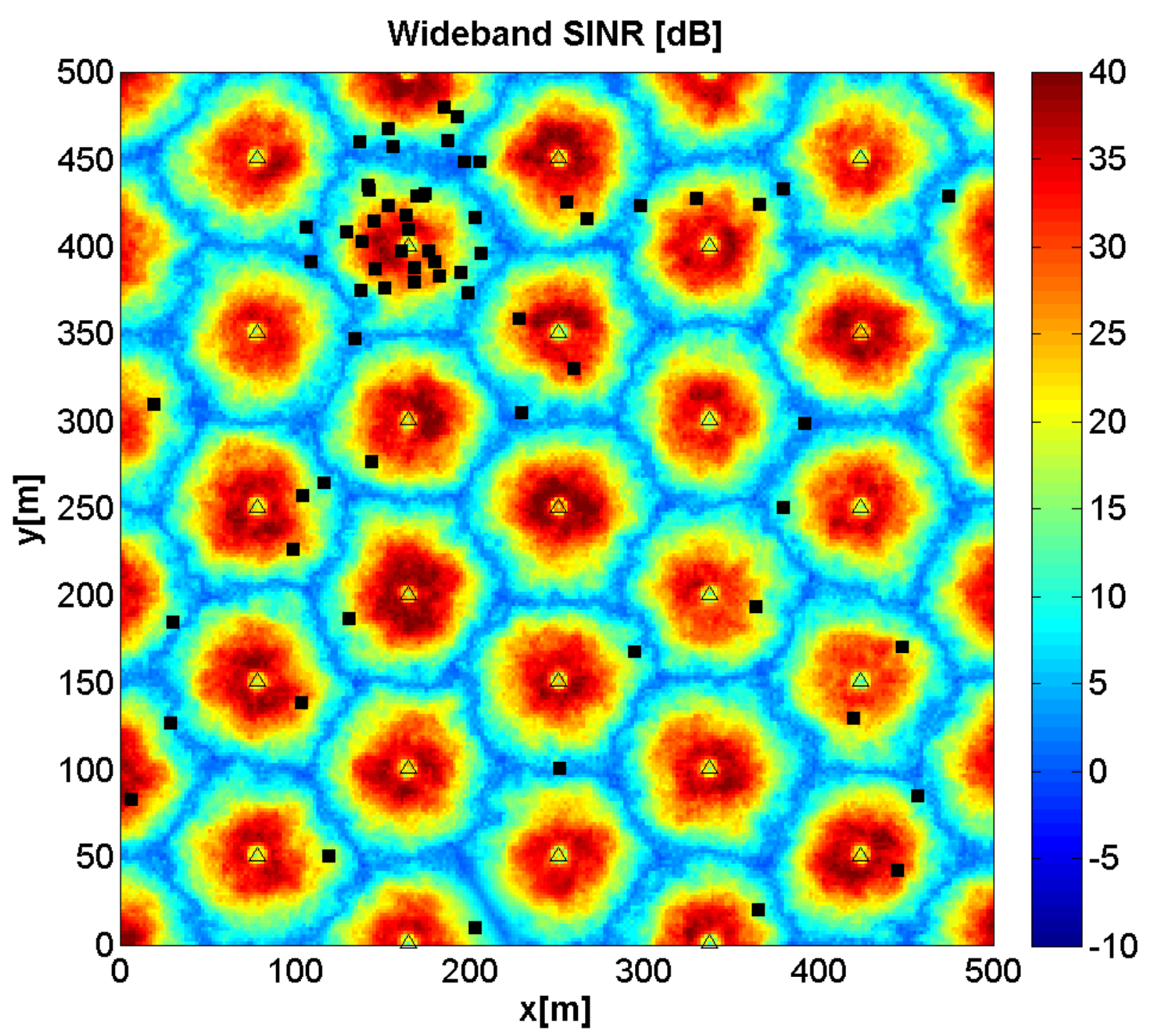}}
	\subfigure[  50 ISD \& idle mode capability deactivated.]{\label{fig:snirPlot50mOFF}
	\includegraphics[width=3.5in]{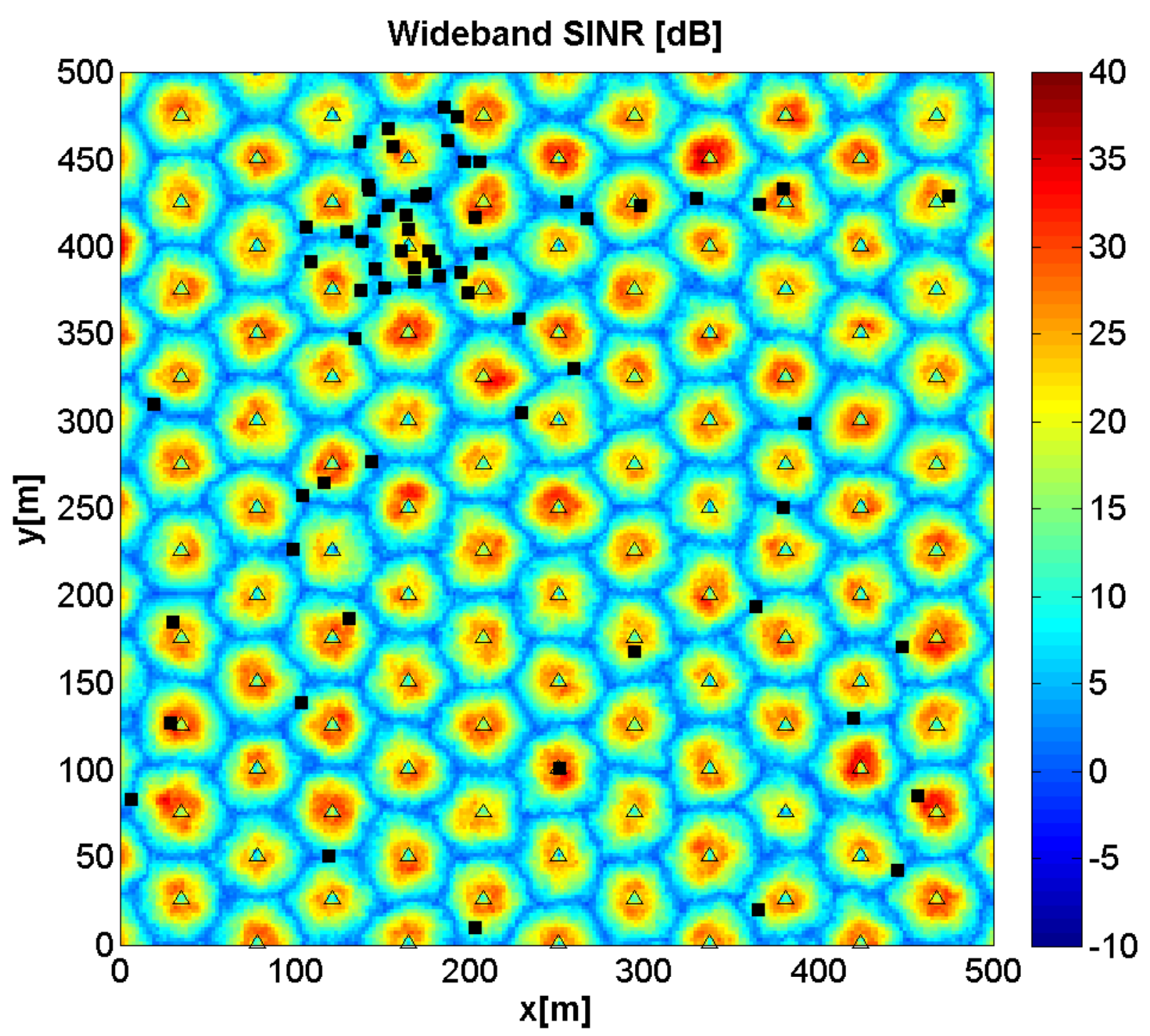}}
	\subfigure[100 ISD \& idle mode capability activated. ]{\label{fig:snirPlot100mON}
	\includegraphics[width=3.5in]{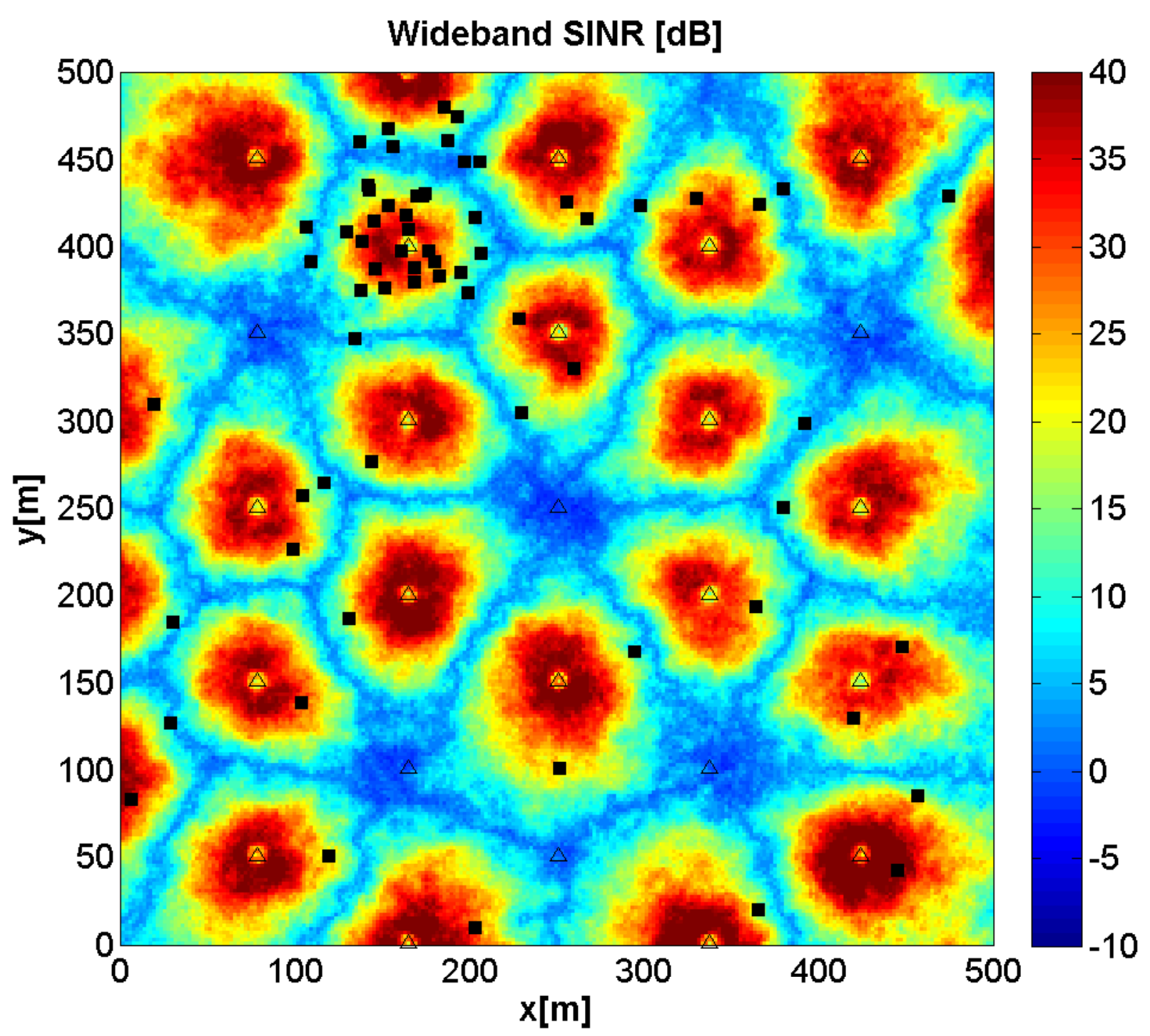}}
	\subfigure[  50 ISD \& idle mode capability activated.]{\label{fig:snirPlot50mON}
	\includegraphics[width=3.5in]{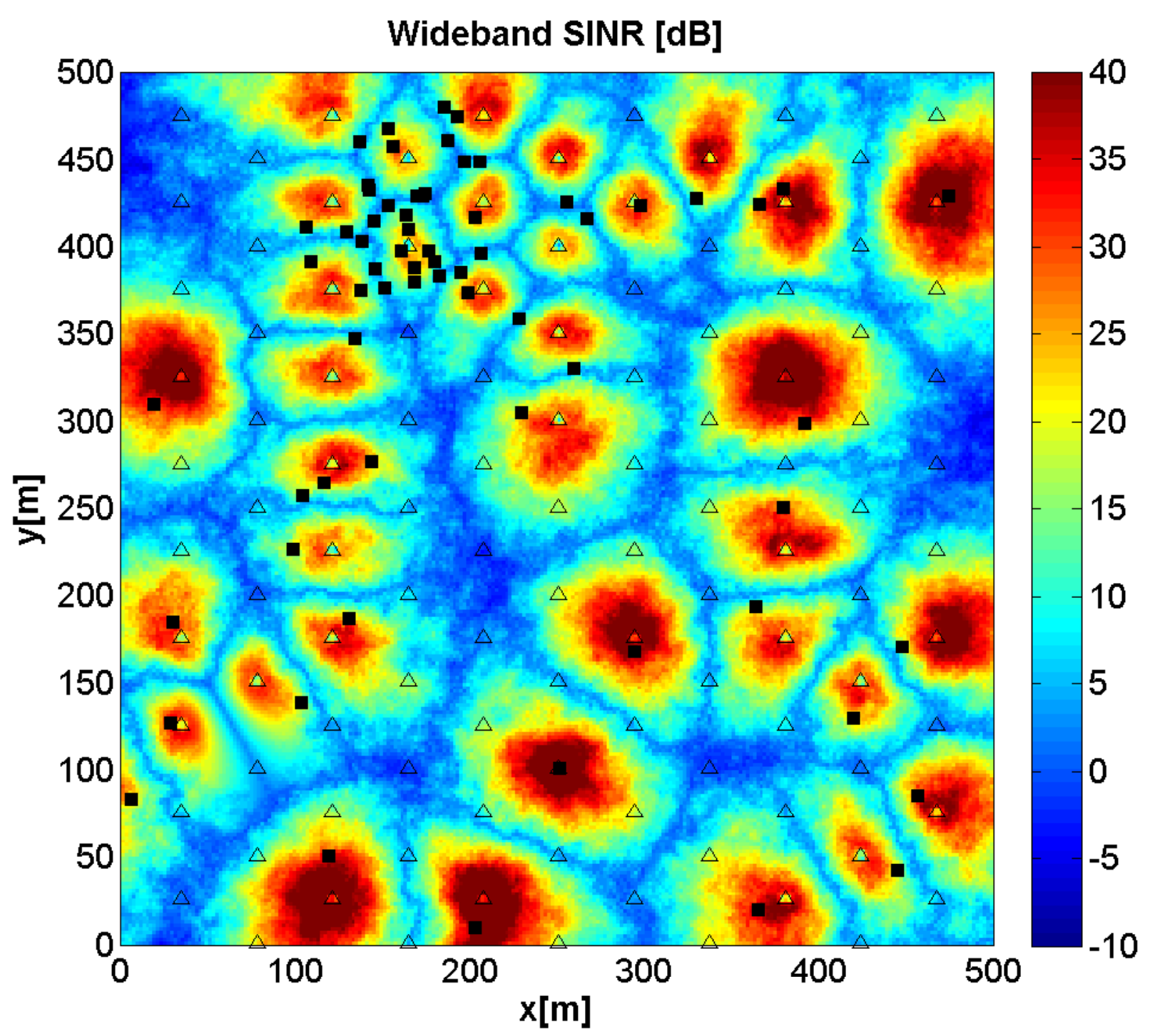}}
  \caption{SINR spatial distributions of several ultra-dense small cell deployments
  with \ac{ISD} among small cell \acp{BS} of 100 and 50\,m
  with or without idle mode capabilities.
  The rest of the parameters are $d=300$\,UE/km$^2$, $ud$=1, $s=1$, $f=2$\,GHz, $a=1$ and $t=$12\,dB.
  The triangles represent \acp{BS} and the squares represent \acp{UE}.}
  \label{fig:sinrDistribution}
\end{figure*}

\begin{figure*}[t]
  \centering
	\subfigure[CDF of SINR for different ISDs with idle mode capability deactivated.]{\label{fig:snirCDFOFF}
	\includegraphics[width=3.5in]{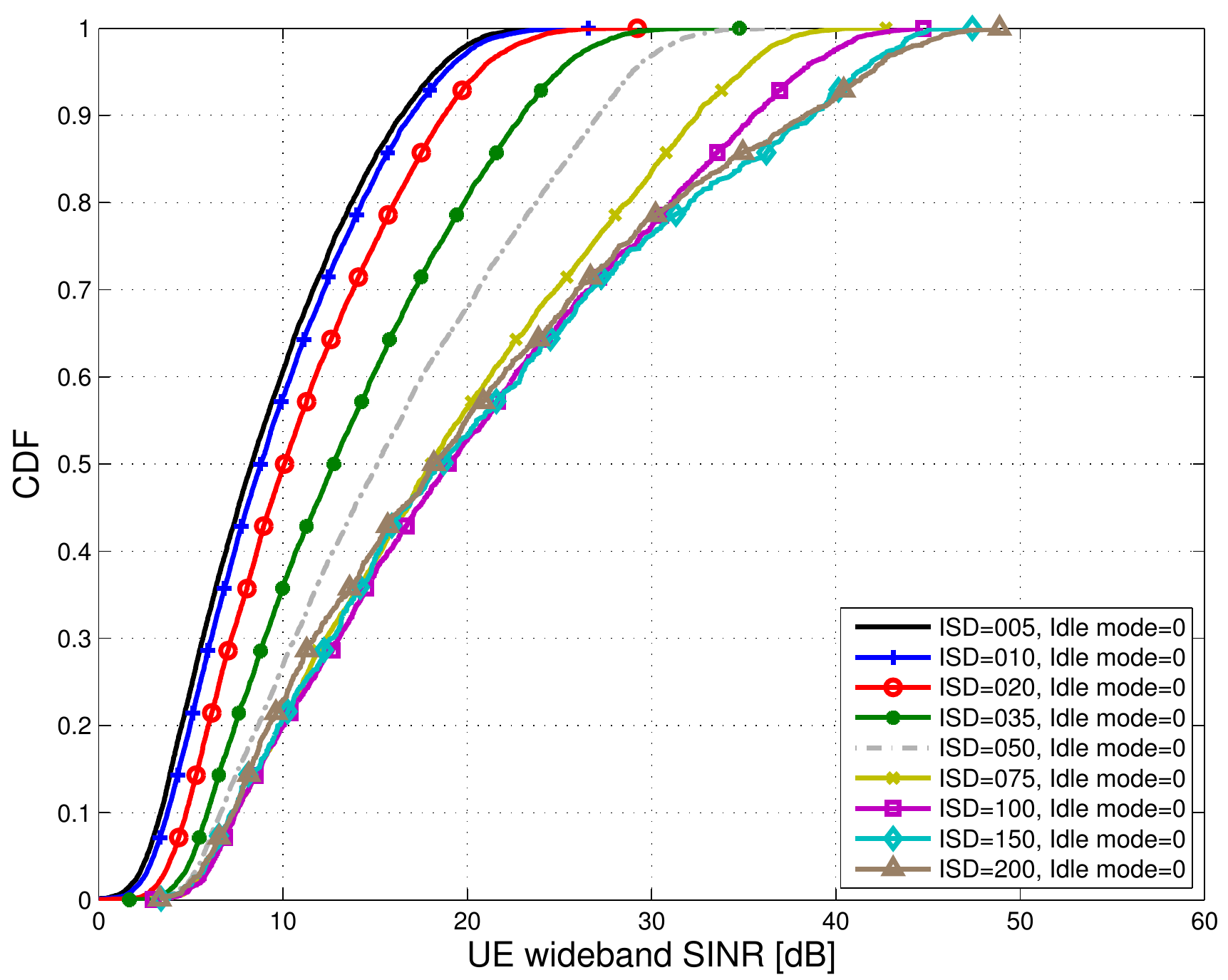}}
	\subfigure[CDF of SINR for different ISDs with idle mode capability activated.]{\label{fig:snirCDFON}
	\includegraphics[width=3.5in]{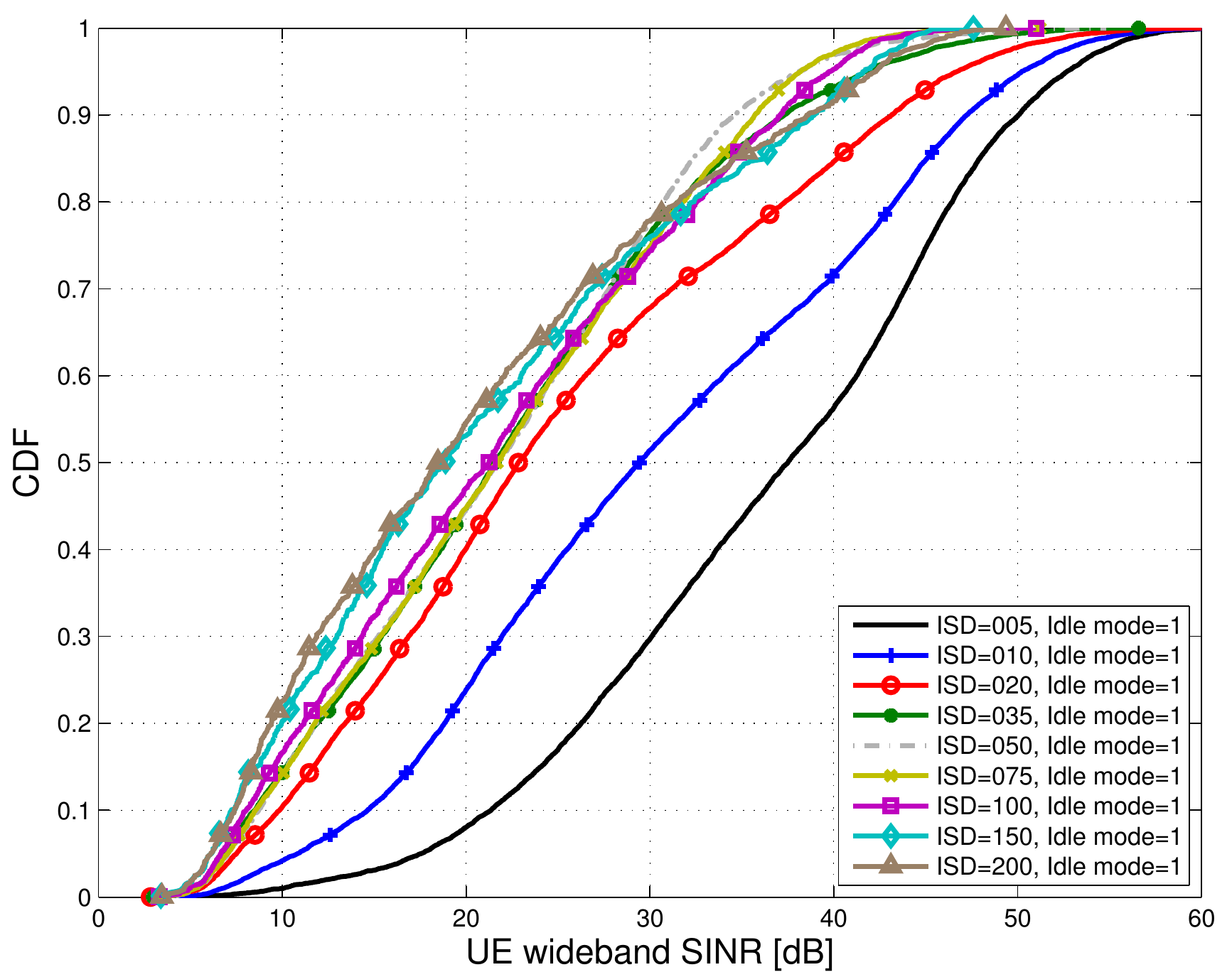}}
  \caption{UE SINR CDF of several ultra-dense small cell deployments
  with \ac{ISD} among small cell \acp{BS} of 200, 100, 75, 50, 35, 20, 10 and 5\,m
  with or without idle mode capabilities.
  The rest of the parameters are $d=300$\,UE/km$^2$, $ud$=1, $s=1$, $f=2$\,GHz, $a=1$ and $t=$12\,dB.}
  \label{fig:sinrCDF}
\end{figure*}

\subsection{Transmit Power and \ac{UE} \ac{SINR} Distribution}

% Reduced transmit power and improved SINR distribution
Combining ultra-dense small cell deployments together with an efficient idle mode capability
has the potential to significantly reduce the transmit power of the network.
This is because active cells transmit to \acp{UE} with a lower power due to their reduced cell size
and empty cells can be put into idle mode until a \ac{UE} appears~\cite{Ashraf:10a}.

Moreover, by turning off empty cells,
the interference suffered by \acp{UE} from always on channels, e.g., synchronisation, reference and broadcast channels,
can also be removed,
neutralising some neighbouring cells and thus improving  \ac{UE} \ac{SINR} distributions.
%[Ming]: Small cell discovery reference signals (DRSs), which should be always on, have just been specified in LTE Release 12. Shall we say a few words about DRSs? I can prepare a paragraph, if needed.
%[David]: Please do :)
%[Ming]: No problem. I will check the final agreement in LTE Release 12 and add one paragraph in my next round of revision.

Working in this direction,
\ac{LTE} Release~12 networks have defined
periodic \acp{DRS} to facilitate \acp{UE} the discovery of small cells that are turned off~\cite{LTE-R12}.
\acp{DRS} are transmitted sparsely in the time domain and they consist of multiple types of \acp{RS},
based on which \acp{UE} are able to perform synchronisation, detect cell identity and acquire coarse \ac{CSI}, etc.
Due to the low periodicity of \acp{DRS},
the impact of \acp{DRS} on \ac{UE} \ac{SINR} distribution is marginal, and thus it can be ignored.
%[Ming]: Done :) Please check the newly added sentences.
% ~\cite{LTE R12}: ETSI MCC, "Draft Report of 3GPP TSG RAN WG1 #77," 3GPP TSG RAN WG1 Meeting #77, Seoul, Korea, May 2014.
%[David]:Sounds good. Citations added.

%Results related to reduced transmit power
\subsubsection{Transmit Power}

In terms of transmit power,
Fig.~\ref{fig:txPowerCells} shows how the transmit power per active \ac{BS} significantly reduces with the small cell \ac{BS} density in the studied scenario.
In this case,
the transmit power of each active \ac{BS} is configured such that it provides a \ac{SNR} of 12\,dB at the targeted coverage range,
which is $\frac{\sqrt{3}}{2}$ of the \ac{ISD}.
Note that here the required transmit powers are significantly lower compared to co-channel deployments where the small cells should exceed the received macrocell power in the intended coverage area.
In addition, Fig.~\ref{fig:txPowerNetwork} shows how the overall transmit power used by the network also significantly reduces with the small cell \ac{BS} density in the studied scenario,
when the idle mode capability is considered.
This is because the reduction of transmit power per cell outweighs the increased number of active cells,
and this fact holds true for both the uniform and non-uniform \ac{UE} distribution,
with network transmit power reductions of up to 43\,dB.
Note that this discussion only considers transmit powers,
and that the overall power consumption of the network when considering the power consumed by the \acp{BS} in idle mode
will be analysed in Section~\ref{Sec:costAndEnergyEfficiency}.

\subsubsection{\ac{UE} \ac{SINR} Distribution}

Traditional understanding has lead to the conclusion that the \ac{UE} \ac{SINR} distribution and thus outage probability is independent of \ac{BS} density.
The intuition behind this phenomenon is that the increase in signal power is exactly counter-balanced by the increase in interference power,
and thus increasing the number of BSs does not affect the coverage probability.
This is  a major result in the literature~\cite{Andrews2011}~\cite{Mukherjee2012},
which only holds under the assumption of a single-slope path loss model more suited for rural areas.
However, conclusions may be different for urban and dense urban scenarios where \ac{NLOS} to \ac{LOS} transitions may occur.

%Results related to improved SINR distribution
In order to show this,
Fig.~\ref{fig:sinrDistribution} and Fig.~\ref{fig:sinrCDF} respectively show the \ac{SINR} spatial distribution and \ac{UE} \ac{SINR} \acp{CDF}
for several ultra-dense small cell deployments
with different small cell densities with and without idle mode capability.
Recall  that our path loss models the \ac{NLOS} to \ac{LOS} transition with the probabilistic function defined for urban microcell environments in~\cite{TR36.814}.
Let us now analyse the results.

When the idle mode capability is deactivated,
our results show that the \ac{UE} \ac{SINR} \ac{CDF} degrades with the small cell density,
contradicting the results in~\cite{Andrews2011}~\cite{Mukherjee2012} where it is independent of \ac{BS} density.
As the network becomes denser,
the \ac{ISD} is reduced and the \ac{LOS} component starts to dominate the path loss model for the interfering signals.
While in traditional networks with large \acp{ISD} the carrier signal may be subject to \ac{LOS} depending on the distance between the \ac{UE} and its serving \ac{BS},
the interfering signal is not usually subject to \ac{LOS} due to the large distance between the serving \ac{BS} and its interfering \acp{BS}.
However, with the smaller \acp{ISD},
\ac{LOS} starts dominating the interfering signal too and this brings down the \ac{UE} \ac{SINR},
thus lowering the \ac{UE} and cell throughputs.
In other words, the interference power increases faster than the signal power with densification due to the transition of the former from \ac{NLOS} to \ac{LOS},
and thus the BS density matters!
This new conclusion should significantly impacts network deployment strategies,
since the network capacity no longer grows linearly with the number of cells, based on previous understanding.
Instead, it seems to increases at a lower pace as the BS density increases,
and thus operators may find themselves paying exponentially more money for diminishing gains when heavily densifying their networks.

However, the good news is that when the idle mode capability is activated,
the trend is just the opposite,
and the \ac{UE} \ac{SINR} \ac{CDF} is significantly boosted with the cell density as a result of interference mitigation.
The denser the \ac{BS} deployments, the larger the capacity increase,
since more \ac{BS} can be turned off,
which reduces interference.
When the \ac{ISD} among \acp{BS} is 35\,m,
the median \ac{SINR} improvement compared to the case when the idle mode capability is deactivated is around 8.76\,dB,
while for an \ac{ISD} of 10\,m
the median \ac{SINR} improvement is around 20.62\,dB.
This is a significant improvement.
% [Ming]: I like Fig. 5 very much! It reveals a lot of intuition by showing the break-up of the cluster of CDF curves in Fig. 5(b)!
% [David]: :)

As conclusion, it is important to note that an optimum idle mode capability not only plays a significant role in transmit power savings,
but also as an interference mitigation technique.

\begin{figure*}[t]
  \centering
	\subfigure[100 active \acp{UE} per km$^2$.]{\label{fig:transitionCDFOFF}
	\includegraphics[width=3.5in]{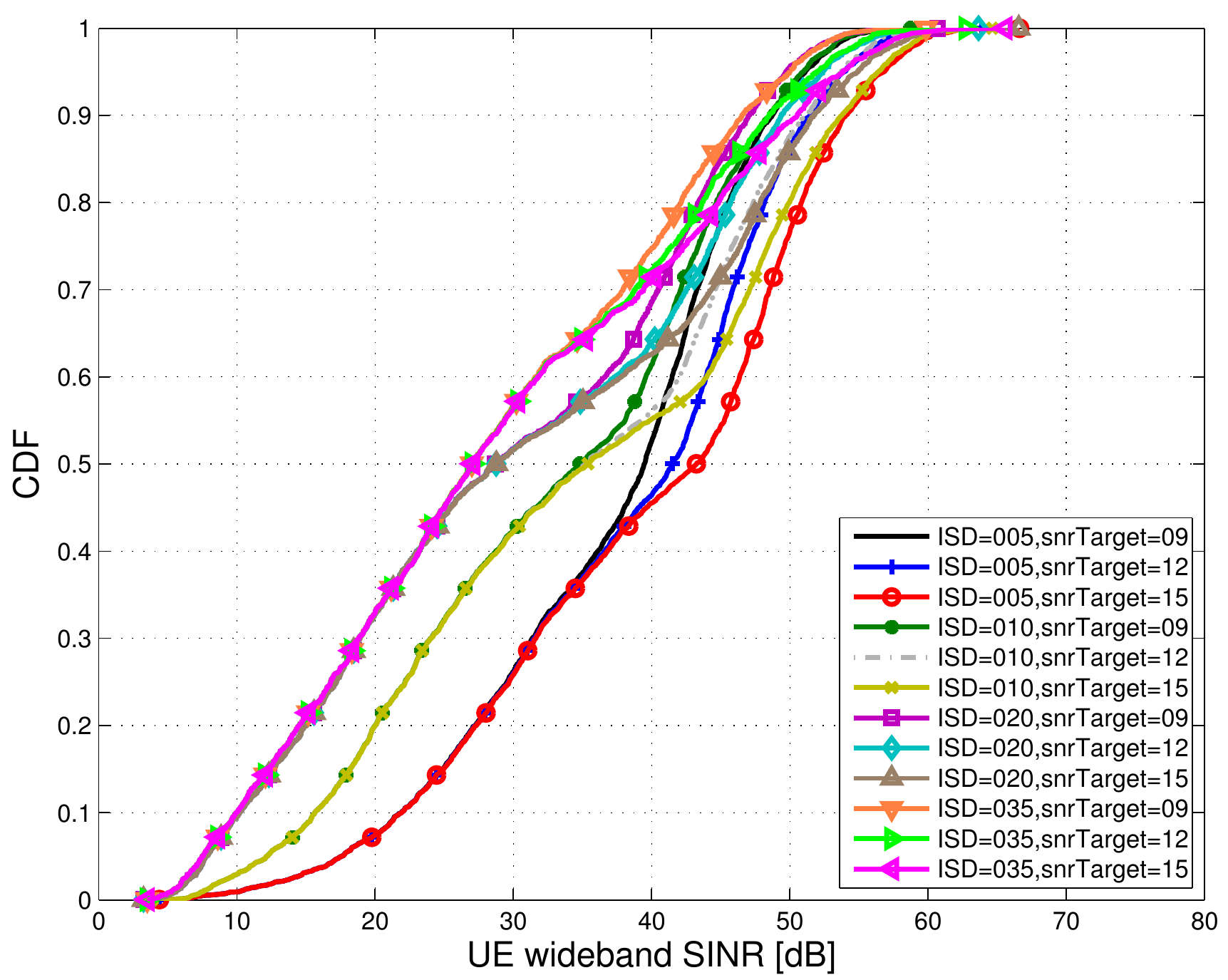}}
	\subfigure[600 active \acp{UE} per km$^2$.]{\label{fig:transitionCDFON}
	\includegraphics[width=3.5in]{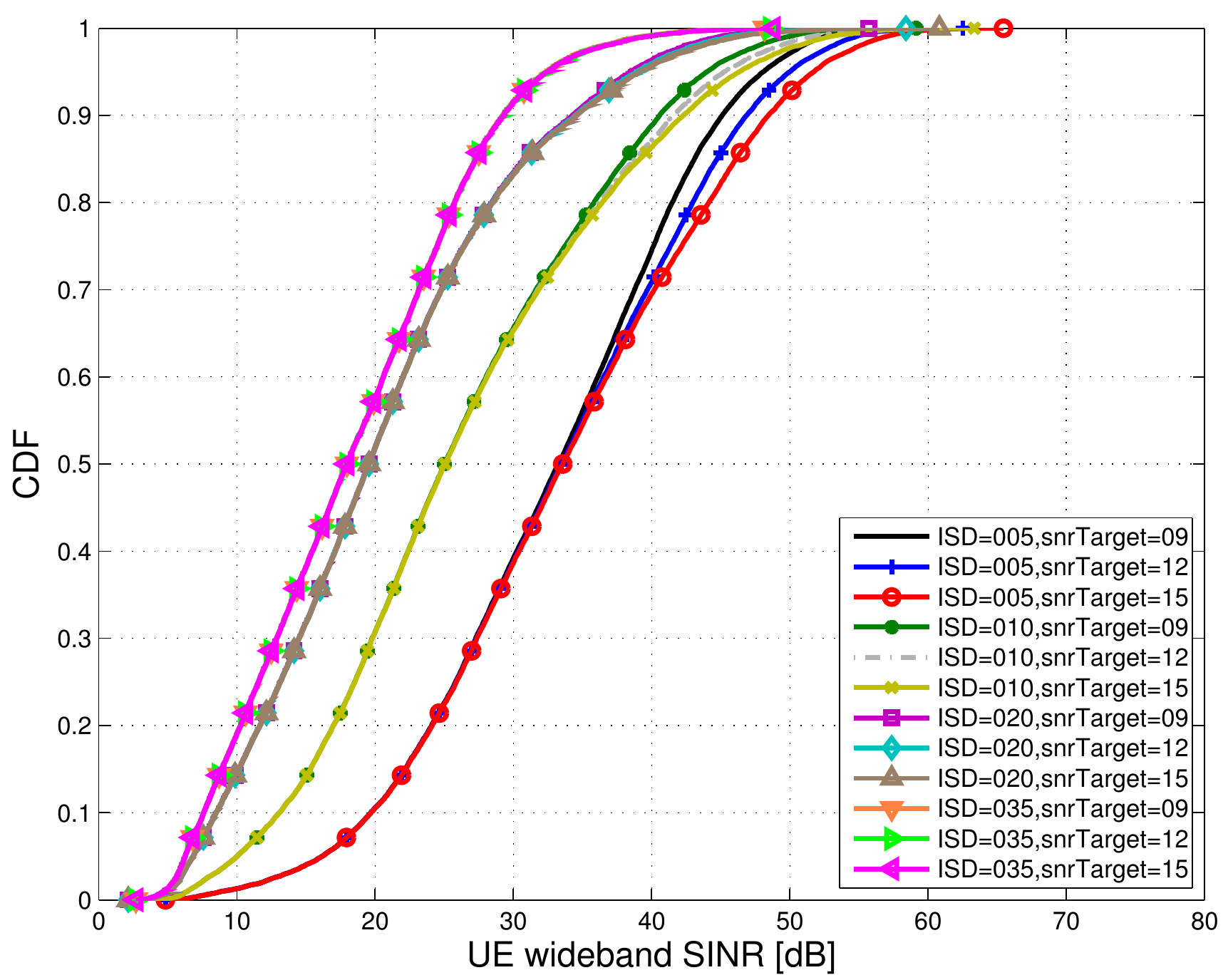}}
  \caption{UE SINR CDF in ultra-dense small cell deployments with
  large \ac{BS} densities, 35, 20, 10 and 5\,m,
  low \ac{UE} densities, 100 and 600 active \acp{UE} per km$^2$
  and \ac{SNR} targets, 15,12 and 9.
   The rest of the parameters are $ud$=1, $s=1$, $f=2$\,GHz, and $a=1$.}
  \label{fig:transition}
\end{figure*}

\subsection{Transition from Interference to Noise Limited Scenarios}

As shown in Fig.~\ref{fig:activeCellsAndUes},
in ultra-dense small cell deployments,
a large number of cells should be switched off
when the \ac{BS} density is large and the \ac{UE} density is low;
this combination leads to the most effective interference mitigation.
However, is this interference mitigation through small cell deactivation large enough
to transition from an interference limited scenario to a noise limited scenario?
In an interference limited scenario,
the signal quality of a \ac{UE} is independent of the transmit power of the serving and the interfering \acp{BS},
provided that they all use the same transmit power.
However, this is not the case in a noise limited scenario
where the signal quality of a \ac{UE} improves with the transmit power of the serving \ac{BS},
as the noise remains constant.
Therefore, if such transition takes place,
the tuning of the small cell \ac{BS} transmit power becomes even more important,
since the transmit power will not only determine the coverage radius of the cell but will also affect the capacity of the network.

In order to answer this question,
Fig.~\ref{fig:transition} shows the \ac{UE} \ac{SINR} \ac{CDF} in different ultra-dense small cell deployments,
while considering different transmit power for the small cell \acp{BS}.
In more detail, the targeted \acp{SNR} at $\frac{\sqrt{3}}{2}$ of the \ac{ISD} are set to 9, 12 or 15\,dB.
Different  \ac{UE} densities are also considered, 100 and 600 UEs per square km.
The results show that
the change in transmit power only has an impact on the \ac{SINR} distribution of the scenario with the larger \ac{BS} densities, ISD$=$5\,m and ISD$=$10\,m,
and the lowest \ac{UE} densities, 100 active \acp{UE} per km$^2$.
Otherwise, the \ac{SINR} distribution is independent of the transmit power,
indicating that this transition does not occur.
It is important to note that even in the indicated cases
the decoupling of the \ac{UE} \ac{SINR} \ac{CDF} only happens at the high \ac{SINR} regime,
whose \acp{SINR} belong to non-cluster \acp{UE} suffering from low interference.
As a result,
since the decoupling only happens for a very extreme \ac{BS} density,
it can be concluded that such transition from interference limited to noise limited does not occur in realistic deployments,
and that the small cell \ac{BS} transmit power should be simply configured to guarantee a targeted range.
%[Ming]: As I've commented in the Appendix, if we consider a non-degraded LoS probability model, I believe that increasing the Tx power should be beneficial to any UE without dominant LoS interferer BSs, i.e., the interference experienced by the UE is below the noise level. And many cell interior UEs when ISD$>$35\,m may enjoy this benefit because the Tx power of interferer BS is not very large (the cell edge UE's SNR is only 9, 12 or 15dB) and the law of large number has not come into play when ISD$>$35 regarding the LoS result of interferer BSs. Therefore, the decoupling phenomenon observed in Fig.6 might happen more often then we think, if we consider a non-degraded LoS probability model.
%[David]: Comments in the Appendix.

\begin{figure*}[t]
  \centering
	\subfigure[Average UE throughput for different frequency bands. ]{\label{fig:throughputAverage}
	\includegraphics[width=3.5in]{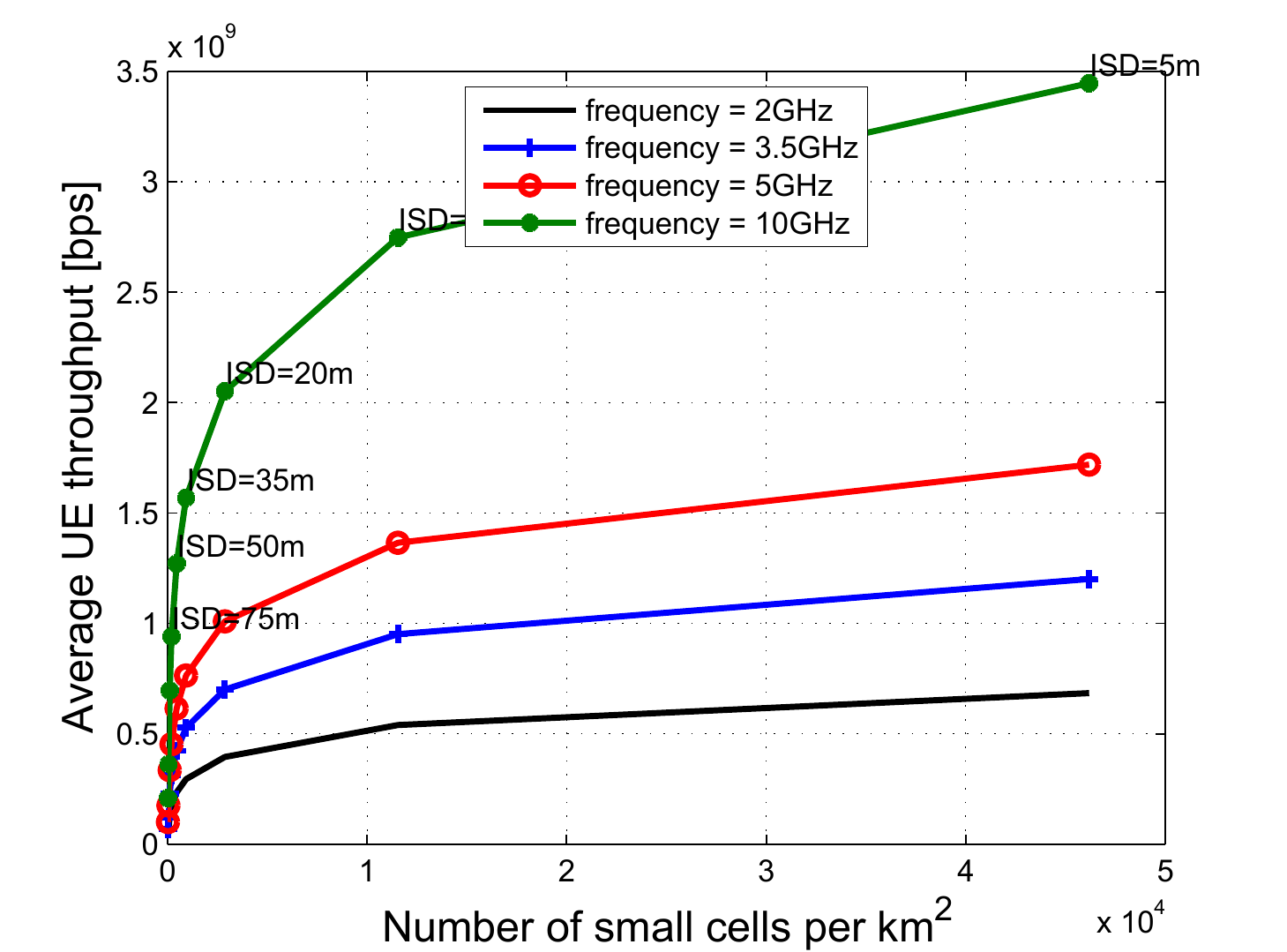}}
	\subfigure[5\,\%-tile UE throughput for different frequency bands.]{\label{fig:throughput5percentile}
	\includegraphics[width=3.5in]{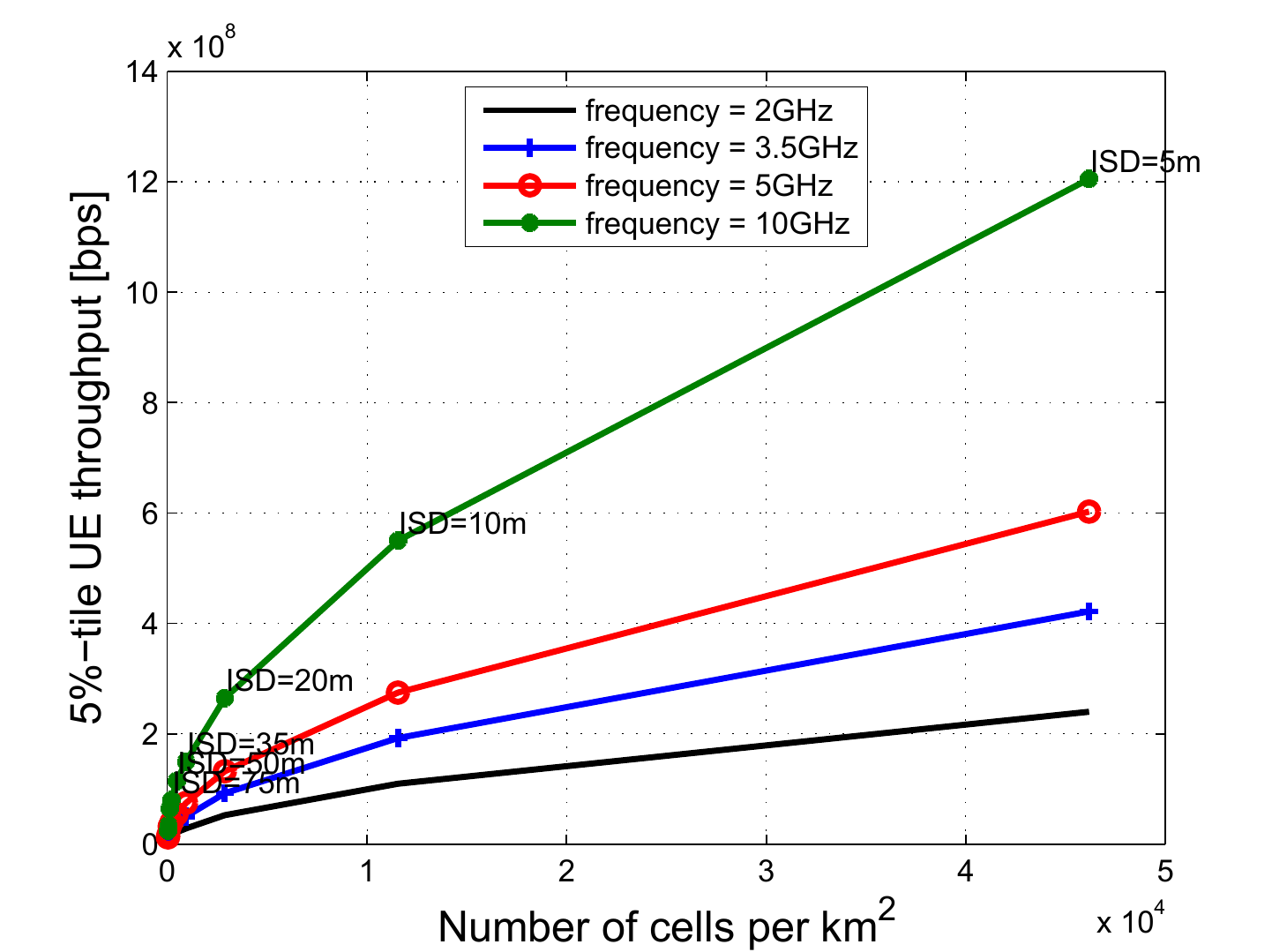}}
  \caption{Average and 5\%-tile UE throughput for four different network configurations where the carrier frequencies are 2.0, 3.5, 5.0 or 10\,GHz.
   The rest of the parameters are $d=300$\,UE/km$^2$, $ud$=1, $s=1$, $a=1$ and $t=$12\,dB.}
 % [Holger]: Check UE density - figure shows 600, file name of figure is 300. check definition of u=1 uniform/non-uniform
 % [David]: 300 is the correct value
  \label{fig:throughput}
\end{figure*}

\begin{figure*}[t]
  \centering
	\subfigure[Transmit power per active \ac{BS} for different frequency bands. ]{\label{fig:txPowerCellsFreq}
	\includegraphics[width=3.5in]{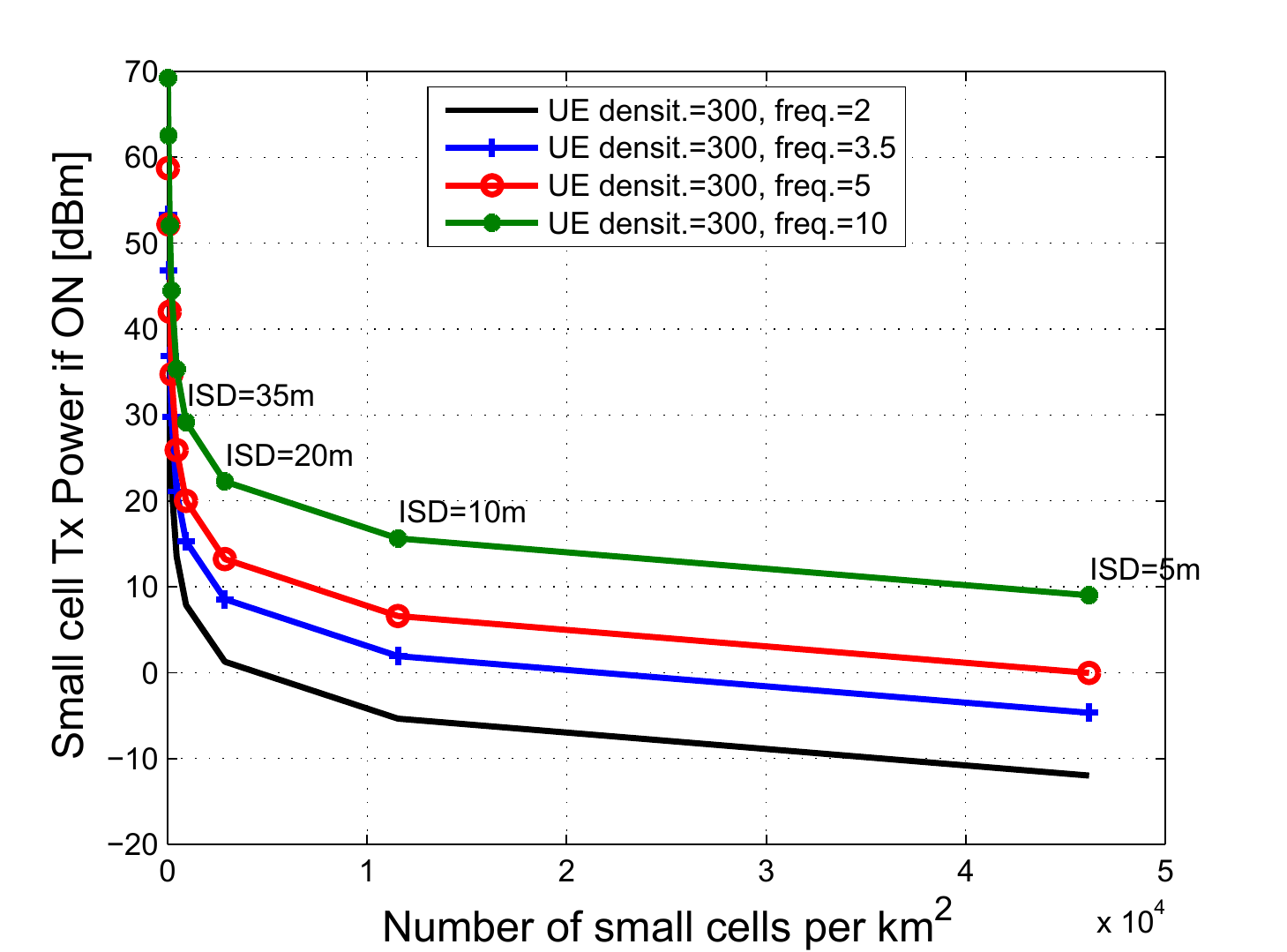}}
	\subfigure[Transmit power of the network per km$^2$ for different frequency bands.]{\label{fig:txPowerNetworkFreq}
	\includegraphics[width=3.5in]{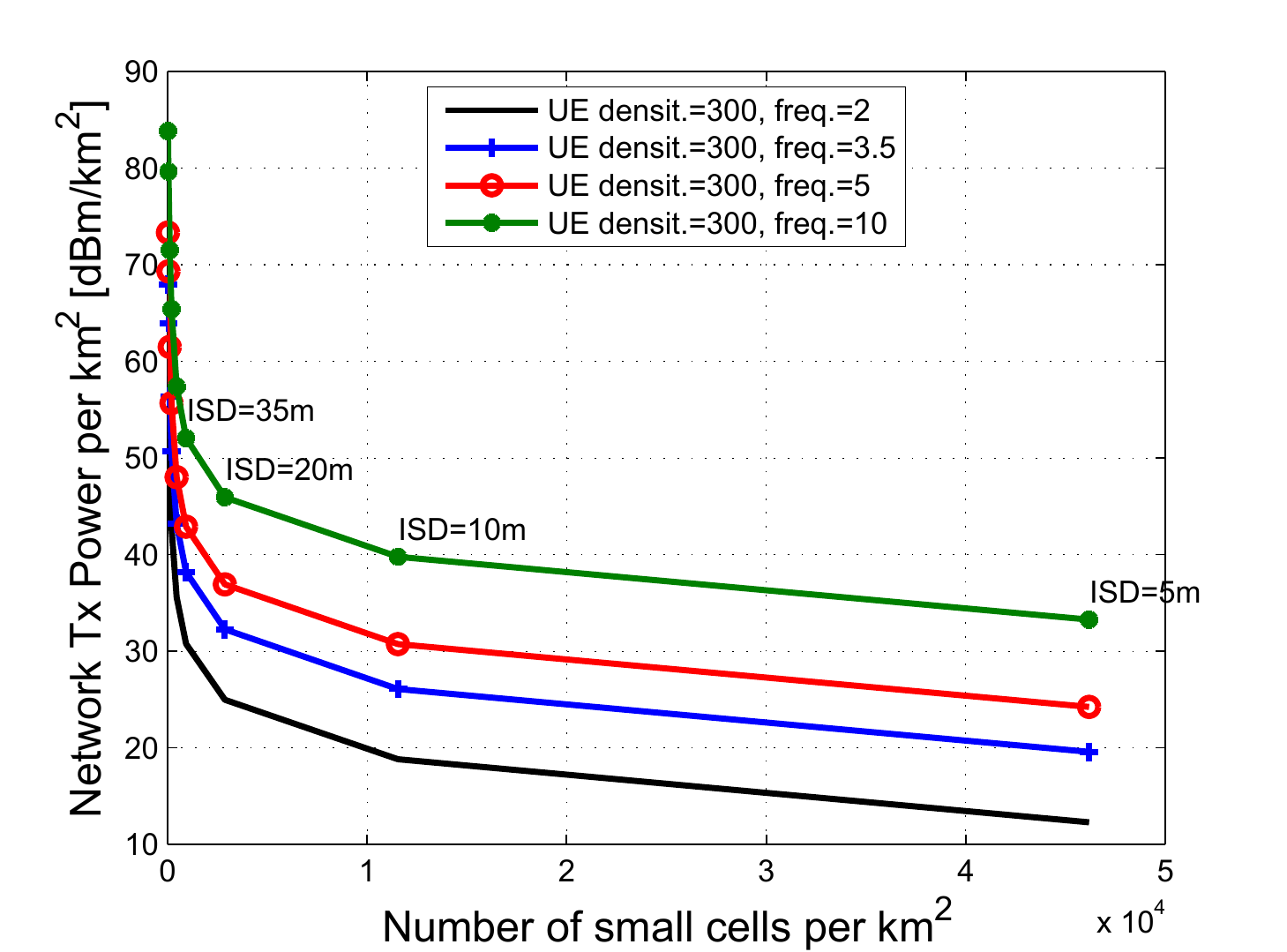}}
  \caption{Transmit power per active \ac{BS} and transmit power of the network per km$^2$ for different network configurations where the carrier frequencies are 2.0, 3.5, 5.0 or 10\,GHz.
  The rest of the parameters are $d=300$\,UE/km$^2$, $ud$=1, $s=1$, $f=2$\,GHz, $a=1$ and $t=$12\,dB.}
  \label{fig:txPowerFreq}
\end{figure*}

\begin{figure*}[t]
  \centering
	\subfigure[Average UE throughput for different number of antennas per \ac{BS} and frequency bands. ]{\label{fig:throughputAverageAllAnt}
	\includegraphics[width=3.5in]{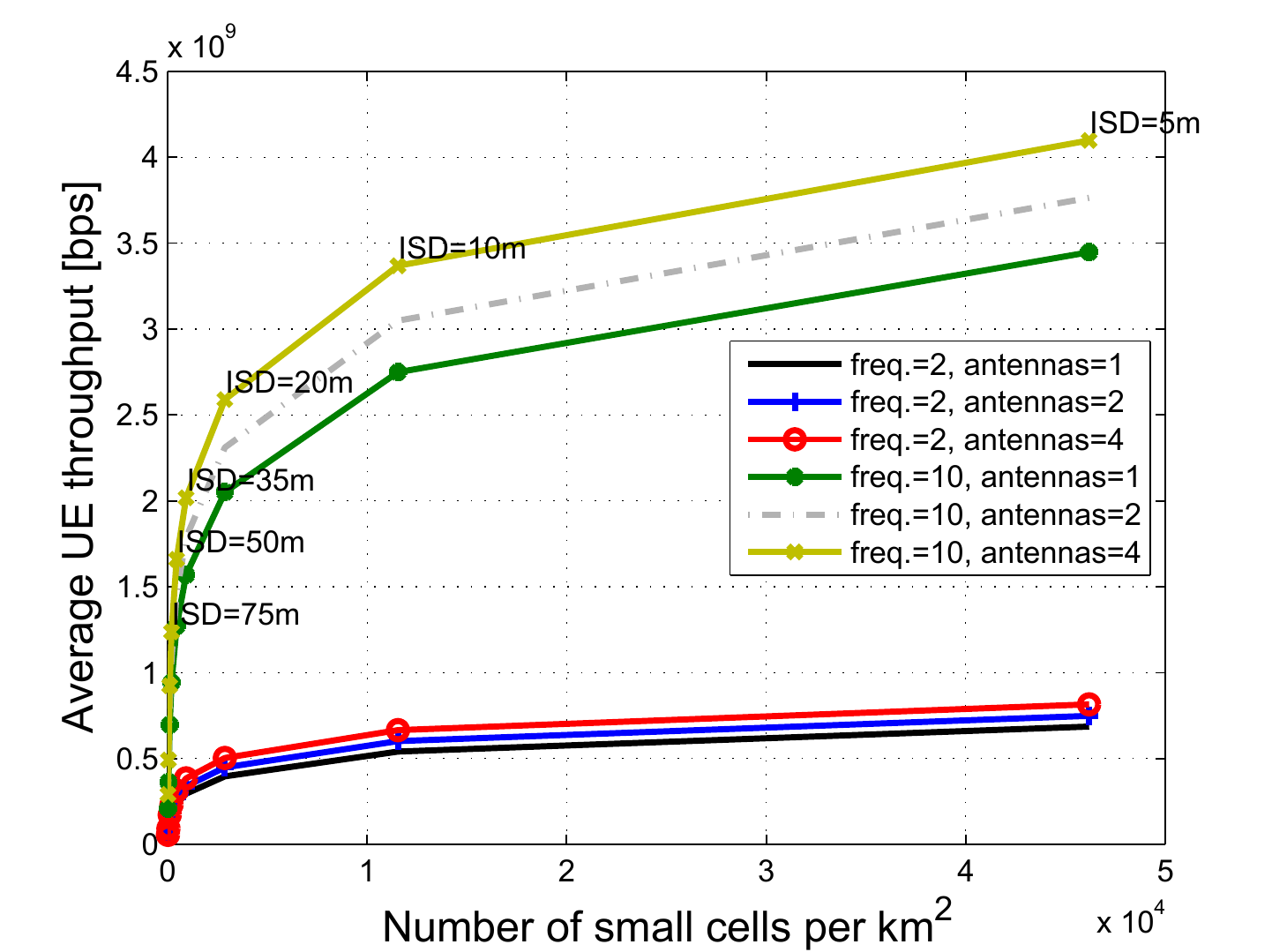}}
	\subfigure[5\,\%-tile UE throughput for different number of antennas per \ac{BS} and frequency bands.]{\label{fig:throughput5percentileAllAnt}
	\includegraphics[width=3.5in]{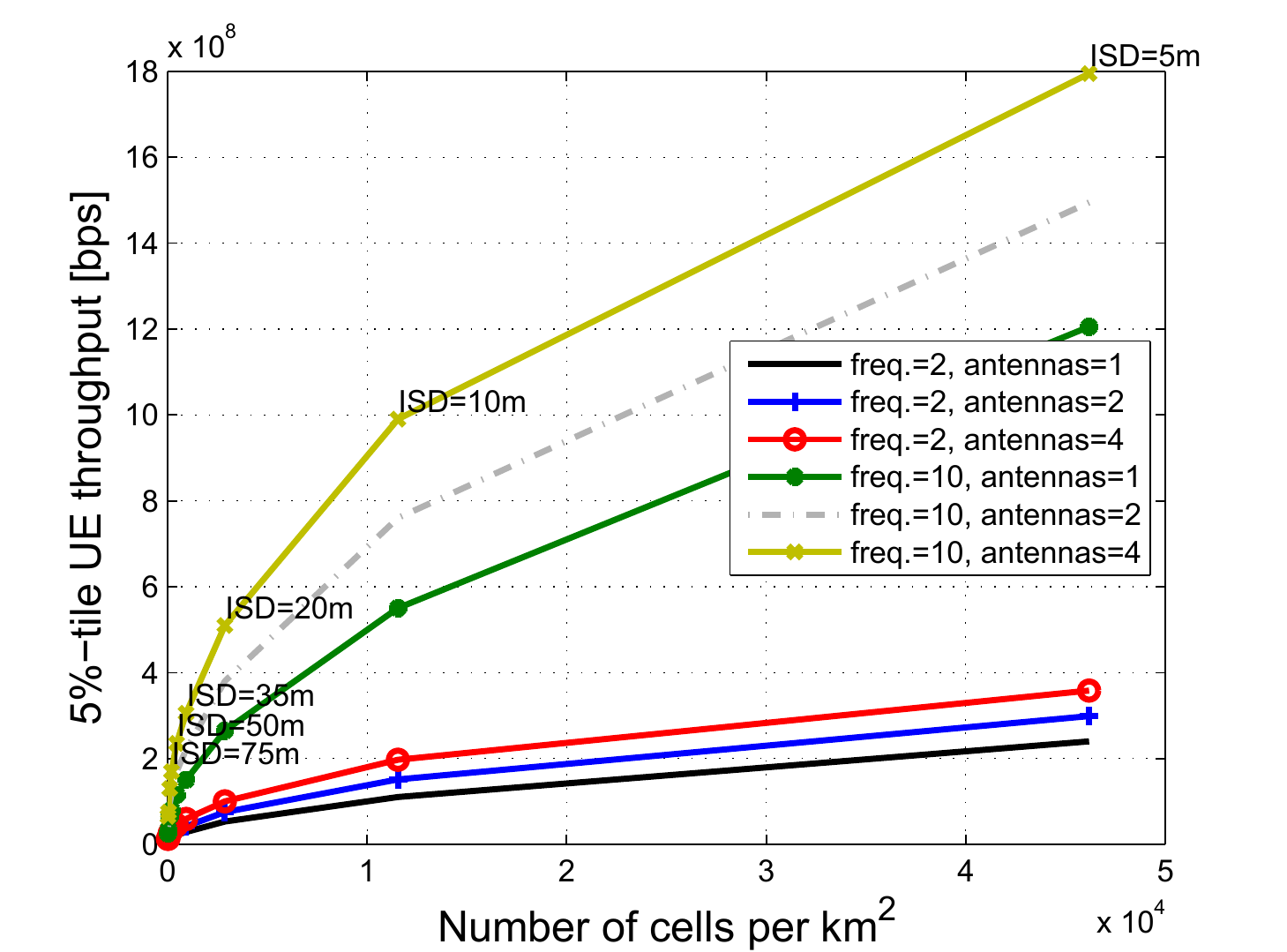}}
  \caption{Average and 5\%-tile UE throughput for different network configurations where the carrier frequencies are 2.0 or 10\,GHz
  and the number of antennas per \ac{BS} are 1, 2 or 4.
  The rest of the parameters are $d=300$\,UE/km$^2$, $ud$=1, $s=1$, $f=2$\,GHz and $t=$12\,dB.}
  \label{fig:throughputAllAnt}
\end{figure*}

%[Holger]: replace gray dashed lines with some other linestyle
%[David]: to be replaced in the camera ready version, other comments on figures may come 

\section{Higher Frequency Bands}
\label{Sec:higherFrequecnyBands}

As shown by the Shannon-Hartley theorem presented in (\ref{eq:Shannon}),
network capacity linearly increases with the available bandwidth.
Therefore, increasing the available bandwidth is an appealing proposition to enhance network capacity.
However, spectrum is a scare resource,
specially at the lower frequency bands, [500-2600]\,MHz,
which are in use today by radio and TV stations as well as the first wireless communication systems due to their good propagation properties.
These frequency bands are thus heavily regulated,
and it is unlikely that large bandwidths become available from them in the near future.
As a result,
in the quest to increase network capacity,
vendors and operators have started to look at the usage of higher frequency bands, $\geq$3500\,MHz,
where large bandwidths $\geq$100\,MHz are available.
%[Ming]: $\leq$3500\,MHz => $\geq$3500\,MHz; $\leq$100\,MHz > $\geq$100\,MHz
%[David]: Sorry for the mistake. This is way a new pair of eyes is handy to review papers.
%[Ming]: Glad to be helpful!
%[David]: Certainly, not just because of this minor issue ;)
Due to the higher path losses at higher frequency bands,
these bands were never appealing for large macrocells but now suit well the operation of small cells,
targeted at short ranges.
These higher path losses should not only be considered as a disadvantage but also as an opportunity,
since they effectively mitigate interference from neighbouring cells,
and thus allow a better spatial reuse.
Moreover, it allows using smaller antennas and packing more of them per unit of area, which benefits multi-antenna techniques.
In the following, the capacity gains provided by and the challenges faced when using higher frequency bands are discussed.

In order to assess the capacity gains provided by ultra-small cell deployments and the use of higher frequency bands,
Fig.~\ref{fig:throughput} shows the average and 5\%-tile UE throughput for different densification levels and four different network configurations
where the carrier frequencies are 2.0, 3.5, 5.0 and 10\,GHz and the available bandwidth is 5\,\% of the carrier frequency,
i.e., 100, 175, 250 and 500\,MHz, respectively.
In this case, the targeted \acp{SNR} by the small cell \acp{BS} at $\frac{\sqrt{3}}{2}$ of the \ac{ISD} is 12\,dB.
thus assuring a constant coverage area regardless of the frequency band.
From the results,
different observations can be made.
Based on Fig.~\ref{fig:throughput}, 
in the following we first analyse the impact of densification in network capacity and then improvements brought by the use of higher frequency bands.
%[Amir]:Can you please rewrite the last sentence?
%[David]: Done
%[Amir]: Thank :)

In terms  of densification,
the average and 5\,\%-tile \ac{UE} throughput do not increase linearly with the number of deployed \acp{BS},
but with diminishing gains.
This is due to the finite nature of \ac{UE} density and the characteristic of its distribution.
In a first phase,
with a low \ac{BS} density and up to an \ac{ISD} of 35\,m,
the UE throughput rapidly grows, almost linearly with the number of cells,
due to cell splitting gains and spatial reuse.
Subsequently, in a second phase,
once the fundamental limit of spatial reuse is reached -- \emph{one \ac{UE} per cell} --
the UE throughput continues growing at a lower pace with the network densification.
This is due to the combined effect of both,
bringing the UE closer to the serving cell BS through densification and further from the interfering BSs through idle modes.
These two effects together results in a \ac{SINR} enhancement.
%[Ming]: Very interesting and useful conclusion!!!
%[David]: This is the argument I did to answer one of your previous comments.
%[Ming]: Agreed!
These transition and two regimes are more obvious for the average than for the 5\,\%-tile UE throughput,
since cell-edge \acp{UE} benefit more from proximity gains and interference mitigation.
%[Holger]: unclear from looking at figures
%[David]: to be fixed in the camera ready version, other comments on figures may come 

Looking at the average and 5\,\%-tile \ac{UE} throughput using the 200\,m \ac{ISD} case with a 100\,MHz bandwidth as a baseline,
the 35\,m \ac{ISD} case can provide an average and cell-edge gain of 7.56$\times$ and 5.80$\times$, respectively,
while the gains provided by the 5\,m \ac{ISD} case are 17.56$\times$ and 48.00$\times$.
This shows that a significant increase in network performance can be achieved through network densification.

In terms of frequency bands,
the average and 5\,\%-tile \ac{UE} throughput increase linearly with the carrier frequency due to the larger available bandwidth,
showing the use of larger bandwidth as a key to achieve larger \ac{UE} throughputs.

Looking at the average and 5\,\%-tile \ac{UE} throughput with the 35\,m \ac{ISD} case with a 100\,MHz bandwidth as a baseline,
a bandwidth of 250\,MHz can provide an average and cell-edge gain of 2.59$\times$ and 2.58$\times$,
while the gains provided by a bandwidth of 500\,MHz are 5.31$\times$ and 5.17$\times$.
Although not as large as the gains provided by network densification,
they also represent a significant increase in network performance.

It is important to highlight that the capacity gains seen through using higher frequency bands do not come for free.
The cost of the \ac{UE} and \ac{BS} equipment increases with the carrier frequency,
as more sophisticated analogue circuit components are needed.
Moreover, a larger transmit power is required for both lighting up the more subcarriers existing in a wider bandwidth
and compensating for the higher path losses at higher frequency bands~\cite{6472194}.
Fig~\ref{fig:txPowerFreq} shows how both the transmit power per active \ac{BS} and the overall transmit power used by the network increases with the wider bandwidth and the larger path losses,
where this increase is not negligible and up to 24.34\,dB.
This transmit power increase is prohibitive in macrocell \acp{BS} where the required power would be up to 70\,dBm,
but it is still well suited to ultra-dense small cell deployments
where \acp{BS} could operate this large bandwidth with less than 20\,dBm of transmit power.
In order to reduce transmit power,
the identification and characterisation of hot spots becomes critical~\cite{1458284}~\cite{KlessigTrafficModelling}
since deploying \acp{BS} where they are most needed,
e.g., right in the middle of a circular hot spot,
will decrease the average \ac{UE} path loss to the serving \ac{BS},
and reduce this transmit power.
Lighting up only those subcarriers with good channel quality that are necessary to achieve the required \acp{UE} throughput targets will also help to reduce transmit power.
%[Ming]: Just a note, the short CIR might stand in the way since the resulting frequency-domain response would be of very low selectivity. We don't need to mention this in the paper.
%[David]: Agreed. Short CIR will mitigate multi-user diversity gains but may simplify scheduling process.  Less cycles needed. I am working on this too. Hope to have a summarising paper soon. Keep you updated. I have updated a bit the above sentence.
%[Ming]: Agreed.

From the results,
one can see that, when combining network densification with increased bandwidth, the targeted average 1\,Gbps per \ac{UE} is reachable with an \ac{ISD} of 50m and 500\,MHz bandwidth,
or 20\,m \ac{ISD} and 250\,MHz bandwidth.
The first and the second combination resulting in an average of 1.27\,Gbps and 1.01\,Gbps per \ac{UE},
respectively.
Thus, it can be concluded that the usage of larger cellular bandwidth than today's 100\,MHz is required
to meet the expected high data rates.

Taking the usage of higher frequency bands to an extreme,
vendors and operators have also started to look at exploiting mmWave with carrier frequencies on 22, 60 and 77\,GHz,
where the available bandwidth is enormous, $\geq$1\,GHz~\cite{5783993}.
However, diffraction and penetration through obstacles are hardly possible at these high frequency bands,
and thus only \ac{LOS} or near-\ac{LOS} links seem feasible.
In addition, the range of the cell may be confined by high atmospheric phenomenas.
Water and oxygen absorption significantly increase path losses,
especially at 22\,GHz and 60\,GHz, respectively.
As a result, providing the required coverage range through larger transmit powers is not feasible anymore,
as it is in the sub-10\,GHz bands.
Thus, it is expected that active antenna arrays and beamforming techniques
become essential to overcome the increased path losses at these high frequency bands~\cite{6736761}.
%[Challenge]: Massive MIMO, mmWave

\section{Multi-antenna Techniques and Beamforming}
\label{Sec:MIMO}

Previous results showed that for a single antenna small cell \ac{BS},
the targeted average 1\,Gbps per \ac{UE} is only reachable with an \ac{ISD} of at least 50\,m and a bandwidth of 500\,MHz (or 35\,m and a bandwidth of 250\,MHz).
In order to enhance \ac{UE} performance and bring down this still large \ac{BS} density,
the usage of multiple antennas at the small cell \ac{BS} is explored in the following.

Multiple antenna systems provide a number of degrees of freedom for transmitting information,
which may vary from one to a number upper bounded by the number of antennas.
The higher the number of degrees of freedom available,
the better spectral efficiency can be expected.
However, how many effective degrees of freedom are available is mostly related to the spatial correlation of the channels~\cite{4274995}.
Given a number of degrees of freedom available,
two multi-antenna techniques stand out.
Beamforming makes use of only one of the degrees of freedom available,
while spatial multiplexing may use all of them.
Beamforming benefits from a low complexity implementation,
together with its ability to extend the cell range by focusing the transmit power in a certain direction.
However, it may be suboptimal in terms of spectral efficiency.
In contrast, spatial multiplexing has the potential to approach the maximum channel capacity,
linearly increasing the capacity of the channel with the minimum of number of transmit and receive antennas.
However, it is significantly more complex to implement~\cite{4274995}.
%[Amir]:Spatial multiplexing increases the capacity by the minimum of number of transmit and receive antennas. Is it necessary to modify?
%[David]: Good point.

This paper focuses on quantised maximal ratio transmission (MRT) beamforming,
using the standardised \ac{LTE} code book beamforming approach~\cite{LTEbook},
partly due to its simplicity and partly due to the fact that the small cell sizes in an ultra-dense deployment may result in a large spatial correlation of the channels,
thus limiting the degrees of freedom available and rendering spatial multiplexing unsuitable.
The effectiveness of spatial multiplexing in ultra-dense small cell deployments is left for future analysis.
%[Challenge]: Understanding spatial multiplexing

Here,
it is assumed that
each \ac{BS} is equipped with a horizontal antenna array comprised of 1, 2, or 4 antenna elements,
and each \ac{UE} has a single antenna.
The existing transmit power is equally distributed among antennas.
The characteristics of the antenna elements used in this analysis are described in Appendix~\ref{sec:appendix}.
Based on measurements over \ac{BS} pilots signals,
the \ac{UE} suggests to the \ac{BS} the precoding weights specified in the LTE code book~\cite{TS36211} that maximise its received signal strength,
and the \ac{BS} follows such suggestion.
This horizontal beamforming helps to shape the horizontal antenna pattern at the \ac{BS} and thus focus the energy towards the \ac{UE} in the horizontal plane.
Interference mitigation is also achieved in an opportunistic manner~\cite{Viswanath:02}.
Real-time inter-\ac{BS} coordination required for cooperation is not supported.

Fig.~\ref{fig:throughputAllAnt} shows the average and 5\%-tile UE throughput for different network configurations
where the carrier frequencies are 2.0, 3.5, 5.0 and 10\,GHz and the number of antennas per \ac{BS} are 1, 2 or 4.
Control channels are not beamformed.
The targeted \acp{SNR} by the small cell \ac{BS} at $\frac{\sqrt{3}}{2}$ of the \ac{ISD} is 12\,dB.
thus assuring a constant coverage area regardless of the frequency band.
The constant coverage is maintained at the expense of larger transmit power,
as explained earlier.
Different remarks can be made based on the results.

Beamforming gains increase in a diminishing manner with the number of antennas.
This is in line with the linear antenna array theory,
which indicates that beamforming antenna gains increase logarithmically with the number of antennas~\cite{antennaTheorybook}.
Looking at the average \ac{UE} throughput for the 35\,m \ac{ISD} case with a 500\,MHz bandwidth,
the average gain of using 2 antennas over 1 antenna is 13.77\,\% (1 more antenna is needed),
while the gain of using 4 antennas over 2 antenna is 13.05\,\% (2 more antennas are needed).
The overall average gain from 1 to 4 antennas is 18.92\,\%.

% [Holger] check gains - combined gain looks strangely low considering the component gains from 1->2 and 2->4 antennas

Beamforming gains are larger at the cell-edge.
This is because of the interference mitigation provided by the beamforming,
which is more noticeable at the cell-edge.
Looking at the cell-edge \ac{UE} throughput for the 35\,m \ac{ISD} case with a 500\,MHz bandwidth,
the cell-edge gain of using 2 antennas over 1 antenna is 46.67\,\% (1 antenna increase),
while the gain of using 4 antennas over 2 antennas is 38.63\,\% (2 antennas increase).
The overall cell-edge gain from 1 to 4 antennas is 48.96\,\%.

% [Holger] check gains - combined gain looks strangely low considering the component gains from 1->2 and 2->4 antennas

Beamforming gains are larger for larger cell sizes.
This is because the beamforming helps improving the received signal strength of \acp{UE},
which may receive only a poor received signal strength if beamforming is not in place.
For the 200\,m \ac{ISD} and the 5\,m \ac{ISD} cases with a 500\,MHz bandwidth,
the average gains of using 2 antennas over 1 antenna are 19.05\,\% and 9.23\,\%, respectively (1 antenna increase),
while the gains of using 4 antennas over 2 antenna are 17.37\,\% and 8.87\,\%, respectively (2 antennas increase).

From the results,
one can see that the targeted average 1\,Gbps per \ac{UE} is now reachable with an \ac{ISD} of 75\,m, 500\,MHz bandwidth and 4 antennas (or 35\,m, 250\,MHz bandwidth and 4 antennas),
showing that multi-antenna techniques enhance \ac{UE} performance and can bring down \ac{BS} density
to meet the targeted \ac{UE} demands.
However, overall, beamforming gains are estimated to be of the order of up to 1.49x,
which are poor compared to the larger gains provided by network densification and use of higher frequency bands.
Therefore, since the larger path loss at the frequency bands between 2 and 10\,GHz could be compensated via larger transmit powers,
the existing antennas at the small cell may be better exploited by using spatial multiplexing,
which has the potential to linearly increase the performance with the number of antennas
provided that the required degrees of freedom exists in the channel.
Spatial multiplexing in ultra-dense small cell networks is one of our future lines of research.

\section{Scheduling}
\label{Sec:scheduling}

In \ac{LTE}, a \ac{RB} refers to the basic time/frequency scheduling resource unit to which a \ac{UE} can be allocated.
Each \ac{RB} expands 180\,KHz in the frequency domain and has a duration of  1\,ms in the time domain.
The \ac{RB} consists of 12 subcarriers of 15\,KHz and its 1\,ms \ac{TTI} is referred to as a subframe.

Unlike other diversity techniques that aim to average the signal variations to mitigate the destructive impact of multi-path fast fading,
channel dependent scheduling takes advantage of multi-path fading
by allocating to each \ac{RB} the \ac{UE} having the best channel conditions.
This is one of the main enhancements of \ac{LTE} over \ac{UMTS}~\cite{Sesia2009}.
Such type of scheduling leads to multi-user diversity gains,
which have been shown to roughly follow a double logarithm scaling law in terms of capacity with regard to the number of \acp{UE} per \ac{BS}
in macrocell scenarios~\cite{4063519}.
It is important to note that in order to aid the channel sensitive scheduling and exploit multi-user diversity gains,
\acp{UE} need to report \ac{DL} \ac{CQI} back to their serving \acp{BS},
which allows the scheduler to asses the \ac{UE} channel quality and perform the scheduling according to a specified metric.

In a network with $U$ \acp{UE},
each \ac{UE} may undergo varying channel conditions
where better channel quality generally refers to higher signal quality and higher throughput.
Sharing the resources fairly between \acp{UE} experiencing different channel qualities is a challenging task~\cite{4146798}.
In the following, a survey of traditional small cell schedulers is provided.

\emph{Opportunistic} schedulers select the \ac{UE} with the best channel quality at each time/frequency resource,
aiming to solely maximise the overall throughput,
whereas \emph{\ac{RR}} schedulers treat the \acp{UE} equally regardless of their channel quality,
giving them the same share of time/frequency resources.
The former scheduler can increase system throughput remarkably compared to the latter at the expense of fairness,
since \acp{UE} with relatively bad channel qualities may be never scheduled~\cite{4489366}.
\emph{\ac{PF} schedulers}, in contrast, exploit multi-user diversity based on \ac{UE} \acp{CQI},
attempting to maximise the throughput while simultaneously reinforcing a degree of fairness in serving all the \acp{UE}.

The \ac{PF} scheduling metric basically aims at weighting the UE's potential instantaneous performance by its average performance,
and \ac{PF} schedulers usually consists of three stages.
In the first stage, according to buffer information,
the schedulable set of \acp{UE} is specified.
The second stage is the time domain scheduling,
which is in charge of reinforce fairness and selects the $N_{max}$ \acp{UE} that will be input to the frequency domain scheduler,
and thus be allocated \acp{RB} in the current subframe.
The last stage corresponds to the frequency domain scheduling,
i.e., allocation of \acp{UE} to \acp{RB}.
The complexity of the frequency domain scheduler highly depends on the number of its input \acp{UE}~\cite{4526113}~\cite{851593},
and thus the time domain scheduler has a major impact on the complexity of the frequency domain scheduler.

Multiple time domain and frequency domain \ac{PF} metrics exist~\cite{4526113}.
A \ac{PF} scheduling metric in time domain can be defined as
\begin{equation}
	M_{m,u}^{\rm PF-TD} = \frac{\hat{D}_{m,u}}{R_{m,u}},
	\label{eq:td}
\end{equation}
where $R_u$ and $\hat{D}_u$ are the past average throughput and potential instantaneous throughput of the $u^{th}$ \ac{UE} connected to the $m^{th}$ \ac{BS}, respectively~\cite{4526113},
and the past average throughput is calculated using a moving average.
The \acp{UE} will be then ranked constantly according to the metric in (\ref{eq:td}),
and the $U_{max}$ \acp{UE} with maximum preference are passed on to the frequency domain scheduler.
A \ac{PF} scheduling metric in frequency domain can be defined as
\begin{equation}
	M_{m,u,k}^{\rm PF-FD} = \frac{{\gamma}_{m,u,k}}{\sum\limits_{k = 1}^{\rm N_{RB}} {\gamma}_{m,u,k}},
	\label{eq:pf}
\end{equation}
where the numerator ${\gamma}_{u,k}$ is the \ac{SINR} of the $u^{th}$ \ac{UE} connected to the $m^{th}$ \ac{BS} on the $k^{th}$ \ac{RB}
and the denominator is the sum of the $u^{th}$ UE SINRs over all \acp{RB},
which represents its average channel quality at specified subframe~\cite{4526113}.
The \acp{UE} will be ranked constantly according to the metric in (\ref{eq:pf}),
and the \ac{UE} with maximum preference is selected to be served on the $k^{th}$ \ac{RB}.

It is important to note that channel gain fluctuations occur from \ac{RB} to \ac{RB} due to multi-path fast fading.
In order to consider the multi-path fast fading effect in this section,
the \ac{BS} to \ac{UE} channel gain  $G_{m,u} {\rm [dB]}$ also comprises the multi-path fast fading gain,
which is modelled using a distant dependent Rician channel.
The lower the \ac{UE} to serving \ac{BS} distance,
the lower the channel fluctuations due to multi-path fading~\cite{2015Jafari}.
This model is presented in Appendix~\ref{sec:appendix2}.
%[David]: The explanation of this model can be removed if we remove the previous sentence and the appendix. Up to you guys? Feel free to comment
%[Ming]: I propose to keep it because it is an essential piece of knowledge to understand our results on the RR and the PF schedulers discussed in the following paragraphs.
%[Amir]: I agree with Ming.
%[David]: Let's leave it.

In the following, the performance of \ac{RR} and \ac{PF} schedulers is analysed under different densification levels.

\begin{figure}[t]
	\centering
	\includegraphics[width=3.8in]{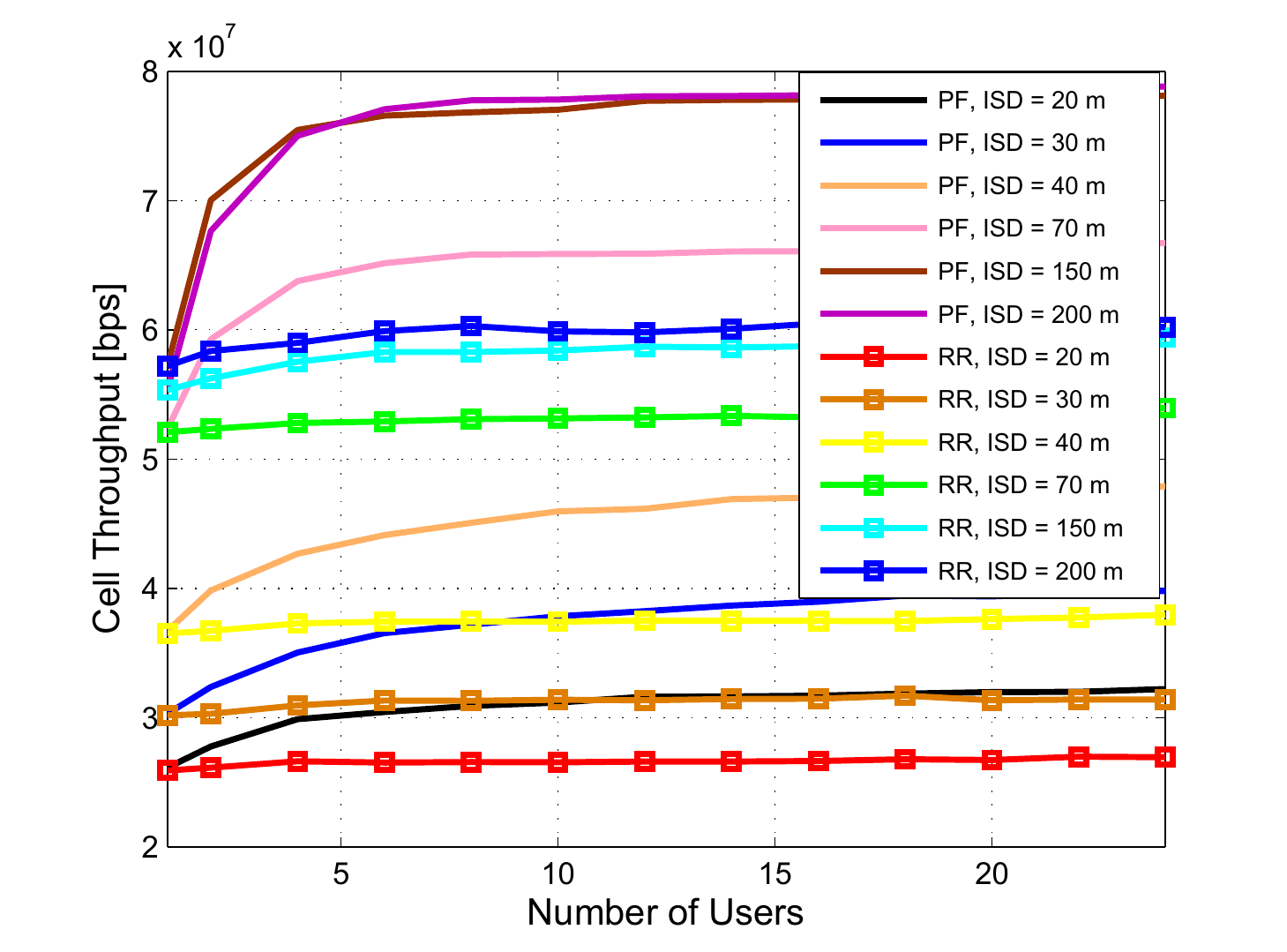}
	\caption{Average cell throughput for various \ac{ISD} with \ac{RR} and \ac{PF} schedulers.}
	\label{fig:cell_ov_thr}
\end{figure}

Fig.~\ref{fig:cell_ov_thr} shows the performances of \ac{RR} and \ac{PF} schedulers in terms of cell throughput with respect to the number of served \acp{UE} per \ac{BS} for different \acp{ISD}.
Let us look first at the trends.
When using \ac{RR},
the number of served \acp{UE} per \ac{BS} does not impact the cell throughput,
since \ac{RR} does not take into account the \ac{UE} channel quality and therefore does not take advantage of multi-user diversity.
In contrast, \ac{PF} is able to benefit from multi-user diversity,
and the cell throughput increases with the number of served \acp{UE} per \ac{BS} up to a given extent.
There is a point in which having more \ac{UE} per \ac{BS} does not bring any cell throughput gains,
and this point is lower with the network density.
For this particular case,
having more than 8 \acp{UE} per cell does not noticeably increase cell throughout for an \ac{ISD} of 200\,m,
while this number is reduced to 6 \acp{UE} for an ISD of 20\,m.
Multi-user diversity gains vanishes with network densification due to stronger \ac{LOS} propagation and less fluctuating channel conditions.
Paying now attention to the overall per cell performance,
although \ac{PF} always outperforms \ac{RR},
it is important to note that the \ac{PF} scheduler starts losing its advantage with respect to \ac{RR} in terms of cell throughput with the reduced cell size.
For a given number of \ac{UE} per \ac{BS}, let's say 4,
the \ac{PF} gain over \ac{RR} reduces from 21.2\,\% to 12.4\,\% and 10.5\,\% for ISDs of 150, 40 and 20\,m, respectively.

\begin{figure}[t]
	\centering
	\includegraphics[width=3.8in]{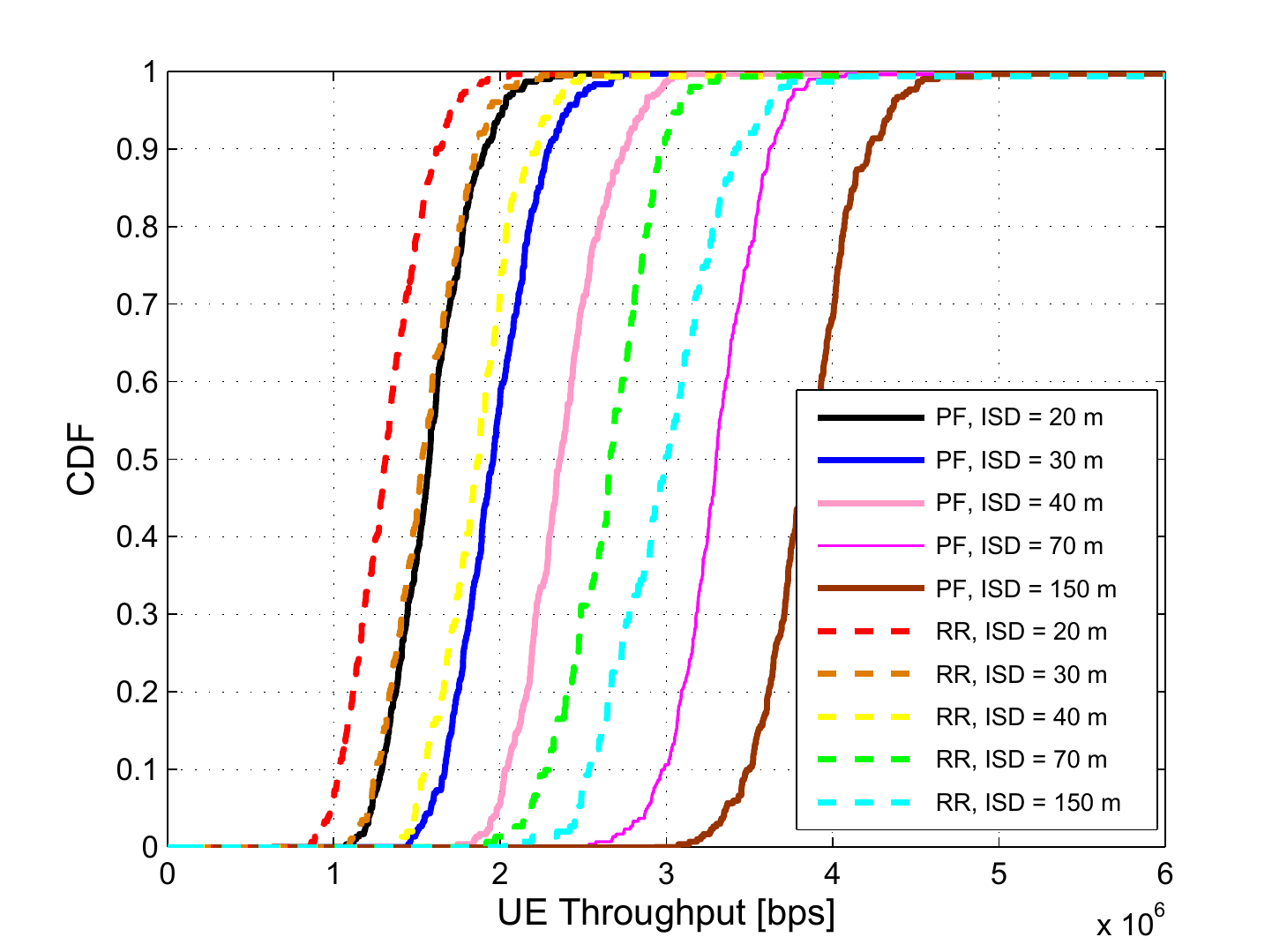}
	\caption{CDF of UE throughput for different ISDs.}
	\label{fig:ue_cdf}
\end{figure}

Shrinking the cell size not only reduces cell throughput but also reduces \ac{UE} throughput.
Applying the \ac{PF} scheduler,
Fig.~\ref{fig:ue_cdf} shows that due to both interference enhancement due to NLOS to LOS transition as well as multi-user diversity loss due to low channel fluctuations,
the \ac{UE} throughput for a given number of served \acp{UE} per \ac{BS} drops down with network densification. 
For instance, reducing the \ac{ISD} from 150\,m to 40\,m and 20\,m, the \ac{UE} 5\,\%-tile throughput drops by $\sim$40.8\,\% and $\sim$36.7\,\%, respectively. 
Comparing the \ac{UE} 5\,\%-tile throughput of \ac{PF} and \ac{RR} for an ISD of 20m,
the gain of the former is almost negligible, around $\sim$9\,\%.

%[Amir]: Shall we add UE_throughput figure here to support the numbers in the text?
%[David]: Shouldn't this be reflected on the CDF? Maybe we should you mean instead of average.
%[Amir]:Done.

The minor gains of \ac{PF} scheduler over the \ac{RR} one at low \acp{ISD} in terms of cell and \ac{UE} throughput suggests that
\ac{RR} scheduler may be a better choice in dense small cell deployments considering its lower complexity.
The complexity of \ac{PF} lies in the evaluation of each \ac{UE} on each \ac{RB} considering a greedy \ac{PF} that operates on a per \ac{RB}.
The \ac{PF} complexity with exhaustive search is considerably higher.
This conclusion may have a significant impact in the manufacturing of small cell \acp{BS}
where the \ac{DSP} cycles saved due to the adoption of \ac{RR} scheduling can be used to enhance the performance of other embedded technologies.

%Notes:
%-Include Rican model in appendix?

\section{Energy-Efficiency}
\label{Sec:costAndEnergyEfficiency}

Deploying a large number of small cells, millions, tens of millions, poses some concerns in terms of energy consumption.
For example, the deployment of 50\,million femtocells consuming 12\,Watts each will lead to an energy consumption of 5.2\,TWh/a,
which is equivalent to half the power produced by a nuclear plant~\cite{Claussen2012P}.
This approach does not scale,
and thus the energy efficiency of ultra-dense small cell deployments should be carefully considered to allow the deployment of sustainable networks.

In Section~\ref{sec:densification},
Fig.~\ref{fig:txPowerNetwork} showed how the overall transmit power used by the network significantly reduces with the small cell \ac{BS} density
when an efficient idle mode capability is used.
This is because the reduction of transmit power per cell outweighs the increased number of active cells.
However, this observation may not hold when considering the total power consumption of each small cell \ac{BS},
since a \ac{BS} in idle mode may still not consume a non-negligible amount of energy,
thus impacting the energy efficiency of the network.
In order to better understand the impact of network densification on the power consumption,
the energy efficiency of ultra-dense small cell networks in terms of throughput per Watt ([bps/W]) is analysed in the following.

This study uses the power model developed by the GreenTouch project~\cite{DessetPowerModelling}.
This power model estimates the power consumption of a cellular \ac{BS},
and is based on tailored modelling principles and scaling rules for each \ac{BS} component
i.e., power amplifier, analog from-end, digital base band, digital control and backhaul interface and power supply.
Moreover, it provides a large flexibility:
multiple \ac{BS} types are available,
they can be configured with multiple parameters (bandwidth, transmit power, number of antenna chains, system load, etc.)
and include different optimised idle modes.
Among the provided idle modes,
we considered the GreenTouch slow idle mode and the GreenTouch shut-down state,
where most or all components of the \ac{BS} are deactivated, respectively.
These two modes are the most efficient ones provided by the model.

Table~\ref{tab:powerConsumption} shows the estimated power consumption for different small cell \ac{BS} types with different transmit power,
where such power consumption is given for the active mode (full load) as well as the slow idle mode and the shut-down state,
and for \acp{BS} with 1, 2 or 4 antennas.
Note that we used the 2020 small cell BS type in the model and that the presented analysis considers a 20\,MHz bandwidth.
In contrast to macrocell \acp{BS},
it is important to realise that the power consumption of a small cell \ac{BS} linearly scales with the number of antenna chains,
since it is the most contributing component to power consumption
in a small cell \ac{BS}.
% [Holger] ??? should not the processing become the most contributing component when the cell size becomes smaller?
% [David]: According to the model, it is not

\begin{figure}[t]
  \centering
	\includegraphics[width=3.8in]{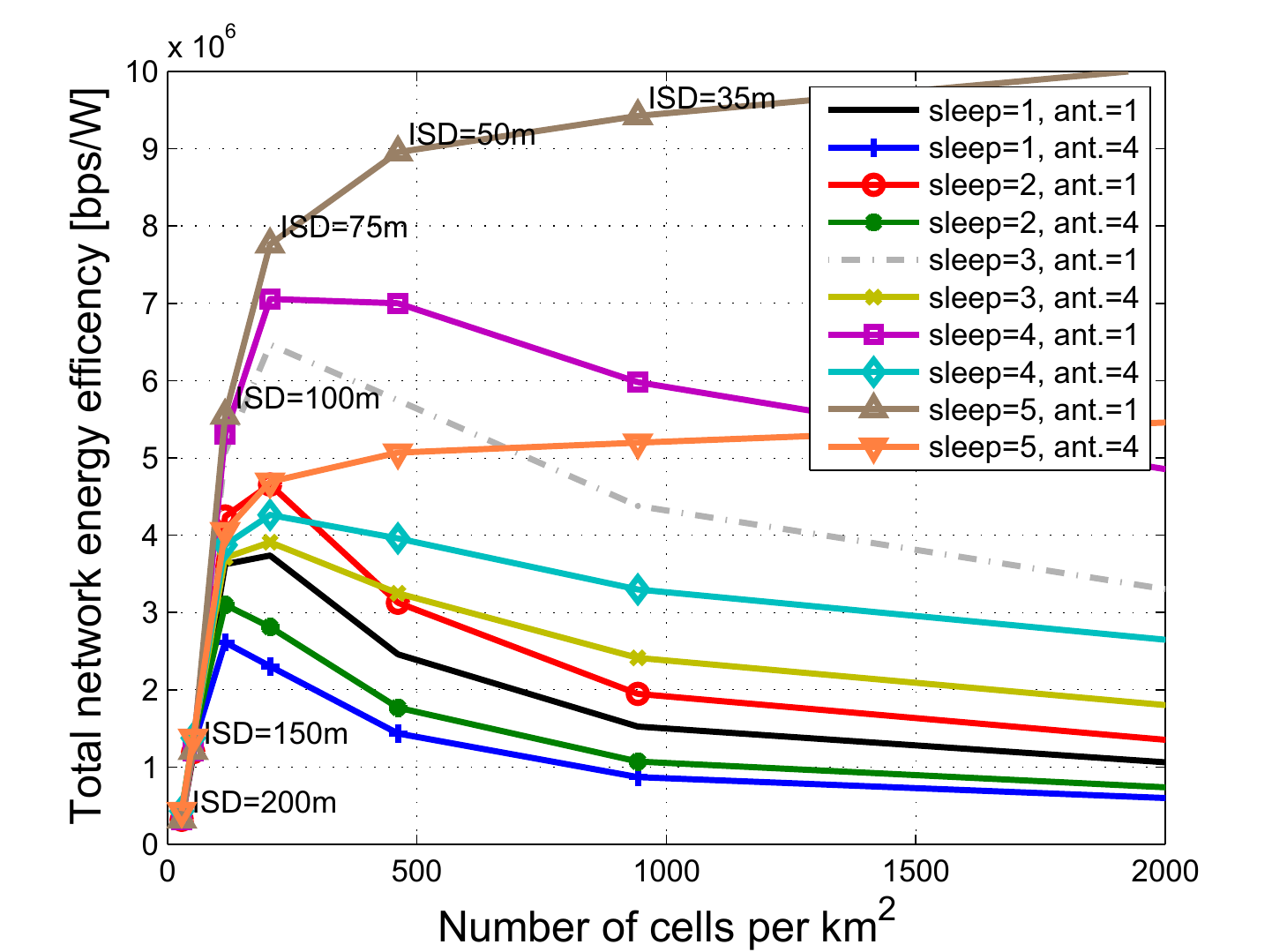}
  \caption{Energy efficiency in bps per Watt for different network configurations where the carrier frequency is  2.0\,GHz,
  the bandwidth is 20\,MHz,
  and the number of antennas per \ac{BS} are 1, 2 or 4.
  The rest of the parameters are $d=300$\,UE/km$^2$, $ud$=1, $s=1$, $f=2$\,GHz and $t=$12\,dB.}
  \label{fig:energyEfficency}
\end{figure}

Based on the values of Table~\ref{tab:powerConsumption} and the throughput analysis in previous sections,
Fig.~\ref{fig:energyEfficency} shows the energy efficiency in bps/W for different network configurations
with 1, 2 or 4 antennas per \ac{BS}.
The idle mode profile is indicated by \emph{sm}.
In addition to the slow idle mode (\emph{sm}=1) and the shut-down state (\emph{sm}=2) models provided by the GreenTouch project,
three futuristic idle modes are considered,
where their energy consumption is 30\,\% (\emph{sm}=3), 15\,\% (\emph{sm}=4), or 0\,\% (\emph{sm}=5 - idle small cell does not consume anything) of the GreenTouch slow idle mode power consumption model (\emph{sm}=1).

For any given idle mode,
results show that increasing the number of antennas at the small cell \ac{BS} always decreases the energy efficiency.
This is because the performance gain provided by adding a new antenna through beamforming
is not large enough to cope with the increase in power consumption resulting from adding a new antenna chain at the small cell \ac{BS}.
%[Holger] Do we assume that the transmit power per antenna remains the same or the total transmit power? This may have an effect on this conclusion.
%[David]: Yes the transmit power is constant per antenna 
A different conclusion may be obtained for the macrocell case,
where adding a new antenna chain does not lead to a large increase in the total power consumption of the \ac{BS},
since other components consume much more energy.
Conclusions for the small cell \ac{BS} case may be different when considering spatial multiplexing instead of beamforming. 
Indeed, provided that the required degrees of freedom exists in the channel, 
the performance of spatial multiplexing in terms of capacity can follow a linear scaling law with the minimum number of transmit and receive antennas,
and this capacity boost will enhance energy efficiency.
%[Amir]: Can we replace above sentence with this? [Conclusions for the small cell \ac{BS} case may also be different when considering spatial multiplexing instead of beamforming. Indeed, providing that the required degrees of freedom exists in the channel, the performance in terms of capacity can follow a linear scaling law with the minimum number of transmit and receive antennas.]
%[David]: Thanks

When comparing the performance of the different idle modes in terms of energy efficiency,
it can be seen that
the lower the power consumption in the idle mode,
the larger the energy efficiency of the network.
This is because less energy is required to transmit the same amount of bits at the network level.
When using the idle mode models provided by the GreenTouch project,
idle modes~1 and~2,
energy efficiency decreases with densification.
This is because the increase in throughput provided by adding more cells is not large enough compared to the increase in their power consumption,
mostly because idle cells following the GreenTouch project model still consume a non-negligible amount of energy.
When considering the energy consumption of the futuristic idle modes~3 and~4,
this trend starts changing.
First, the energy efficiency increases with densification,
and then starts decreasing when the number of deployed cells becomes large and many cells are empty.
%[Ming]: (sm=5) might seem to be a too strong assumption. I'm wondering whether there exists a (sm=x) that the energy efficiency would show a flat trail after it peaks. For example, how about 5% or 3% energy consumption of (sm=1)?
%[David]: I agree this is a strong assumption but it is given as a benchmark and perfect case. The solution you are talking about will depend on the scenario and thus would require a more careful analysis. 
It is important to notice that when considering the energy consumption of the new idle mode~5,
idle cells do not consume anything,
the trend significantly changes.
Energy efficiency always increases with densification,
and it does it in a significant manner.
This is because the increase in throughput provided by adding more cells comes now at a no or very low energy cost
since idle cells do not consume power from the energy grid.

This last observation shows the need for the development of more advanced idle mode capabilities,
where the power consumption of a small cell \ac{BS} from the energy grid in idle mode is zero.
This can be optimally realised by using energy harvesting approaches
that are able to supply sufficient power to keep the small cell \ac{BS} alive when it is in idle mode.
For example, assuming the use of idle mode~2 and 4 antennas at the small cell BS,
the energy harvesting mechanism would need to provide the affordable amount of 0.4073\,W
to make ultra-dense small cell deployments energy efficient.
In this area, researchers are looking at both thermoelectric as well as mechanical energy harvesting techniques.
An example of the latter is the research in~\cite{ODonoghue2014a} and~\cite{Nico2014a},
based on vibration energy harvesting techniques for powering wireless sensors and small cell infrastructure.
Moreover, an issue with all ambient environment energy sources, to varying degrees, is their intermittent nature.
For example, harvesting useable levels of solar and wind power is dependent on the time of the day, the season and the local weather conditions.
For this reason, an efficient and cost-effective energy storage system is also key~\cite{Divya2009}.
These are important areas of research,
whose results are vital to deploy energy efficient and green telecommunications networks that have a minimal impact in the ecosystem.

%[David]: Adding energy harvesting in the challenge section.
%[Holger]:  Shall we use ''idle mode`` or ''idle mode`` ?
%[David]: Done, we are using idle mode now

%[Holger] use "sm" instead of "sleep" in figure as in text
%[David]: to be replaced in the camera ready version, other comments on figures may come 

\begin{table*}
\begin{centering}
{\footnotesize \caption{\label{tab:powerConsumption} Power consumption.}}
\vspace{-0.1cm}
\par\end{centering}{\footnotesize \par}
\centering{}{\footnotesize }%
\scalebox{0.86}{
\begin{tabular}{ c c c c c c c c c c c c }
\hline
\bf{Small cell \ac{ISD} ([m])} & \bf{Tx Power ([dBm/mW])}  & \multicolumn{9}{c}{\bf{Consumed Power ([W])}}   \\

\hline
 &  &   \multicolumn{3}{c}{\bf{Full load}} &  \multicolumn{3}{c}{\bf{idle mode 1 (\emph{sm}=1)}}  & \multicolumn{3}{c}{idle mode 2 \bf{(\emph{sm}=2)}}  \\

\hline
 &  &   \bf{1 antenna} & \bf{2 antenna}  & \bf{4 antenna} &   \bf{1 antenna} & \bf{2 antenna}  & \bf{4 antenna} &   \bf{1 antenna} & \bf{2 antenna}  & \bf{4 antenna}  \\

\hline
 200&  23.27/212.32  & 1.8923& 2.5848 &4.4560 & 0.2324&0.3105&0.4959  & 0.1881&0.2478&0.4073\\

 150&  20.52/112.72  & 1.3405&2.0316&3.9015  &  0.2191&0.2971&0.4825  &  0.1748&0.2345&0.3939\\

 100&  16.64/46.13  & 0.9793&1.6696&3.5386  &  0.2104&0.2884&0.4738  &  0.1661&0.2257&0.3852\\

 75&  13.90/24.55  &  0.8643&1.5544&3.4231  &  0.2076&0.2856&0.4710  & 0.1633&0.2230&0.3824\\

 50&  10.02/10.05  &  0.7853&1.4752&3.3437  &  0.2057&0.2837&0.4691  & 0.1614&0.2210&0.3804\\

 35&  6.61/4.58  &  0.7558&1.4456&3.3141  &  0.2050&0.2830&0.4683  &  0.1607&0.2203&0.3797\\

 20&   1.27/1.34 &  0.7383&1.4281&3.2965  &  0.2046&0.2826&0.4679  & 0.1603&0.2199&0.3793\\

 10&  -5.20/0.30  &  0.7326&1.4224&3.2908  & 0.2044&0.2824&0.4678  &  0.1601&0.2198&0.3792\\

  5&  -11.89/0.06  &  0.7314&1.4211&3.2895  & 0.2044&0.2824&0.4678  & 0.1601&0.2197&0.3791\\
\hline
\end{tabular}}
\vspace{-0.1cm}
\end{table*}

\section{What is Different in Ultra-Dense Small Cell Deployments}
\label{sec:Differences}

In this section, the main differences between regular HetNets and  ultra-dense HetNets are highlighted.

{\bf Difference~1 - BS to UE density:}
UE density is larger than  BS density in regular HetNets, 
while UE density is smaller than BS density in ultra-dense HetNets
BSs with no active UEs should be powered off to reduce unnecessary interference and to conserve power in ultra-dense HetNets, 
which is considered in the numerical/analytical results for regular HetNets.

{\bf Difference~2 - Propagation conditions:} 
NLoS interferers count for most cases in regular HetNets, 
while LoS interferers count for most cases in ultra-dense HetNets.
More sophisticated and practical path loss models should be considered for ultra-dense HetNets, 
while the simple single-slope path loss model is usually assumed to obtain the numerical/analytical results for regular HetNets.

{\bf Difference~3- Diversity loss:} 
Rich UE diversity in regular HetNets, 
while limited UE diversity in ultra-dense HetNets.
Independent shadowing and multi-path fading among UEs in one cell is usually assumed in regular HetNets to obtain the numerical/analytical results. 
Such assumption is not true in the ultra-dense HetNets. Thus, new results for the ultra-dense HetNets should be derived. 

As a result of these differences, the numerical/analytical results for regular HetNets cannot be directly applied to ultra-dense HetNets due to different assumptions for the fundamental characteristics of the networks. 

\section{Challenges in Ultra-Dense Small Cell Deployments}
\label{sec:Challenges}

In this section, the main challenges faced on the way to ultra-dense small cell deployments are highlighted.
%[Ming]: The following paragraphs are thought-provoking. Please find my comments for every one of them.
%[David]: I would like to make this section ours and that we discuss the top issues that we think are important for ultra-dense small cell deployments. Please feel free to add or remove stuff and modify.
%[Ming]: I have revisited all the paragraphs in this section. Please have a look at them.

{\bf Challenge~1 - Backhaul:}

Recent surveys show that 96\,\% of the operators consider backhaul as one of the most important challenges to small cell deployments,
and this issue is exacerbated in ultra-dense ones~\cite{2013Nicoll}~\cite{2014SCF}.
% [Holger]: ref missing
% [Ming]: I didn't find this reference. An alternative one could be: [Small Cell Forum, "SCF049: Backhaul technologies for small cells (Release 4)," Feb. 2014.]
%[Amir]: Shall we reference to our backhaul paper? Addede above!
%[David]: It is not published yet.
%[Amir]: Backhaul paper removed.
While today's wireless backhaul solutions may be a choice in higher network tiers,
we anticipate that they will only be a solution up to a certain extent in ultra-dense small cell deployments.
Due to the high capacity requirements,
wired backhaul will likely be a key requirement for these deployments,
and thus this type of network only makes sense in dense urban scenarios,
where the traffic demands are the highest and the city infrastructure can provide dense fibre and/or \ac{DSL} connectivity to operators.
The existing backhaul capability may well influence the deployment of the small cell \acp{BS}.
Massive \ac{MIMO} multicast is also a promising solution to provide backhaul to a large number of underlaid small cells~\cite{6736761}.
However, this technology still faces its own challenges,
architecture and hardware impairments~\cite{Gustavsson2014}, pilot contamination~\cite{5898372} and accurate \ac{CSI} acquisition~\cite{Truong2013} may be still an issue to provide a global solution.
% [Ming]: I suppose two-way/four-way relaying together with network coding might be an alternative solution too.
% [David]: Feel free to comment.
% [Ming]: Let me insert a new paragraph next time.
% [David]: It would be great if you can revisit this paragraph.
% [Ming]: Done.
Other wireless backhaul technologies include two-way relaying~\cite{two-way-relay},
or even four-way relaying~\cite{four-way-relay}.
However, as explained earlier,
such solution might not be scalable to ultra-dense small cell networks.
Hence, their usage is envisaged to be limited in practice.
%~\cite{two-way-relay}: L. Chun-Hung, X. Feng, "Network Coding for Two-Way Relaying: Rate Region, Sum Rate and Opportunistic Scheduling," IEEE International Conference on Communications (ICC '08), Beijing, P.R.C., pp. 1044-1049, May 2008.
%~\cite{four-way-relay}: L. Huaping, P. Popovski, E. de Carvalho, Y. Zhao, F. Sun, "Four-Way Relaying in Wireless Cellular Systems," IEEE Wireless Communications Letters, vol. 2, No. 4, pp. 403-406, Aug. 2013.
%[David]: Citations added.

{\bf Challenge~2 - Mobility:}

As indicated in Section~\ref{sec:hetnet},
we anticipate a future network architecture comprised of different small cell tiers with different types of small cell \acp{BS},
targeted at different types of environments and traffic,
where  dedicated channel mid-frequency small cell deployments with the macrocell tier may be ultra-dense to enhance network capacity.
Within this architecture,
mobile \acp{UE} should be kept in the macrocell tier,
while static \acp{UE} should be handed over to the ultra-dense small cell tier.
In order to realise this,
a new mobility management approach is needed,
in which \acp{UE} only take measurements and access the cells of the appropriate network tier according to their velocity.
Accurate mobility state estimation is key to realise this~\cite{Puttonen_ITNG_2009}.
Splitting the transmission of the \ac{UE} control and data planes will also provide mobility robustness,
allowing the larger cells to transmit and manage control/mobility related information,
while the smaller cells provide the majority of the data traffic~\cite{6477646}.
However, this new architecture poses challenges in the management of data bearers towards the core network,
since in order to realise the mobility benefits,
the data bearer should be anchored at the cell in the higher network tier (e.g. macrocells),
and this one forward the information to the cell in the lower network tier (e.g. small cells)~\cite{TR36842}.
If many small cells anchor to the same macrocell,
the latter may become a bottleneck.
% [Ming]: I believe 3GPP RAN2 guys are investigating the issues in the framework of dual connectivity. Current agreement is that the split of macrocell connection and small cell connection can happen at the PDCP layer (Option 1A) or RLC layer (Option 3C). Your description explains Option 1A, i.e., split of control plane and user plane. Option 3C is also viable. The basic idea of 3C is radio bear offloading; while that of 1A is radio bear splitting.
%[David]: I am aware of them. Actually, I believe that both architectures make sense and we should adopt one or the other depending on the velocity of the UEs. Bearer off loading makes sense for static UEs while bearer splitting makes sense for mobile UEs. The second one anchors the bearer at the macro layer and thus there is no need to update MME when changing from one small cell to another. There may be some patentable ideas here. What do you think? If so, we should not comment much.
% [Ming]: I think both options have been adopted in RAN2. You are right that we may not want to comment it too much due to patent considerations. I will revisit this paragraph next time.
% [David]: It would be great if you can revisit this paragraph. Let's think about patentable issues in this matter in October.
% [Ming]: Agreed.

{\bf Challenge~3 - Reducing costs:}

In order to improve the cost efficiency of network deployments,
it is important that small cell \ac{BS} are developed in enormous numbers to take advantage of an economy of scale.
Moreover, \ac{CAPEX} and \ac{OPEX} should be brought to a minimum.
In order to reduce \ac{CAPEX},
the price of small cell \ac{BS} should be considered,
which may require the use of cheap filters, power amplifiers and other components.
Small cell \acp{BS} may also be deployed by the end users themselves to leverage existing power and back-haul infrastructure,
thus reducing \ac{OPEX}~\cite{Claussen:07a}.
Moreover, in order to deploy a larger number of cells cost effectively in a short period of time,
it is required that the cells can be deployed in a plug \& play manner and do not require any human involvement for optimisation.
Self-organisation capabilities are thus key to manage an ultra-dense small cell deployment.
For example, the approach in~\cite{Claussen:09d} ensures that the small cell coverage is confined within the targeted household
through power control and antenna switching techniques,
thus minimising necessary handovers.
The approach in~\cite{6879477} maximises capacity by using machine learning techniques at the small cell to learn where the \acp{UE} are clustered
and beamforming to point its antenna beam towards such hotspot.
Neighbouring cell list optimisation is also a key issue in ultra-dense small cell networks where neighbouring cells are switched on and off in a dynamic manner~\cite{4531860}.

{\bf Challenge~4 - Small cell location planning:}

Finding hotspots and characterising their traffic is of crucial importance in order to deploy the right number of small cells in the right positions according to \ac{UE} needs.
Misunderstanding this information can result in an under- or over-estimated number of deployed cells,
which will affect both energy and cost efficiency as well as network performance.
However, finding and characterising hot spots is not an easy task since \ac{UE} traffic is not usually geolocated,
and network operators mostly rely today on inaccurate proximity or triangulation approaches.
Fingerprinting approaches are a possible solution to enhance the accuracy of geolocation~\cite{1458273}.
%[Ming]: Maybe we should revise the name of this bulletin to "Small cell deployment". I thought we were talking about how to let UEs discover the densely deployed small cell BSs. You know, most of the small cell BSs' signals should be turned off to save power. So it is not a easy task for a UE to know the existence of a BS. Besides, a suddenly powered-on BS might be detrimental to the network since many UEs may re-select their camp cells according to the best-RSRP rule and then go back to their previous cells when the small cell BS is turned off, typically after tens of milliseconds. I think we should use another paragraph to talk about the challenge of "Discovery of small cells". What do you think?
%[David]: Done. I agree. We can add such challenge after the Optimal idle mode capabilities since they are related.
%[Ming]: Agreed!

{\bf Challenge~5 - Smart idle mode capabilities:}

As shown in the paper,
the availability of an efficient idle mode capability at the small cell \acp{BS} is key in order to mitigate inter-cell interference and save energy.
In idle mode,
it is important that small cell \acp{BS} minimise signalling transmissions and consume as little power as possible.
Indeed, our previous results show that in order to achieve energy efficient deployments,
a small cell \ac{BS} in idle mode should not consume any power.
To realise this,
aside from the LTE solution based on \acp{DRS}~\cite{LTE-R12} (see Section~III-B),
another feasible solution would be to equip the small cell \ac{BS} with a sniffing capability,
as proposed in~\cite{Ashraf:10a},
such that the small cell \acp{BS} is idle and switches off most of the small cell \ac{BS} modules when it is not serving active \acp{UE}.
When the small cell is idle, it transmits no signalling,
and wakes up upon the detection of uplink signalling from the \ac{UE} towards the macrocell tier.
However, this solution does not allow selective wake-ups,
where only the most adequate cell in a cluster of idle cells wakes up to serve the incoming \ac{UE}.
%In order to achieve a selective wake up,
%small cell \ac{BS} cannot be completely switch off and the transmission of some signalling
%-- light weight pilot signals --
%may be still required in order to allow \acp{UE} to detect its present and aid the a network assisted wake up~\cite{TR36872}.
%The transmission of such signalling does not allow to completely switch off the cell and threaten the energy efficiency of ultra-dense small cell deployments.
%This posses some challenges in the design of the small cell \ac{BS} architecture and the wake up radio system.
Moreover, it is important to note that smart idle mode capabilities imply \ac{UE}-oriented operations,
and thus should be performed in a dynamic manner,
considering the traffic dynamics of large variety of \acp{UE}.
Dynamic small cell idle mode control also poses new challenges such as the ping-pong cell re-selection.
To be more specific,
a suddenly powered-on \ac{BS} might confuse idle \acp{UE} since they need to re-select cells according to the best-\ac{RSRP} rule,
and then go back to their previous cells when the small cell \ac{BS} returns to idle mode.
The ping-pong cell re-selection greatly consumes \acp{UE} battery life and it should be avoided.

%[Ming]: New comment, "Optimal" might be a bit too strong word here. I have revised it to "Smart".
%[David]: Agreed.

%[Ming]: If interested, we can consider a joint work on this particular topic :) It is basically a 0-1 programming and it is NP-hard. So the playground is wide open.
%[David]: Let us think about it. We would need to check first what has been previously done. The issue of adjusting transmit power also pops up. In order to cover few users it may be better to wake up one cell with more transmit power than 3 cells with lower transmit power. This depends on the fix consumption of the cells and the traffic demands of the UEs. Definitively, there is stuff to do here, if it is not done by others. Again, some patentable ideas may come up.
%[Ming]: Agreed. Let's make this our future work and talk about it later.

{\bf Challenge~6 - Modulation and coding schemes:}

Deploying higher order modulation and coding schemes
is critical to take advantage of the high \acp{SINR} resulting from ultra-dense small cell deployments.
%[Ming]: I guess the problem might be minor since we only consider stationary or quasi-stationary UEs in ultra-dense small cell networks and hence the CSI may change very slowly. However, on a related subject, it seems that 256QAM is really pushing the limit of the industry. Can we do better than 256QAM in the future? If not, it would be kind of sad to see the waste of very high SINR (e.g., 40dB) in ultra-dense small cell networks. You know, doing SU MIMO is difficult and spectrum is not unlimited. What else can we do to exploit the very high SINR? Hope the spectral efficiency of 8bps/Hz (provided by 256QAM) is not the cap of the future wireless communication systems. I guess we may want to consider an enhanced OFDM transmission allowing inter-carrier interference as well as non-linear receiver structures in the future.
%[David]: You open up a nice discussion here. It would be great if you discuss why 256QAM is really pushing the limit of the industry, and then lead the discussion to new wave form, transmission techniques and non-linear receiver structures. This would be great. Generalised OFDMA  and other wave form are being look at by some colleagues for M2M communications.
%[Ming]: Sure, no problem. I will check the potentially issues such as EVM and PAPR, and insert some comments in the text next time.
% [David]: It would be great if you can revisit this paragraph.
% [Ming]: Done.
Even higher modulation schemes than currently used in \ac{LTE} and \ac{WiFi}, i.e., 256-QAM, may be required, e.g., 1024-QAM.
However, this brings about the need for accurate channel state information for coherent de-modulation.
However, the implementation feasibility of 1024 or higher QAMs is still unclear due to the \ac{EVM} issues at transmitters~\cite{EVM}.
In addition, the \ac{PAPR} problem should also be re-considered for 1024 or higher QAMs.
% ~\cite{EVM}: ETSI MCC, Draft Report of 3GPP TSG RAN WG1 #75. Nov. 2013.

{\bf Challenge~7 - Radio resource management:}

In terms of radio resource management,
current scheduling and other network procedures have to be revisited
since due to the lower number of \acp{UE} per cell,
the current approaches used in macrocell may not be optimum anymore.
For example, proportional fair scheduling  may not be the best solution for very small cells,
since there are not many \acp{UE} to be fairly served and channel fluctuation may be low due to \ac{LOS} channel conditions.
Simpler solutions may be more appealing such as round robin.
%[Ming]: As I previously commented, the large CDD technology (i.e., transmission with different cyclic shifts on different antennas) can artificially creates channel fluctuation. I generally agree with this bulletin. No need to bring up the CDD thing in this paper.
%[David]: This is interesting and actually something to consider with Amir, PhD student starting to look at similar issues. Not sure if we should bring this up here since the paper is quite generic and CDD technology is specific.
%[Ming]: No need to bring it up here. We can talk about it later.

{\bf Challenge~8 - Spatial multiplexing:}

Using spatial multiplexing techniques, multiple stream of data can be transmitted simultaneously and successfully decoded at the receiver,
provided that the channels corresponding to every pair of transmit and receive antennas are independent. 
%[Amir]: The last two sentences have been repeatedly used before! Maybe it is better to rewrite.
%[David]: Thanks!
However, the small cell sizes in an ultra-dense small cell deployment may result in a large spatial correlation of the channels,
thus limiting the degrees of freedom available and rendering spatial multiplexing less useful.
Therefore, further research is needed in order to understand
which is the optimum number of antennas per small cell \acp{BS} according to the small cell density,
so that beamforming and spatial multiplexing can be exploited in a cost effective manner.
Multiuser \ac{MIMO} is another avenue that should be explored in order to benefit from spatial multiplexing,
which as a byproduct may bring down BS power consumption.
%[Ming]: I'm with you that SU-MIMO brings limited gain due to the rank-deficient channels. What's your opinion of MU-MIMO? One extreme approach is one BS per UE. Another extreme approach could be one BS per pair of UEs and using the MU-MIMO to double the network capacity by activating half the BSs compared with the first approach. What do you think? :) Recently I realize that 802.11ac goes so far to implement the MU-MIMO technology and it makes me re-think about its true value in the future. BTW, by MU-MIMO I'm referring to two or more UEs with the inter-UE distance larger than 10*\lambda, e.g., 0.5m.
%[David]: This is in line with my thoughts. I believe the optimum cell size is not the smallest one. Indeed, according to the number of antennas and channel characteristics, I believe that there is an optimum cell size that allows to avoid rank-deficient channels and take advantage of spatial multiplexing. For example if there are 2 antennas at the BS, we can make the cell size larger so as two capture 2 UEs and ensure a rank-2 channel. Again, this is something that we are planning to look in Amir's PhD. MU-MIMO is of course another avenue to this. Since you are the expert on this. It would be nice to sync up with you and collaborate in this front. Something nice can be done.
%[Ming]: Totally agree! Let's talk about this later.
%[David]: I have added a new line on MU-MIMO. Please feel free to upgrade.
%[Ming]: The added general comment on MU-MIMO seems nice. No more detail is needed in this paper.

{\bf Challenge~9 - Dynamic TDD transmissions:}
%[Amir]: There were two challenge 10 :)

It can be envisaged that in future networks,
small cells will prioritise \ac{TDD} schemes over \ac{FDD} ones
since \ac{TDD} transmissions are particularly suitable for hot spot scenarios with traffic fluctuations in both link directions~\cite{TR36828}.
In this line, a new technology has recently emerged,
referred to as dynamic \ac{TDD},
in which \ac{TDD} \ac{DL} and \ac{UL} subframes can be dynamically configured in small cells to adapt their communication service to the fast variation of \ac{DL}/\ac{UL} traffic demands in either direction.
The application of dynamic \ac{TDD} in homogeneous small cell networks has been investigated in recent works with positive results~\cite{6353682}~\cite{DavidLopez2014homoDynamicTdd}~\cite{DavidLopez2015homoDynamicTdd}.
Gains in terms of \ac{UE} packet throughput and energy saving have been observed,
mostly in low-to-medium traffic load conditions.
However, up to now, it is still unclear whether it is feasible to introduce the dynamic \ac{TDD} transmissions into \acp{HetNet}
because it will complicate the existing \ac{CRE} and \ac{ABS} operations and its advantage in the presence of macrocells in terms of UE packet throughput is currently being investigated~\cite{DavidLopez2014hetnetDynamicTdd}~\cite{TR36828}.
%[3GPP, TR 36.828 (V11.0.0): Further enhancements to \ac{LTE} TDD for \ac{DL}-\ac{UL} interference management and traffic adaptation, Jun. 2012.].
Some pioneering work on the application of dynamic TDD in small cell HetNets can be found in~\cite{DavidLopez2014hetnetDynamicTdd}.
% [Ming]: A new Challenge, i.e., dynamic TDD, has been added based on our previous work. Please check.

{\bf Challenge~10 - Co-existence with WiFi:}

In order to gain access to more frequency resources,
\ac{LTE} small cells may be deployed in unlicensed bands, where they are required to coexist with \ac{WiFi} networks.
However, when \ac{LTE} and \ac{WiFi} nodes are deployed in the same frequency band,
\ac{WiFi} nodes may tend to stay in listening mode waiting for a channel access opportunity,
due to their courteous \ac{CSMA}/Collision Avoidance (CA) protocol and the high-power interference from the \ac{LTE} network.
Simulation results have shown that when co-existing with \ac{LTE} nodes,
if these ones do not implement any co-existence mechanism,
\ac{WiFi} nodes in some indoor scenarios may spend even up to 96\,\% of the time in listening mode due to inter-radio access technology interference and the poor performance of \ac{CSMA}/CA mechanisms.
This significantly degrades \ac{WiFi} performance~\cite{6692702}.
New co-existence solutions have to be devised to enhance \ac{LTE} and \ac{WiFi} coexistence and ensure that those two networks share the unlicensed bands in a fair manner.
Moreover, it is desirable that the coexistence scheme should be as much frequency agnostic as possible
so that a global solution can be achieved.
In the \ac{3GPP}, a prevailing view is that in the near future,
we should first consider unlicensed operation in 5\,GHz and focus on \ac{DL}-only operations assisted by licensed carriers,
the so called \ac{LAA}~\cite{TR36889}~\cite{LTE-U}.

\section{Conclusion}
\label{sec:conclusion}

In this paper, the gains and limitations of network densification, use of higher frequency bands and multi-antenna techniques have been analysed.
Network densification has been shown to provide the most of UE throughput gains of up to 48x at the cell-edge,
followed by a linear scale up with the available bandwidth of up to 5x,
provided by the use of higher frequency bands.
Beamforming gains of around 1.49 at cell-edge have been shown to be low compared to network densification and use of higher frequency bands,
and thus the use of spatial multiplexing seems more appealing to exploit the multiple antennas in a \ac{BS}
at the studied frequency carriers,
provided that sufficient channel decoration among antennas exists.

Efficient idle mode capabilities at the small cells has been shown to be key to mitigate interference and save energy.
Moreover, one \ac{UE} per cell has been shown to be the fundamental limit of spatial reuse and may drive the cost-effectiveness of ultra-dense small cell deployments.
As a result, \ac{UE} density and distribution have to be well understood before any roll-out.

Network densification has also been shown to reduce multi-user diversity,
and thus \ac{PF} alike schedulers lose their advantages with respect to \ac{RR} ones.
For an \ac{ISD} of 20\,m, the difference in average \ac{UE} throughput of \ac{PF} over \ac{RR} is only of around 5\,\%,
indicting that \ac{RR} may be more suitable for ultra-dense small cell deployments due to their reduced complexity.

As a bottom line, simulation results have shown that for a realistic non-uniformly distributed UE density of 300 active \ac{UE} per square km,
it is possible to achieve an average 1\,Gbps per UE with roughly an \ac{ISD} of 35\,m, 250\,MHz bandwidth and 4 antennas per small cell \ac{BS}.
Today, these deployments are neither cost-effective nor energy efficient.
However, this may change in the future with new deployment models, lower cost and more efficient equipment.
The top ten challenges to realise cost-effective ultra-dense small cell deployments have also been discussed in this paper.

\appendix
\section{Simulation Setup}
\label{sec:appendix}

%*** Overall Gain
The simulated environment and its channel gains are calculated in form of matrices,
representing a two dimensional gain (loss) map with a given resolution~\cite{Claussen2012}.
The overall gain is calculated as a sum of individual gains (losses) in decibels for each \ac{BS} $m$ and each location as
\begin{equation}
	\label{env_gain}
	G_m {\rm [dB]}  = G_{\mathrm{A},m} {\rm [dB]}  + G_{\mathrm{P},m} {\rm [dB]}  + G_{\mathrm{S},m} {\rm [dB]} , %+ G_{\mathrm{E},m}
\end{equation}
where $G_{\mathrm{A},m} {\rm [dB]} $ is the antenna gain,
$G_{\mathrm{P},m} {\rm [dB]}$ is the path gain (loss)
%$G_{\mathrm{E},m}$ is the environment gain (loss) of buildings,
and $G_{\mathrm{S},m} {\rm [dB]}$ is the shadow fading gain (loss).
%In this paper, multi-path fading is not considered
%since we  envision that because of the use of mid/high-frequency frequency bands and due to the importance of the \ac{LOS} component in    small cells,
%the time spread of the \ac{CIR} will be very small,
%typically in the order of several $\mu$s,
%and hence the impact of multi-path fading will become less significant in the future.

\begin{figure}[t]
  \centering
  \def\svgwidth{8.1cm}
  
\immediate\write18{inkscape -z -D --file=figures/antennas.svg
--export-pdf=figures/antennas.pdf --export-latex}
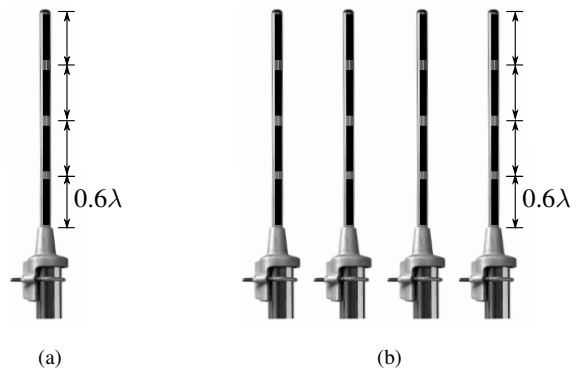

  \caption{(a) Vertical half-wave dipole array with four elements spaced by 0.6\,$\lambda_c$ (b) Horizontal array of vertical half-wave dipoles with four elements spaced by 0.6\,$\lambda_c$ used for beamforming.}
  \label{fig:dipole}
\end{figure}

%*** Antenna Gain
The antenna gain $G_{\mathrm{A},m} {\rm [dB]}$ represents the gain resulting from focussing the antenna beam towards one direction.
In this case, a vertical half-wave dipole array with four elements spaced by 0.6\,$\lambda_c$ is used (see Fig.~\ref{fig:dipole}\,(a)),
% [Ming]: An illustration would be nice.
% [David]: Done
% [Ming]: Great!
where the combined gain of the horizontal and vertical antenna patterns together with the vertical array factor gain is calculated as:
\begin{align}
	G^{\rm a}(\varphi,\theta) {\rm [dB]} & =
	\notag\\
	G^{\rm a}_{\rm M} [dBi] & + G^{\rm a}_{\rm H}(\varphi) [dB] + G^{\rm a}_{\rm V}(\theta) [dB] + G_{\rm V}^{\rm a,array}(\theta) [dB],
	\label{eq:3DantennaPattern-DipoleArray}
 \end{align}
where $\varphi$  and $\theta$  are the angles of arrival
in the horizontal and vertical planes with respect to the main beam direction, respectively,
 $G^{\rm a}_{\rm M}$, $G^{\rm a}_{\rm H}(\varphi)$ and $G^{\rm a}_{\rm V}(\theta)$
are the maximum antenna gain and the attenuation offsets of one antenna element of the array, respectively,
and $G_{\rm V}^{\rm a,array}(\theta)$ is the vertical array factor gain.

%[Holger]: remove duplicated text after (a) and (b) in figure
%[David]: Done

The horizontal and vertical attenuation offsets of one antenna element of the array can be respectively modelled as:
\begin{eqnarray}
	G^{\rm a}_{\rm H}(\varphi) {\rm [dB]} &&
	= 0 ,
	\label{eq:antennaPatternH-DipoleArray}
\end{eqnarray}
and
\begin{eqnarray}
	G^{\rm a}_{\rm V}(\theta) {\rm [dB]} &&=
	20 \log_{10}  \left( \frac{\cos \left( \frac{\pi}{2}\cos \left(\theta + \frac{\pi}{2} \right) \right)}{\sin \left(\theta + \frac{\pi}{2} \right)} \right) ,
	\label{eq:antennaPatternV-DipoleArra}
\end{eqnarray}
%[Ming]: A small figure plotting G^{\rm a}_{\rm V}(\theta) {\rm [dB]} might be nice.
%[David]: Done
%[Ming]: Great!
while the vertical array factor gain $G_{\rm V}^{\rm a, array}(\theta)$ can be modelled as:
 \begin{eqnarray}
	G_{\rm V}^{\rm a, array}(\theta) =
	\sum_{n=1}^{N_t} a(n) {\rm e}^{(j (n-1) \times 2 \pi d_{\epsilon} (-\sin(\theta)) + \delta_{\rm phase})},
	\label{eq:antennaPatternV-DipoleArray}
\end{eqnarray}
where $N_t$ is the number of antenna elements in the vertical array,
$d_{\epsilon}$ is the spacing between antenna elements in wavelength,
$a(n)$ is the normalised voltage of antenna element $n$,
% [Ming]: I don't quite understand why $a(n)~=1$. Is it because of non-ideal PAs that different a(n) are assumed for difference antenna elements?
% [David]: I don't think it is related to the PA issue, but because changing the power in each antenna element also changes the beam forming pattern.
% [Ming]: Understood.
and $\delta_{\rm phase}$ is the phase increment within antenna elements in radian
(see Table~\ref{tab:dipoleParameters}).
The horizontal array factor gain in the horizontal plane depends on the beamforming weights used,
and those are specified by the used standardised \ac{LTE} code book beamforming~\cite{LTEbook}.
%[Holger]: add equation for horizontal gain for array shown in fig 11b?
%[David]: The gain in the horizontal plane depends on the beamforming weights used and those are specified in the LTE beamforming code book used.

%Fig.~\ref{fig:dipoleArray} shows  the resulting horizontal and vertical attenuation offsets as well as the spatial antenna gains of the adopted vertical half-wave dipole array with four elements spaced by 0.6\,$\lambda_c$.

%\begin{figure*}[t!]
%  \centerings
%	\subfigure[Horizontal and vertical attenuation offsets.]{\label{fig:dipoleArrayPattern}
%	\includegraphics[width=3.5in]{f_dipoleArray_antennaPatterns}}
%	\subfigure[Spatial antenna gains.]{\label{fig:dipoleArrayGains}
%	\includegraphics[width=3.5in]{f_dipoleArray_spatialGain}}
%	\caption{Radiation characteristics of the 4 element vertical dipole array.}
% \label{fig:dipoleArray}
%\end{figure*}

\begin{table}[t]
\centering
\caption{Typical parameters of a dipole array}{
\begin{tabular}{@{}lcc@{}}
\toprule
\bf{Parameter} & \bf{Value}\\
\midrule
$N_t$         		& 4\\
$G^{\rm a}_{\rm M}$ & 2.15\,dBi\\
$d_{\epsilon}$         	& 0.6\,wavelength\\
$a(n)$        		& $[0.97 \,\,\,\, 1.077 \,\,\,\, 1.077 \,\,\,\, 0.86]$\\
$\delta_{\rm phase}$& 1.658\,radian\\
\midrule
\end{tabular}}
\label{tab:dipoleParameters}
\end{table}

It is important to note that the proposed configuration was optimised for a given \ac{ISD} and antenna height,
and results in a given certain downtilt.
In order to keep the resulting downtilt fixed,
the height of the antenna is changed with the \ac{ISD}.
The longer the \ac{ISD}, the higher the antenna height.

%*** Path gain
The path gain (loss) $G_{\mathrm{P},m} {\rm [dB]}$ represents the gain between a transmitter and a receiver located in this case outdoors for the given environment.
The path gain (loss) is modelled as a expected gain value composed of line-of-sight gain, non-line-of-sight gain and a line-of-sight probability as
\begin{align}
\label{path_gain}
	G_{\mathrm{P},m} {\rm [dB]} & =
	\notag\\
	P_\mathrm{l}(d  {\rm [m]}) {\rm [\cdot]}  & \times G_{\mathrm{Pl}}(d  {\rm [m]}) {\rm [dB]} + \left( 1-P_\mathrm{l}(d  {\rm [m]}) \right) \times G_{\mathrm{Pn}}(d) {\rm [dB]}.
\end{align}

The line-of-sight probability $P_\mathrm{l}(d)  {\rm [\cdot]}$  for a propagation distance $d$ is based on~\cite{TR36.814},
with an additional spline interpolation for a smooth transition,
The resulting line-of-sight $G_{\mathrm{Pl}}(d) {\rm [dB]}$ and non-line-of-sight $G_{\mathrm{Pn}}(d) {\rm [dB]}$ path gains for a propagation distance $d$
are calculated using the \ac{3GPP} urban micro (UMi) models~\cite{TR36.814}.
$d$ is computed in the 3D space.

%*** Environment gain
%The environment gain (loss) $G_{\mathrm{E},m}$ represents the gain resulting from walls for indoor locations not captured in the path gain (\ref{path_gain}).
%Since ray-tracing models are very computational expensive for simulating larger areas,
%a simple indoor penetration model is used with a fixed additional loss of 20\,dB for indoor locations~\cite{3gpp.36.814}.
%Building locations were extracted from a map for the centre of Dublin, Ireland.

%*** Shadow fading gain
The shadow fading gain (loss) $G_{\mathrm{S},m}  {\rm [dB]}$ represents the variability of the path loss due to obstruction effects that are not explicitly modelled.
The shadow fading gain values are spatially correlated and generated with a log-normal distribution based on~\cite{Gudmundson:91, Claussen:05b}
with a standard deviation of 6\,dB~\cite{TR36.814}.
The inter-small cell \ac{BS} correlation of shadow fading gain values is 0.5.
The small cell \ac{BS} transmit power $P_{\mathrm{tx},m}  {\rm [mW]}$ is calculated to achieve an average targeted \ac{SNR} at the cell-edge of  $\gamma^{\mathrm{edge}} {\rm [dB]}=$ 9, 12 or 15\,dB as
\begin{equation}
\label{received_power}
P_{\mathrm{tx},m}  {\rm [dBm]} = P_{\mathrm{N}} {\rm [dBm]} + G_{\mathrm{P},m,{\rm edge}} {\rm [dB]}  + \gamma^{\mathrm{edge}} {\rm [dB]}. %\\+ \psi_u {\rm [dB]}
\end{equation}
where $P_{\mathrm{N}} {\rm [dBm]}$ is the noise power in dBm,
and $G_{\mathrm{P},m,{\rm edge}} {\rm [dB]}$ is the path gain (loss) from \ac{BS} $m$ to its cell-edge,
which is $\frac{\sqrt{3}}{2}$ of the \ac{ISD}.
%and $\psi_u {\rm [dB]}$ is the noise figure of the \ac{UE}.
%[Ming]: An equation to define $P_{\mathrm{tx},m}  {\rm [mW]}$ would be nice, i.e., $P_{\mathrm{tx},m}  {\rm [dBm]}=P_{\mathrm{N}} {\rm [dBm]}+SNR^{\mathrm{edge}} {\rm [dB]}$
%[David]: Done. The power transited should be the one required for overcoming the channel gain (losse) + noise figure + snr target
%[Ming]: Right, the edge PL and the noise figure should be considered. I further add a noise power (dBm) in this equation.
%[David]: Maybe it would be better to consider that the noise figure is contained in the noise power and then only consider the noise power in the equation.
%[Ming]: Agreed! Let's keep this equation neat. Shall we present another equation for $P_{\mathrm{N}} {\rm [dBm]}$ and explain the noise figure in that equation?

%[Ming]: Another question. Shadow fading equals to 0dB in $G_{m,{\rm edge}} {\rm [dB]}$?
%[David]: Updated. I assume that the shadow fading is on its mean value 0dB.
%[Ming]: Understood.

The received power from each small cell \ac{BS} $m$ at each \ac{UE} $u$ is calculated using $P_{\mathrm{tx},m}  {\rm [mW]}$ and the overall gain $G_{m,u}$ as
\begin{equation}
\label{received_power}
P_{\mathrm{rx},m,u}  {\rm [dBm]} = P_{\mathrm{tx},m}  {\rm [dBm]} + G_{m,u}  {\rm [dB]}.
\end{equation}

%*** SINR
The \ac{SINR} of \ac{UE} $u$ when served by small cell \ac{BS} $m$ is
\begin{equation}
\label{SINR}
\small
\gamma_{m,u}  {\rm [dB]} = \frac{P_{\mathrm{rx},m,u}  {\rm [mW]}}{\left( \sum_{n=1}^{M} P_{\mathrm{rx},n,u}  {\rm [mW]} \right) - P_{\mathrm{rx},m,u}  {\rm [mW]} + P_\mathrm{N}  {\rm [mW]}}, % + \psi_u {\rm [dB]}
\end{equation}
\normalsize
where $M$ is the total number of small cell \acp{BS} transmitting in the studied frequency band.
%[Ming]: The noise figure is reflected in this equation.
%[David]: I have removed the noise figure according to the new defineition of noise power containing the noise figure.
%[Ming]: Agreed.

%*** User throughput
From the SINRs and assuming a round robin scheduling,
the throughput of \ac{UE} $u$ can be calculated as
\begin{equation}
\label{throughput}
C_u = \frac{C(\gamma_{m,u})}{U},
\end{equation}
where $C(\gamma_{m,u})$ is the SINR to throughput mapping and $U$ is the number of \acp{UE} served by small cell \ac{BS} $m$.
The mapping used here is the Shannon-Hartley theorem presented in (\ref{eq:Shannon}) with an operation point 3.5\,dB from the optimum capacity,
and there is no cap for the \ac{MCS}.
%[Ming]: Nice simplification! I agree that no need to consider quantized CQI in this paper.
%[David]: Agreed.
%[Holger]: I think it would be better to assume an operation point x dB from the Shannon capacity where x mathes roughly the difference for typical receivers today (e.g. look at LTE performance)
%[David]: Thanks for catching this. I forgot to mention that we were operating 3.5dB away from the Shannon capacity

%*** Collection of results
Throughput performance statistics are collected over 150 simulation runs from \acp{UE} connected to the small cell \acp{BS} in the scenario,
with independent \ac{UE} location and shadow fading realisations.
Extra tiers of small cell BSs outside the scenario were added to avoid border effects.

\section{Distance dependent multi-path fast fading}
\label{sec:appendix2}

Considering the cell size and the relative proximity of \acp{UE} to their serving \acp{BS},
there is a high probability of \ac{LOS} in dense small cell networks,
which indicates that Rician fading channel models may be more appropriate than Rayleigh ones to model multi-path channels in this type of deployment.
The Rician fading model considers a dominant, non-fluctuating strong path in addition to a number of reflections and scatterings,
referred to as \ac{LOS} and \ac{NLOS} components, respectively~\cite{HetNetbook}.

In the distance dependent multi-path fast fading model used in this paper~\cite{2015Jafari},
the $K$ factor of the traditional Rician model is derived according to the probability of \ac{LOS} in~\cite{TR36.814}.
This probability of \ac{LOS} as a function of distance for micro urban scenarios is given as
\begin{equation}
	\vspace{-0.5mm}
	P_{\rm LOS}= \min(\frac{18}{d},1) \times (1-e^{\frac{-d}{36}}) + e^{\frac{-d}{36}},
	\label{eq:plos}
\end{equation}
where $d$ is the distance between the \ac{UE} and its serving \ac{BS}.
According to this model,
the $P_{\rm LOS}$ is equal to $1$ within 18\,m of the \ac{BS},
meaning that \acp{UE} that are positioned up to 18\,m from the serving \ac{BS} location have a guaranteed strong \ac{LOS} component.

In order to comply with the $P_{\rm LOS}$ of 1,
in this distant dependent Rician channel,
the value of 32 ($\sim$15 dB) is assigned to the $K$ factor to secure the existence of a strong \ac{LOS} component within the \ac{LOS} zone.
This value is selected since it results  in a standard deviation of the Rician fading  smaller than 0.5~dB, flat fading.
For \acp{UE} locate further away from the \ac{BS},
the $P_{\rm LOS}$ exponentially decays and so does the $K$ factor due to their one-to-one correspondence.
This is modelled by approximating the K factor,
$\rm K = \frac{\rm P_{\rm LOS}}{\rm P_{\rm NLOS}}$,
by an exponentially decaying function.
This is interpreted as a distance dependant transition from Rician to Rayleigh fading for \acp{UE} that are located further away from their \acp{BS}
where the \ac{LOS} component gradually fades.
%Fig.~\ref{fig:los_rician} shows the derived $K$ factor and the corresponding $P_{\rm LOS}$.
%[Amir]: No figure is here for the K_factor. We should either remove the above sentence or add the figure.
%[David]: I have removed the sentence. Readers can refer to the conference paper.

\balance
\bibliographystyle{IEEEtran}
\bibliography{IEEEabrv,claussen-library,claussen-publications,david-library}
\begin{acronym}[AAAAAAAAA]

\acro{3D}{3-Dimensional}
 \acro{3GPP}{3rd Generation Partnership Project}
 \acro{4G}{Fourth Generation}
 \acro{AAA}{Authentication, Authorization and Accounting}
 \acro{ABS}{Almost Blank Subframe}
 \acro{ACIR}{Adjacent Channel Interference Rejection ratio}
 \acro{ACL}{Allowed CSG List}
 \acro{ACLR}{Adjacent Channel Leakage Ratio}
 \acro{ACPR}{Adjacent Channel Power Ratio}
 \acro{ACS}{Adjacent Channel Selectivity}
 \acro{ADSL}{Asymmetric Digital Subscriber Line}
 \acro{AF}{Amplify and Forward}
 \acro{AGCH}{Access Grant Channel}
 \acro{AH}{Authentication Header}
 \acro{AKA}{Authentication and Key Agreement}
 \acro{AMC}{Adaptive Modulation and Coding}
 \acro{ANR}{Automatic Neighbour Relations}
 \acro{AoA}{Angle of Arrival}
 \acro{AoD}{Angle of Departure}
 \acro{API}{Application Programming Interface}
 \acro{APP}{A Posteriori Probability}
 \acro{ARQ}{Automatic Repeat Request}
 \acro{AS}{Access Stratum}
 \acro{AS}{Access Stratum}
 \acro{ASE}{Area Spectral Efficiency}
 \acro{ASN}{Access Service Network}
 \acro{ATM}{Asynchronous Transfer Mode}
 \acro{ATSC}{Advanced Television Systems Committee}
 \acro{AUC}{Authentication Centre}
 \acro{AWGN}{Additive White Gaussian Noise}
 \acro{BCCH}{Broadcast Control Channel}
 \acro{BCH}{Broadcast Channel}
 \acro{BE}{Best Effort}
 \acro{BER}{Bit Error Rate}
 \acro{BLER}{Block Error Rate}
 \acro{BPSK}{Binary Phase-Shift Keying}
 \acro{BR}{Bit Rate}
 \acro{BS}{Base Station}
 \acro{BSC}{Base Station Controller}
 \acro{BSIC}{Base Station Identity Code}
 \acro{BSP}{Binary Space Partitioning}
 \acro{BSS}{Blind Source Separation}
 \acro{BTS}{Base Transceiver Station}
 \acro{CA}{Carrier Aggregation}
 \acro{CAC}{Call Admission Control}
 \acro{CAS}{Cluster Angular Spread}
 \acro{CAPEX}{Capital Expenditure}
 \acro{CAZAC}{Constant Amplitude Zero Auto-Correlation}
 \acro{CC}{Chase Combining}
 \acro{CCC}{Common Control Channel}
 \acro{CCCH}{Common Control Channel}
 \acro{CCPCH}{Common Control Physical Channel}
 \acro{CCRS}{Coordinated and Cooperative Relay Systems}
 \acro{CCTrCH}{Coded Composite Transport Channel}
 \acro{CDF}{Cumulative Distribution Function}
 \acro{CDMA}{Code Division Multiple Access}
 \acro{CDS}{Cluster Delay Spread}
 \acro{CFL}{Courant-Friedrichs-Lewy}
 \acro{CGI}{Cell Global Identity}
 \acro{CID}{Connection Identifier}
 \acro{CIO}{Cell Individual Offset}
 \acro{CIR}{Channel Impulse Response}
 \acro{C-MIMO}{Cooperative MIMO}
 \acro{CN}{Core Network}
 \acro{CoMP}{Coordinated Multi-Point }
 \acro{CPCH}{Common Packet Channel}
 \acro{CPC}{Cognitive Pilot Channel}
 \acro{CPE}{Customer premises Equipment}
 \acro{CPICH}{Common Pilot Channel}
 \acro{CQI}{Channel Quality Indicator}
 \acro{CR}{Cognitive Radio}
 \acro{CRC}{Cyclic Redundancy Check}
 \acro{CRE}{Cell Range Expansion}
 \acro{C-RNTI}{Cell Radio Network Temporary Identifier}
 \acro{CRS}{Cell-Specific Reference Symbol}
 \acro{CRP}{Cell Re-selection Parameter}
 \acro{CSCC}{Common Spectrum Coordination Channel}
 \acro{CSG ID}{CSG Identity}
 \acro{CSG}{Closed Subscriber Group}
 \acro{CSI}{Channel State Information}
 \acro{CSIR}{Receiver-Side Channel State Information}
 \acro{CSMA}{Carrier Sense Multiple Access}
 \acro{CSO}{Cell Selection Offset}
 \acro{CTCH}{Common Traffic Channel}
 \acro{CTS}{Clear-To-Send}
 \acro{CWiND}{Centre for Wireless Network Design}
 \acro{DAB}{Digital Audio Broadcasting}
 \acro{DAS}{Distributed Antenna System}
 \acro{DC}{Dual Connectivity}
 \acro{DCCH}{Dedicated Control Channel}
 \acro{DCF}{Decode-and-Forward}
 \acro{DCH}{Dedicated Channel}
 \acro{DCM}{Directional Channel Model}
 \acro{DCP}{Dirty-Paper Coding}
 \acro{DCS}{Digital Communication System}
 \acro{DECT}{Digital Enhanced Cordless Telecommunications}
 \acro{DeNB}{Donor eNB}
 \acro{DFP}{Dynamic Frequency Planning}
 \acro{DFS}{Dynamic Frequency Selection}
 \acro{DFT}{Discrete Fourier Transform}
 \acro{DL}{Downlink}
 \acro{DMC}{Dense Multipath Components}
 \acro{DMF}{Demodulate-and-Forward}
 \acro{DMT}{Diversity and Multiplexing Tradeoff}
 \acro{DoA}{Direction-of-Arrival}
 \acro{DoD}{Direction-of-Departure}
 \acro{DoS}{Denial of Service}
 \acro{DPCCH}{Dedicated Physical Control Channel}
 \acro{DPDCH}{Dedicated Physical Data Channel}
 \acro{D-QDCR}{Distributed QoS-based Dynamic Channel Reservation}
 \acro{DRS}{Discovery Reference Signal}
 \acro{DRX}{Discontinuous Reception}
 \acro{DSA}{Dynamic Spectrum Access}
 \acro{DSCH}{Downlink Shared Channel}
 \acro{DS}{DownStream}
 \acro{DSL}{Digital Subscriber Line}
 \acro{DSP}{Digital Signal Processor}
 \acro{DTCH}{Dedicated Traffic Channel}
 \acro{DVB}{Digital Video Broadcasting}
 \acro{DXF}{Drawing Interchange Format}
 \acro{E2E}{End to End}
 \acro{EAGCH}{Enhanced uplink Absolute Grant Channel}
 \acro{EAP}{Extensible Authentication Protocol}
 \acro{ECGI}{Evolved Cell Global Identifier}
 \acro{ECR}{Energy Consumption Ratio}
 \acro{ECRM}{Effective Code Rate Map}
 \acro{EDCH}{Enhanced Dedicated Channel}
 \acro{EESM}{Exponential Effective SINR Mapping}
 \acro{EF}{Estimate-and-Forward}
 \acro{EGC}{Equal Gain Combining}
 \acro{EHICH}{EDCH HARQ Indicator Channel}
  \acro{eICIC}{enhanced Intercell Interference Coordination}
 \acro{EIR}{Equipment Identity Register}
 \acro{EMS}{Enhanced Messaging Service}
 \acro{eNB}{evolved NodeB}
 \acro{eNodeB}{evolved NodeB}
 \acro{EoA}{Elevation of Arrival}
 \acro{EoD}{Elevation of Departure}
 \acro{EPC}{Evolved Packet Core}
 \acro{EPLMN}{Equivalent PLMN}
 \acro{EPS}{Evolved Packet system}
  \acro{E-RAB}{E-UTRAN Radio Access Bearer}
 \acro{ERGCH}{Enhanced uplink Relative Grant Channel}
 \acro{ertPS}{Extended real time Polling Service}
 \acro{ESF}{Even Subframe}
 \acro{ESP}{Encapsulating Security Payload}
 \acro{ETSI}{European Telecommunications Standards Institute}
 \acro{EUTRA}{Evolved UTRA}
 \acro{EUTRAN}{Evolved UTRAN}
 \acro{EVDO}{Evolution-Data Optimised}
 \acro{EVM}{Error Vector Magnitude}
 \acro{FACCH}{Fast Associated Control Channel}
 \acro{FACH}{Forward Access Channel}
 \acro{FAP}{Femtocell Access Point}
 \acro{FCC}{Federal Communications Commission}
 \acro{FCCH}{Frequency-Correlation Channel}
 \acro{FCFS}{First-Come First-Served}
 \acro{FCH}{Frame Control Header}
  \acro{FCI}{Failure Cell ID}
  \acro{FD}{Frequency-Domain}
 \acro{FDD}{Frequency Division Duplexing}
 \acro{FDM}{Frequency Division Multiplexing}
 \acro{FDTD}{Finite-Difference Time-Domain}
 \acro{FER}{Frame Error Rate}
 \acro{FFRS}{Fractional Frequency reuse Scheme}
 \acro{FFT}{Fast Fourier Transform}
 \acro{FGW}{Femto Gateway}
 \acro{FIFO}{First In First Out}
 \acro{FMC}{Fixed Mobile Convergence}
 \acro{FPGA}{Field-Programmable Gate Array}
 \acro{FRS}{Frequency reuse Scheme}
 \acro{FRS}{Frequency reuse Scheme}
 \acro{FSU}{Flexible Frequency Usage}
 \acro{FTP}{File Transfer Protocol}
 \acro{FUSC}{Full Usage of Subchannels}
 \acro{GAN} {Generic Access Network}
 \acro{GANC}{Generic Access Network Controller}
 \acro{GERAN}{GSM EDGE Radio Access Network}
 \acro{GGSN}{Gateway GPRS Support Node}
 \acro{GMSC}{Gateway Mobile Switching Center}
 \acro{GPRS}{General Packet Radio Service}
 \acro{GPS}{Global Positioning System}
 \acro{GPU}{Graphics Processing Unit}
  \acro{GNSS}{Global Navigation Satellite System}
 \acro{GSCM}{Geometry-based Stochastic Channel Models}
 \acro{GSM}{Global System for Mobile communication}
 \acro{GTD}{Geometry Theory of Diffraction}
 \acro{GTP}{GPRS Tunnel Protocol}
 \acro{HA}{Hybrid Access}
 \acro{HARQ}{Hybrid Automatic Repeat reQuest}
 \acro{HBS}{Home Base Station}
 \acro{HCN}{Heterogeneous Cellular Network}
 \acro{HCS}{Hierarchical Cell Structure}
 \acro{HDFP}{Horizontal Dynamic Frequency Planning}
 \acro{HeNB}{Home eNodeB}
 \acro{HetNet}{Heterogeneous Network}
 \acro{HLR}{Home Location Register}
 \acro{HNB}{Home NodeB}
 \acro{HNBAP}{Home NodeB Application Protocol}
 \acro{HNBGW}{Home NodeB Gateway}
 \acro{HO}{Handover}
 \acro{HOM}{Handover Hysteresis Margin}
 \acro{HPLMN}{Home PLMN}
 \acro{HRD}{Horizontal Reflection Diffraction}
 \acro{HSDPA}{High Speed Downlink Packet Access}
 \acro{HSDSCH}{High-Speed DSCH}
 \acro{HSPA}{High Speed Packet Access}
 \acro{HSS}{Home Subscriber Server}
 \acro{HSUPA}{High Speed Uplink Packet Access}
 \acro{HUA}{Home User Agent}
 \acro{HUE}{Home User Equipment}
 \acro{IC}{Interference Cancellation}
 \acro{ICI}{Inter-carrier Interference}
 \acro{ICIC}{Intercell Interference Coordination}
 \acro{ICNIRP}{International Commission on Non-Ionizing Radiation Protection}
 \acro{ICS}{IMS Centralised Service}
 \acro{ICT}{Information Communication Technology}
 \acro{ID}{Identifier}
 \acro{IDFT}{Inverse Discrete Fourier Transform}
 \acro{IEEE}{Institute of Electrical and Electronics Engineers}
 \acro{IETF}{Internet Engineering Task Force}
 \acro{IFFT}{Inverse Fast Fourier Transform}
 \acro{IKE}{Internet Key Exchange}
 \acro{IKEv2}{Internet Key Exchange version 2}
 \acro{ILP}{Integer Linear Programming}
 \acro{IMEI}{International Mobile Equipment Identity}
 \acro{IMS}{IP Multimedia Subsystem}
 \acro{IMSI}{International Mobile Subscriber Identity}
  \acro{IMT}{International Mobile Telecommunications}
 \acro{InH}{Indoor Hotspot}
 \acro{IP}{Internet Protocol}
 \acro{IPsec}{Internet Protocol Security}
 \acro{IR}{Incremental Redundancy}
 \acro{IRC}{ Interference Rejection Combining}
 \acro{ISD}{Inter Site Distance}
 \acro{ISI}{Inter Symbol Interference}
 \acro{ITU}{International Telecommunication Union}
 \acro{Iub}{UMTS Interface between RNC and Node B}
 \acro{JFI}{Jain's Fairness Index}
 \acro{IWF}{IMS Interworking Function}
 \acro{KPI}{Key Performance Indicator}
 \acro{L1}{Layer One}
 \acro{L2}{Layer Two}
 \acro{L3}{Layer Three}
 \acro{LA}{Location Area}
 \acro{LAA}{Licensed Assisted Access}
 \acro{LAC}{Location Area Code}
 \acro{LAI}{Location Area Identity}
 \acro{LAU}{Location Area Update}
 \acro{LPN}{Low Power Node}
 \acro{LLR}{Log-Likelihood Ratio}
 \acro{LLS}{Link-Level Simulation}
 \acro{LMDS}{Local Multipoint Distribution Service}
 \acro{LMMSE}{Linear Minimum Mean-Square-Error}
 \acro{LOS}{Line Of Sight}
 \acro{LPN}{Low Power Node}
 \acro{LR}{Likelihood Ratio}
 \acro{LTE}{Long Term Evolution}
  \acro{LTE/SAE}{Long Term Evolution/System Architecture Evolution}
 \acro{LSP}{Large-Scale Parameter}
 \acro{LUT}{Look Up Table}
 \acro{MAC}{Medium Access Control}
 \acro{MAP}{Media Access Protocol}
 \acro{MaxI}{Maximum Insertion}
 \acro{MaxR}{Maximum Removal}
 \acro{MBMS}{Multicast Broadcast Multimedia Service}
 \acro{MBS}{Macrocell Base Station}
 \acro{MBSFN}{Multicast-Broadcast Single-Frequency Network}
 \acro{MC}{Modulation and Coding}
 \acro{MCM}{Multi-Carrier Modulation}
 \acro{MCP}{Multi-Cell Processing}
 \acro{MCS}{Modulation and Coding Scheme}
  \acro{MDT}{Minimisation of Drive Tests}
 \acro{MGW}{Media Gateway}
 \acro{MIB}{Master Information Block}
 \acro{MIC}{Mean Instantaneous Capacity}
 \acro{MIMO}{Multiple Input Multiple Output}
 \acro{MinI}{Minimum Insertion}
 \acro{MinR}{Minimum Removal}
 \acro{MIP}{Mixed Integer Program}
 \acro{MISO}{Multiple-Input Single-Output}
 \acro{ML}{Maximum-Likelihood}
  \acro{MLB}{Mobility Load Balancing}
 \acro{MM}{Mobility Management}
 \acro{MME}{Mobility Management Entity}
 \acro{MMSE}{Minimum Mean Square Error}
 \acro{MNC}{Mobile Network Code}
 \acro{MNO}{Mobile Network Operator}
 \acro{MPC}{Multipath Components}
 \acro{MR}{Measurement Report}
 \acro{MRC}{Maximal Ratio Combining}
 \acro{MR-FDPF}{Multi-Resolution Frequency-Domain ParFlow}
 \acro{MRT}{Maximal Ratio Transmission}
 \acro{MRO}{Mobility Robustness Optimisation}
 \acro{MS}{Mobile Station}
 \acro{MSC}{Mobile Switching Center}
 \acro{MSISDN}{Mobile Subscriber Integrated Services Digital Network Number}
 \acro{MU}{Multi-User}
 \acro{MUE}{Macrocell User Equipment}
 \acro{NAS}{Non Access Stratum}
 \acro{NAS}{Non-Access Stratum}
 \acro{NAV}{Network Allocation Vector}
 \acro{NCL}{Neighbour Cell List}
 \acro{NGMN}{Next Generation Mobile Networks}
 \acro{NIR}{Non Ionisation radiation}
 \acro{NLOS}{Non Line Of Sight}
 \acro{nrtPS}{non-real-time Polling Service}
 \acro{NSS}{Network Switching Subsystem}
 \acro{NTP}{Network Time Protocol}
 \acro{ntp}{Network Time Protocol}
 \acro{NWG}{Network Working Group}
 \acro{OA}{Open Access}
  \acro{OAM}{Operation, Administration and Maintenance}
 \acro{OC}{Optimum Combining}
 \acro{OCXO}{Oven Controlled Oscillator}
 \acro{OFDM}{Orthogonal Frequency Division Multiplexing}
 \acro{OFDMA}{Orthogonal Frequency Division Multiple Access}
 \acro{OPEX}{Operational Expenditure}
  \acro{OSF}{Odd Subframe}
 \acro{OSI}{Open Systems Interconnection}
 \acro{OSS}{Operation Support Subsystem}
 \acro{P2MP}{Point to Multi-Point}
 \acro{P2P}{Point to Point}
 \acro{PAPR}{Peak-to-Average Power Ratio}
 \acro{PC}{Power Control}
 \acro{PCCH}{Paging Control Channel}
 \acro{PCCPCH}{Primary Common Control Physical Channel}
 \acro{PCH}{Paging Channel}
 \acro{PCI}{Physical Layer Cell Identity}
 \acro{PCPCH}{Physical Common Packet Channel}
 \acro{PCPICH}{Primary Common Pilot Channel}
 \acro{PDCP}{Packet Data Convergence Protocol}
 \acro{PDF}{Probability Density Function}
 \acro{PDSCH}{Physical Downlink Shared Channel}
 \acro{PDU}{Packet Data Unit}
 \acro{PeNB}{Pico eNodeB}
 \acro{PF}{Proportional Fair}
 \acro{PGW}{Packet Data Network Gateway}
 \acro{PhD}{Doctor of Philosophy}
 \acro{PHY}{Physical}
 \acro{PIC}{Parallel Interference Cancellation}
 \acro{PKI}{Public Key Infrastructure}
 \acro{PLMN ID}{PLMN Identity}
 \acro{PLMN}{Public Land Mobile Network}
 \acro{PML}{Perfectly Matched Layer}
 \acro{PN}{Pseudorandom noise}
  \acro{PP}{Ping-Pong}
 \acro{PRACH}{Physical Random Access Channel}
 \acro{PRB}{Physical Resource Block}
 \acro{PSC}{Primary Scrambling Code}
 \acro{PSTN}{Public Switched Telephone Network}
 \acro{PUE}{Picocell User Equipment}
 \acro{PUSC}{Partial Usage of Subchannels}
 \acro{QAM}{Quadrature Amplitude Modulation}
 \acro{QCI}{QoS Class Identifier}
 \acro{QoS}{Quality of Service}
 \acro{QPSK}{Quadrature Phase Shift Keying}
 \acro{RAB}{Radio Access Bearer}
 \acro{RACH}{Random Access Channel}
 \acro{RADIUS}{Remote Authentication Dial-In user Services}
 \acro{RAN}{Radio Access Network}
 \acro{RANAP}{Radio Access Network Application Part}
 \acro{RAT}{Radio Access Technology}
 \acro{RAXN}{Relay-Aided X Network}
 \acro{RB}{Resource Block}
  \acro{RCI}{Re-establish Cell ID}
 \acro{RE}{Range Expansion}
 \acro{REB}{Range Expansion Bias}
 \acro{RF}{Radio Frequency}
 \acro{RFP}{Radio Frequency Planning}
 \acro{RI}{Random Insertion}
 \acro{RLC}{Radio Link Control}
 \acro{RLF}{Radio Link Failure}
 \acro{RMa}{Rural Macro}
 \acro{RMSE}{Root Mean Square Error}
 \acro{RN}{Relay Node}
 \acro{RNC}{Radio Network Controller}
  \acro{RNL}{Radio Network Layer}
 \acro{RNS}{Radio Network Subsystem}
 \acro{RNTP}{Relative Narrowband Transmit Power}
 \acro{RPSF}{Reduced-power Subframes}
 \acro{RPLMN}{Registered PLMN}
 \acro{RR}{Round Robin}
 \acro{RRC}{Radio Resource Control}
 \acro{RRH}{Remote Radio Head}
 \acro{RRM}{Radio Resource Management}
 \acro{RS}{Reference Signal}
 \acro{RS-CS}{Resource-Specific Cell-Selection}
 \acro{RSQ}{Reference Signal Quality}
 \acro{RSRP}{Reference Signal Received Power}
  \acro{RSRQ}{Reference Signal Received Quality}
 \acro{RSS}{Reference Signal Strength}
 \acro{RTP}{Real Time Transport}
 \acro{rtPS}{real-time Polling Service}
 \acro{RTS}{Request-To-Send}
  \acro{SA}{Simulated Annealing}
 \acro{S1-AP}{S1 Application Protocol}
 \acro{S1-MME}{S1 for the control plane}
\acro{S1-U}{S1 for the user plane}
 \acro{SACCH}{Slow Associated Control Channel}
 \acro{SAE}{System Architecture Evolution}
 \acro{SAEGW}{System Architecture Evolution Gateway}
 \acro{SAIC}{Single Antenna Interference Cancellation}
 \acro{SAP}{Service Access Point}
 \acro{SAS}{Spectrum Allocation Server}
 \acro{SBS}{Super Base Station}
 \acro{SCC}{Standards Coordinating Committee}
 \acro{SCCPCH}{Secondary Common Control Physical Channel}
 \acro{SCFDMA}{Single Carrier FDMA}
 \acro{SCH}{Synchronisation Channel}
 \acro{SCM}{Spatial Channel Model}
 \acro{SCP}{Single Cell Processing}
 \acro{SCTP}{Stream Control Transmission Protocol}
 \acro{SDCCH}{Standalone Dedicated Control Channel}
 \acro{SDMA}{Space-Division Multiple-Access}
 \acro{SDR}{Software Defined Radio}
 \acro{SDU}{Service Data Unit}
 \acro{SE}{Spectral Efficiency}
 \acro{SFID}{Service Flow Identifier}
 \acro{SFBC}{Space Frequency Block Coding}
 \acro{SG}{Signalling Gateway}
 \acro{SGSN}{Serving GPRS Support Node}
 \acro{SGW}{Serving Gateway}
 \acro{SI}{System Information}
 \acro{SIB}{System Information Block}
 \acro{SIB1}{SystemInformationBlockType1}
 \acro{SIB4}{SystemInformationBlockType4}
 \acro{SIC}{Successive Interference Cancellation}
 \acro{SIGTRAN}{Signalling Transport}
 \acro{SIM}{Subscriber Identity Module}
 \acro{SIMO}{Single Input Multiple Output}
 \acro{SN}{Serial Number}
 \acro{SINR}{Signal to Interference plus Noise Ratio}
 \acro{SIP}{Session Initiated Protocol}
 \acro{SISO}{Single Input Single Output}
 \acro{SLAC}{Stochastic Local Area Channel}
 \acro{SLNR}{Signal to Leakage Interference and Noise Ratio}
 \acro{SLS}{System-Level Simulation}
 \acro{SMS}{Short Message Service}
 \acro{SNMP}{Simple Network Management Protocol}
 \acro{SNR}{Signal to Noise Ratio}
 \acro{SOCP}{Second-Order Cone Programming}
 \acro{SOHO}{Small Office/Home Office}
 \acro{SON}{Self-Organising Network}
 \acro{SoT}{Saving of Transmissions}
 \acro{SPS}{Spectrum Policy Server}
 \acro{SSL}{Secure Socket Layer}
 \acro{SSMA}{Spread Spectrum Multiple Access}
  \acro{SUE}{Small Cell User Equipment}
 \acro{SUI}{Stanford University Interim}
  \acro{TA}{Timing Advance}
 \acro{TAC}{Tracking Area Code}
 \acro{TAI}{Tracking Area Identity}
 \acro{TAU}{Tracking Area Update}
 \acro{TCH}{Traffic Channel}
 \acro{TCP}{Transmission Control Protocol}
 \acro{TCXO}{Temperature Controlled Oscillator}
 \acro{TDD}{Time Division Duplexing}
 \acro{TDM}{Time Division Multiplexing}
 \acro{TDMA}{Time Division Multiple Access}
 \acro{TEID}{Tunnel Endpoint Identifier}
 \acro{TLS}{Transport Layer Security}
 \acro{TNL}{Transport Network Layer}
 \acro{TP}{ThroughPut}
 \acro{TPC}{Transmit Power Control}
 \acro{TPM}{Trusted Platform Module}
 \acro{TR}{Transition Region}
 \acro{TS}{Tabu Search}
 \acro{TSG}{Technical Specification Group}
 \acro{TTG}{Transmit/Receive Transition Gap}
 \acro{TTI}{Transmission Time Interval}
  \acro{TTT}{Time-to-Trigger}
 \acro{TV}{Television}
 \acro{Tx}{Transmission}
  \acro{TU}{Typical Urban}
  \acro{TWXN}{Two-Way Exchange Network}
 \acro{UARFCN}{UTRA Absolute Radio Frequency Channel Number}
 \acro{UDP}{User Datagram Protocol}
 \acro{UE}{User Equipment}
 \acro{UGS}{Unsolicited Grant Service}
 \acro{UICC}{Universal Integrated Circuit Card}
 \acro{UK}{United Kingdom}
 \acro{UL}{Uplink}
 \acro{UMa}{Urban Macrocell}
 \acro{UMA}{Unlicensed Mobile Access}
 \acro{UMTS}{Universal Mobile Telecommunication System}
 \acro{US}{UpStream}
 \acro{USIM}{Universal Subscriber Identity Module}
 \acro{UTD}{Theory of Diffraction}
 \acro{UTRA}{UMTS Terrestrial Radio Access}
 \acro{UTRAN}{UMTS Terrestrial Radio Access Network}
 \acro{UWB}{Ultra Wide Band}
 \acro{VD}{Vertical Diffraction}
 \acro{VDFP}{Vertical Dynamic Frequency Planning}
 \acro{VeNB}{Virtual eNB}
 \acro{VLR}{Visitor Location Register}
 \acro{VoIP}{Voice over IP}
 \acro{VPLMN}{Visited PLMN}
 \acro{VR}{Visibility Region}
 \acro{WCDMA}{Wideband Code Division Multiple Access}
 \acro{WEP}{Wired Equivalent Privacy}
 \acro{WG}{Working Group}
 \acro{WHO}{World Health Organisation}
 \acro{WiFi}{Wireless Fidelity}
 \acro{WiMAX}{Wireless Interoperability for Microwave Access}
 \acro{WNC}{Wireless Network Coding}
 \acro{WLAN}{Wireless Local Area Network}
 \acro{WMAN}{Wireless Metropolitan Area Network}
 \acro{WRAN}{Wireless Regional Area Network}
 \acro{X2-AP}{X2 Application Protocol}
\end{acronym}
%
%\clearpage
\end{document}